\documentclass[twoside,12pt]{article}
\usepackage{epsfig}
\usepackage{amsmath,amssymb}
\usepackage{color}
\usepackage{hyperref}

\newcommand{\be}{\begin{equation}}
\newcommand{\ee}{\end{equation}}
\newcommand{\bea}{\begin{eqnarray}}
\newcommand{\eea}{\end{eqnarray}}

\begin{document}

\title{ \vspace{1cm} Hadron matter in neutron stars\\ 
in view of gravitational wave observations.}

\author{Felipe J. Llanes-Estrada and Eva Lope-Oter,\\ 
Departamento de F\'{\i}sica Te\'orica and \\ Institute of Physics of Particles and the Cosmos (IPARCOS),\\ 
Universidad Complutense de Madrid, 28040, Spain.
}
\maketitle
\begin{abstract}
In this review we highlight a few physical properties of neutron stars and their theoretical treatment
inasmuch as they can be useful for nuclear and particle physicists concerned with matter at finite density (and newly, temperature). \\
Conversely, we lay out some of the hadron physics necessary to test General Relativity with binary mergers including at least one neutron star, in view of the event GW170817: neutron stars and their mergers reach the highest matter densities known, offering access to the matter side of Einstein's equations.\\
In addition to minimum introductory material for those interested in starting research in the field of neutron stars, we dedicate quite some effort to a discussion of the Equation of State of hadron matter in view of gravitational wave developments; 
we address phase transitions and how the new data may help;
we show why transport is expected to be dominated by turbulence instead of diffusion through most if not all of the star, in view of the transport coefficients that have been calculated from microscopic hadron physics; and we relate many of the interesting physics topics in neutron stars to the radius and tidal deformability.
\end{abstract}

\newpage
\tableofcontents
\vspace{2cm}
\newpage

\section{Introduction: learning from gravitational waves}

\subsection{Motivation}
This succinct review is meant to give a manageable overview of developments in neutron star physics 
following the first binary neutron star merger event, GW170817, detected by Gravitational Wave observatories
and across the electromagnetic spectrum. The field has become vast as attested by the original discovery letter~\cite{TheLIGOScientific:2017qsa} recording about two thousand citations in a year and a half. Reviewing this body of literature even superficially is an effort unattainable to many interested colleagues in nuclear and particle physics, and graduate students in their groups, that could contribute to many aspects relevant for current discussions but require a quick overview of the much recent progress.
Thus, we have accepted the challenge of the journal's editor to compile, to the best of our ability, a bundle of the many open threads that lead to advanced research in the field. 

Our intention is not to be exhaustive; such would be an impossible endeavour now (80 pages of the journal would be filled just with the references). We have selected a number of topics that clearly cross-pollinate nuclear and particle physics with astrophysics, but sometimes the omission or inclusion of a certain piece of physics is a matter of taste. We do hope the document will be useful to bring interested colleagues to speed. Of particular note is that, with the exception of a handful of classic works, when faced with the choice of an older or a newer reference we have quoted the newest (up to a cutoff date on July 2019). 

We do not intend to be rigorous either; we have rather attempted that the entire review be pedagogically written, including several relevant figures from the literature and modifying them when necessary to note facts or interpretations that are obvious to practitioners but need to be pointed out to outsiders.

What we however want to convey to the reader is our enthusiasm for a truly multidisciplinary field where new, fundamental physics is being probed and where colleagues educated in nuclear and particle physics can find ample room to pursue their scientific interests.

The rest of this introduction is dedicated to understanding a typical GW signal (subsection~\ref{subsec:anatomy}),
to mention some efforts trying to constrain General Relativity with neutron stars (subsection~\ref{subsec:fR}) which is a motivation for later studies of these systems from hadron physics; and finally, to inquire into how many merger events one should hope to collect (subsection~\ref{subsec:howmany}) which will impact the ultimate precision of the measurements made. We think this is the thrust that is making the field evolve rapidly.

In section~\ref{sec:static} we briefly review selected static observables of neutron stars, a decades--old effort to show the strength of the field even before the appearance of the merger event. This is the most basic material on top of which the new developments are being constructed.

We have divided section~\ref{sec:static} in four parts. Subsection~\ref{subsec:TOV} addresses the basic theoretical treatment of static neutron stars through the Tolman--Oppenheimer-Volkoff equations. Then we proceed to two classical experimental aspects, the distribution and extremes of neutron star masses (subsection~\ref{subsec:NSmasses}), constraints on neutron star radii (subsection~\ref{subsec:radii}) and finally, we discuss the newly assessed tidal deformability (subsection~\ref{subsec:tidal}), the static-like property that can be read off from binary mergers.

The heart of the manuscript is section~\ref{sec:EoS}, dedicated to the Equation of State of cold hadron matter. This is the most straightforward and important point of contact between Hadron Physics and General Relativity and Astrophysics.
In addition to general principles and a connection to laboratory observables on Earth, this section concentrates on reporting model independent information from hadron physics and QCD in subsection~\ref{subsec:limits} so that it can be used with confidence in astrophysical applications. A short discussion on small--temperature extensions has been appended at the end of  section~\ref{sec:EoS}.

We have then dedicated an independent section~\ref{sec:Phases} to a contained discussion of possible phase transitions in cold hadron matter. While much of the discussion is a continuation of section~\ref{sec:EoS} if one is just interested in a tabulated equation of state of hadron matter to use in a relativity application, we feel that the critical importance of this topic for the nuclear and hadron physics community deserves separate treatment. 
The last major piece of the work is section~\ref{sec:otros} where we have collected several other areas of research currently pursued by the neutron star community. The choice of specific topics has been driven by our wish to connect hadron physics (hence a discussion of transport, damping and cooling) with astrophysics and the possibility of detecting further types of gravitational waves (thus, the discussion of the vibration modes of the star, and the related discussion of rotation phenomena that lead to a time--varying matter quadrupole).

An outlook recapitulation and two short appendices complete the manuscript.

Among the many topics that we have chosen to leave out is a summary of the historical development of the current paradigm on neutron star physics: a short walk through this development and some additional topics is to be found in~\cite{Vidana:2018lqp}).

     \subsection{Anatomy of a GW signal in a NS binary merger}\label{subsec:anatomy}

The discovery of neutron star mergers came about with the gravitational--wave event GW170817.
Therefore, the first task we should address is to illustrate the meaning of this new data; but we face a difficulty in that a good analysis of the signal for that neutron star--neutron star merger event has not been provided. Therefore, we delay discussing that particular event a couple of paragraphs, while we show 
the physical content of a gravitational--wave signal with one that has been very clearly reported and explained in the literature.

\begin{figure}
\begin{center}
\includegraphics[width=0.8\columnwidth]{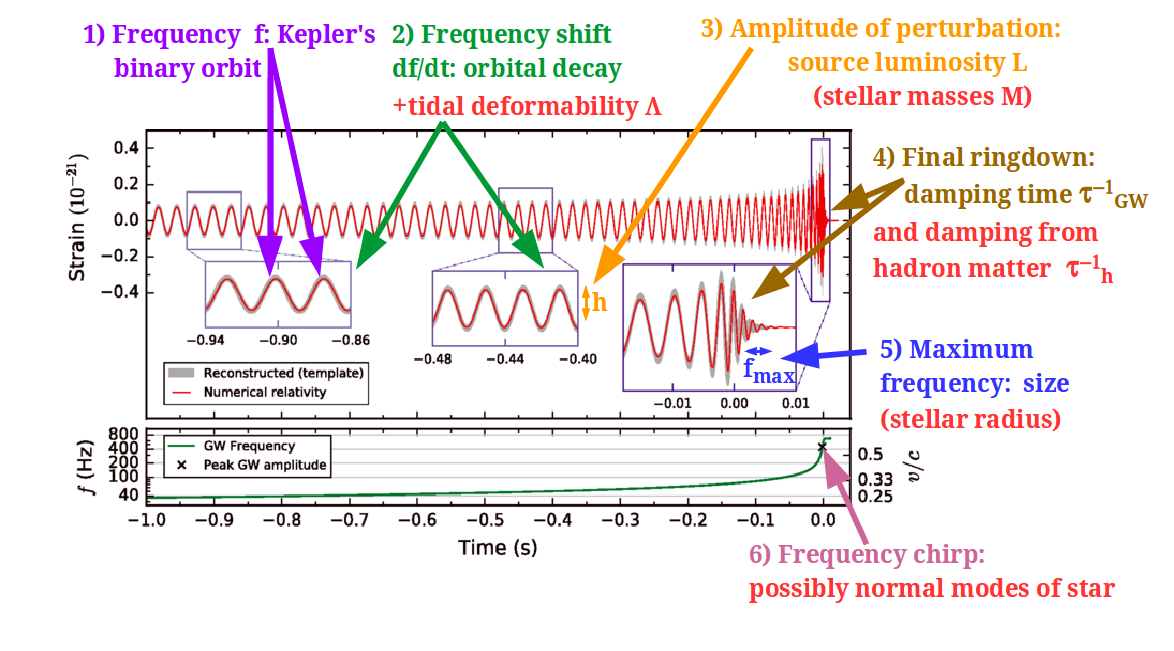}
\caption{\label{fig:GWpulse}
Various pieces of information that can be (or could be) extracted from gravitational wave signals of double neutron-star mergers. (Here, the actual signal corresponds to a black hole-black hole merger event, since cleaned NS-NS signal shapes have not, to our knowledge, been publicly released).
\emph{Figure adapted (and modified) from~\cite{Abbott:2016nmj} under the terms of the} \href{https://creativecommons.org/licenses/by/3.0/}{\tt Creative Commons 3.0 License}.}
\end{center}
\end{figure}

The second gravitational wave signal ever detected, that of GW151226, is shown in figure~\ref{fig:GWpulse}. This event had no electromagnetic counterpart~\footnote{The GBM camera on board of the Fermi satellite claimed a $\gamma$-ray burst associated to GW150917 that no other experiment detected; accepting it deteriorates the limits that can be imposed on emission power by BH-BH events~\cite{Perna:2019pzr}. Still, the neutron--star merger remnant is supposed to glow in a solid angle larger than any jets along the axis perpendicular to the merger that a Black-Hole pair might emit.} and by the sheer mass of the objects involved (in solar masses of about $2\times 10^{30}$kg, $\sim 7$ and $\sim 14$ $M_\odot$ respectively) it was early-on identified to have been produced by two colliding black-holes (neutron stars in general relativity are believed to have a maximum mass somewhat about $2M_\odot$, see subsection~\ref{subsec:NSmasses}).

Circulating clockwise from the top-left corner of the figure we find~\cite{Llanes-Estrada:2016oom}  the following notes (in red online, those of special interest for hadron physics).
\begin{enumerate}
\item The frequency of the GW signal detected at the interferometer: it is twice the orbital frequency, $\color{black}f\color{black}=2f_{\rm Kepler}$ (because the radiation is quadrupolar, a 180$^{\rm o}$ half-turn rotation of the source does not affect the pulse as the quadrupole $Q$ is left invariant) and just informs us of the two-body orbital problem.
\item The frequency decrease $\color{black} \frac{df}{dt}\color{black}$ over many periods 
(that revealed early on the existence of GWs in the Hulse-Taylor PSR B1913+16 binary pulsar): 
this can be used to reconstruct the binary's ``chirp'' mass, 
\be \label{chirpmass}
{\mathcal M} = \left( \frac{(m_1m_2)^3}{m_1+m_2} \right)^{1/5}
            = \frac{c^3}{G} \left(\frac{5\pi^{-8/3}}{96} \color{black} f^{11/3} \frac{df}{dt} 
\color{black} \right)^{3/5}  \ .
\ee
This chirp mass, the best measured combination of the masses $M_1$ and $M_2$ of the merging objects, provides an absolute normalization (unlike the reconstruction from Kepler's orbital problem yielding $m_1+m_2)$) that allows reconstruction of the individual masses. Additionally, the loss of orbital energy provides us with a GW luminosity calibration.
Finally, separations of the phase from the numerical simulation for the inspiral of two point masses, allows in principle to discern their structure; for NS events, the first number is the tidal deformability $\Lambda$ discussed in subsection~\ref{subsec:tidal} below.
\item The amplitude at the detector, $\color{black}h\color{black}\sim 10^{-21}$, gives the distance to the GW source because the luminosity $L$ is known from the orbital decay. This BH-BH merger happened at $440\pm190$ Megaparsec. It also assists in the reconstruction of the two masses, which helps the identification of possible NS mergers. 
\item The final ringdown in GW150914 was attenuated with characteristic time $\color{black}\tau\color{black} \simeq 4$ milliseconds (commensurate with the Schwarzschild-radius expressed as a light-crossing time, $2\times 370 $km$/c\simeq 2.5$ ms). 
In this BH event, the damping is caused by the emission of gravitational waves; but in NS mergers,  viscous damping might be detected, so that $\tau^{-1}=\tau^{-1}_{GW}+\tau^{-1}_{\eta}$. Transport coefficients in neutron stars are very briefly discussed in subsection~\ref{subsec:transport}.
\item Once the two objects have merged and the final ringdown occurs, the maximum signal frequency $\color{black} f_{\rm max} \color{black}$ roughly reveals the size of the resulting compound (for orientation, a relativistic ellipsoid can spin at a rate consistent with  $2\pi R_{\rm NS}f_{\rm max}\sim c$; for $R\sim10-100$ km, $f_{\rm  max}$ is in the audible kHz frequency range).
\item Finally, detailed analysis of the last instants before merger might be able to detect the normal-mode oscillations of the star (saliently, the $f-$ and possibly $p-$, $w-$modes discussed in subsection~\ref{subsec:vibration}) as the strong tidal forces can transfer energy from the orbital potential to stellar vibrations, thus accelerating the merger.
\end{enumerate}

This last point deserves an additional comment. In studying the nucleon, accelerator probes have played a large role by penetrating increasingly deeper layers of their structure; they are complementary to bound state studies in which the nucleon is bound, for example, in a deuteron. In neutron stars, the analogous ways of exciting various vibration modes of the star~(see subsection \ref{subsec:vibration}) are binary systems, especially if highly eccentric ones can be found~\cite{Chirenti:2016xys}) and the impact of accreted material in the surface. The analogy is reversed respect to microscopic studies of the nucleon: the "collision" of accreted material only excites superficial modes of the star, whereas the final collapse of a binary system provides information about deeper layers.

\begin{figure}
\begin{center}
\includegraphics[width=0.4\columnwidth]{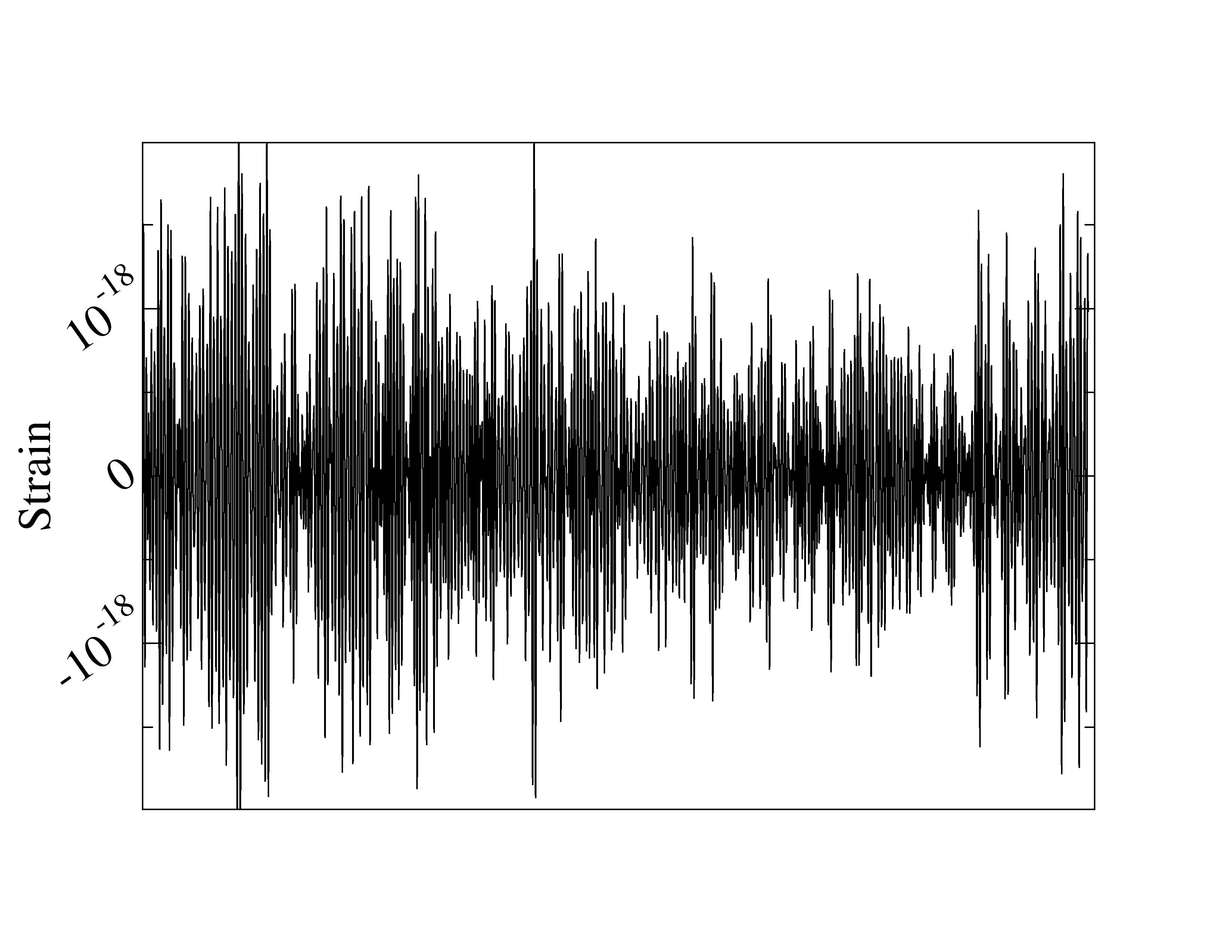}
\hspace{2cm}
\includegraphics[width=0.4\columnwidth]{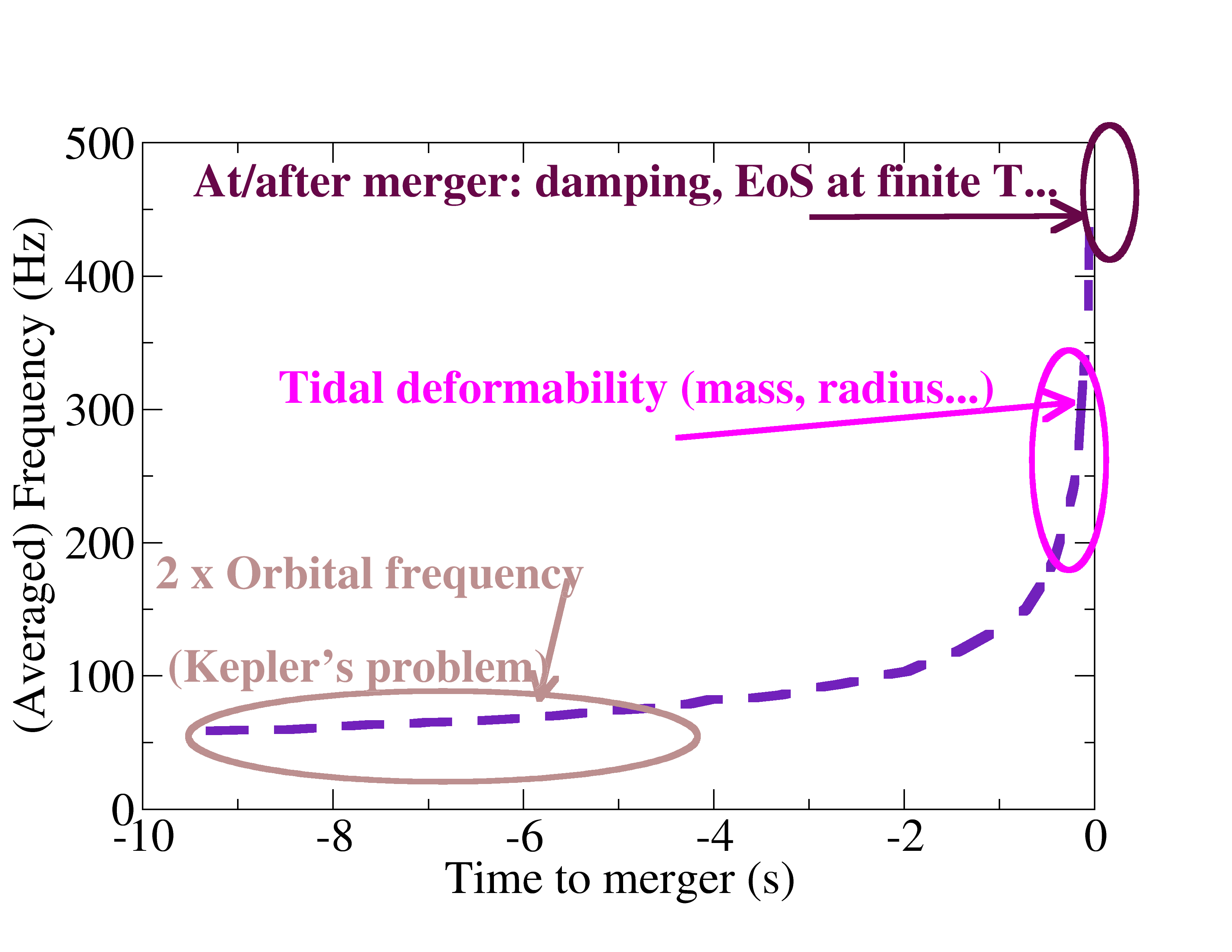}
\caption{\label{fig:chirp}
Left: example data for the best candidate of a binary neutron-star merger, GW170817 over 32 seconds at 4 kHz.
Right: time-frequency map signal (data from~\cite{Monitor:2017mdv}).  All the information about neutron star physics is concentrated in the last moments or ``chirp'' when the two stars come close and finally merge.}
\end{center}
\end{figure}

The actual analysed signal shape of the NS merger event GW170817 is not easily found, but the raw data from 
{\tt https://www.gw-openscience.org/catalog/GWTC-1-confident/single/GW170817/ } is re-plotted in figure~\ref{fig:chirp} (for the frequency band of 4 kHz and the 32 second most relevant period).

The pulse shape $h(t)$ is compared to a theoretical waveform $h^{\rm th}(t;\theta_i)$ where the $\theta_i$ are a set of parameters. They include~\cite{Choi:2018zbi} intrinsic dynamical parameters from the binary system that control the frequencies detected, and extrinsic parameters (distance to the source, sky position, inclination of the orbital axis respect to Earth...) that affect the amplitude but not the signal shape (see fig.\ref{fig:GWpulse} above). 
The best match to the parameter set is found by  maximizing the likelihood function
\begin{equation}
L(x\arrowvert \{\theta_i\}) = e^{-<h-h^{\rm th}\arrowvert h-h^{\rm th}>}
\end{equation}
in terms of the scalar product (calculated from the Fourier transforms) weighted with the detector's noise power spectral density $S(f)$,
\begin{equation}
<x|y> = 2 \int df \frac{
\hat{x}^*(f)\hat{y}(f) +\hat{x}(f)\hat{y}^*(f) 
}{S(f)}\ .
\end{equation}

As that data in the left plot of figure~\ref{fig:chirp} is not particularly informative to the eye, we give in the same figure the frequency of the GW170817 signal as function of time (averaged over small time intervals; strictly speaking, $f(t)$ cannot be plotted because $f$ and $t$ are Fourier conjugate variables, so an uncertainty principle applies). Several notes remind us of what the different intervals can teach us.

This renowned data has stimulated interest in neutron star physics. There are two avenues of research that can be pursued with it, in combination with all the other extant NS observables. The bulk of the community is assuming that General Relativity is the correct theory of gravity, also under extreme NS conditions, and has therefore taken on using the astrophysical data to constrain hadron physics quantities. Most of this article is dedicated to this interplay between both fields.

Alternatively, one can use the state of the art theory predictions from hadron physics in combination with the astrophysical data to constrain modifications of the theory of General Relativity itself. As this is a less extended approach, we will deal with it first in the next subsection~\ref{subsec:fR}.

\subsection{Constraints on modifications of General Relativity}\label{subsec:fR}

\subsubsection{Models of gravity beyond GR}

General Relativity is very well tested at Solar System scales; and if it should come successfully out of modern cosmological tests, then the balance of the matter content of the universe is of a hitherto unknown form. The possibility that it is the theory itself that needs to be modified has maintained interest in studies of theories of gravity beyond General Relativity. Dark energy, dark matter, the sticky issue of quantizing gravity, or the prediction of spacetime singularities, are often quoted unsatisfactory features of GR motivating work in alternative schemes (massive gravity, scalar-tensor and $f(R)$ theories, Chern-Simons, Horava-Lifschitz and many others, see~\cite{Joyce:2016vqv}). 

Moreover, early universe is believed to have undergone a rapid phase of expansion, ``inflation''; and after  Planck's satellite data~\cite{Ade:2015lrj}, one of the theories thereof that remains in agreement with its observations of CMB mode polarisation is Starobinsky's. This enhances the Hilbert-Einstein Lagrangian of General Relativity from $R$ to $R+aR^2$ (the Gravity Probe B constraint on $a$ is quite weak, $a<5\times 10^5$km$^2$). Further generalising this expression, the community is actively investigating  $f(R)$ theories.
In a Laurent series expansion of $f$, the quadratic $R^2$ term and higher order ones become important at higher field, and neutron stars are extremely compact objects, not far from the Schwarzschild radius, so that their fields are extreme, surpassed only at black holes. 
Though the observational astronomy of black holes is making great strides and the recent detailed imaging of a galactic one~\cite{Akiyama:2019cqa,Akiyama:2019eap} offers the possibility of designing large--field tests of General Relativity, neutron stars remain the only system where Einstein's equations can be checked in the presence of a significant density of energy-matter, so that both sides of $G_{\mu\nu}=8\pi G T_{\mu\nu}$ can be simultaneously addressed.
For example, a family of theories in which the Einstein equations are modified at the level of the matter tensor $T$, known as $f(T)$ theories, is also under study; for these, neutron stars offer the densest possible matter and hence highest $T$.

Pulsar orbital measurements are a staple of GR tests; for example, those in the J0348+0432 2$M_\odot$ pulsar with a white dwarf companion~\cite{Antoniadis:2013pzd} constrain the scalar-tensor coupling ratio of Brans-Dicke theories from the measured orbital decay frequency shift $df/dt$.  But again, the exterior tests are not sensitive to the $T^{\mu\nu}$ part of Einsteins equation.

\begin{figure}[h]
\begin{center}
\includegraphics[width=7cm]{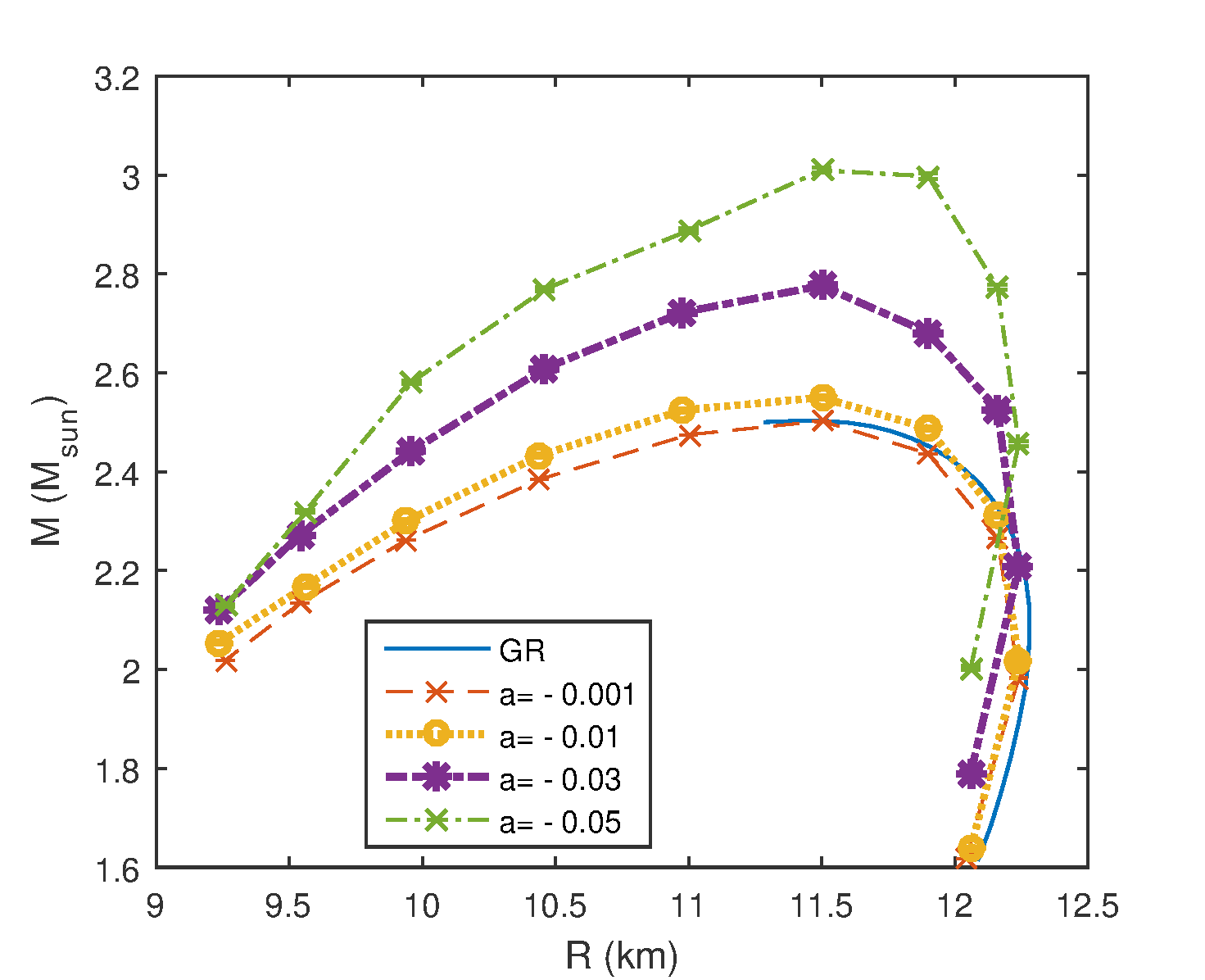}
\caption{\label{fig:Resco} Given the neutron matter EoS, the maximum mass of a neutron star can take different values depending on the parameter $a$ extending General Relativity in $f(R)$ theories, here $R+aR^2$~\cite{Resco:2016upv}.}
\end{center}
\end{figure}

For example, in General Relativity, there is a maximum mass that any neutron star can take. This and the mass-radius diagram will be discussed shortly in section~\ref{sec:static}. But in different  modified theories of gravity the maximum can be different, or there can be no maximum at all~\cite{Resco:2016upv}, as shown in figure~\ref{fig:Resco}, since one can slide a new parameter (for example, the $a$ of $R+aR^2$) to meet an arbitrary mass. (See figure 1a of~\cite{Carvalho:2019ert} for a similar computation within a more generic $f(R,L_{\rm matter})$ coupling the matter Lagrangian in a nonminimal way.)

Likewise, the parameter $a$ of $R+aR^2$ theory can be constrained with further data from the maximum mass measured in gravitational wave events  (see subsection~\ref{subsec:NSmasses}),  following~\cite{Astashenok:2018iav}, or from the threshold mass for collapse of the postmerger object (see table II in~\cite{Astashenok:2018iav}).

Actually, the definition of these masses as seen from an observer at infinity is quite technical~\cite{Resco:2016upv,Yazadjiev:2014cza,Yazadjiev:2015xsj,Capozziello:2015yza} in the modified theories of gravity as the solution in the Starobinsky model is slightly different from being asymptotically flat~\cite{Yu:2017uyd} and matching the Schwarzschild solution is far from trivial~\cite{Yazadjiev:2014cza}. (A recent preprint~\cite{Sbisa:2019mae} opts for using the quantity of baryonic matter as the gravitational mass even in modified gravity theories, but it is not clear to us how it can be disentangled from observations outside the star.)

Current bounds~\cite{Naf:2010zy} are not so constraining, $a<10^3\times (10{\rm km})^2$ (10km being the characteristic neutron star scale). And it appears that current gravitational wave detectors will have a hard time extracting a meaningful bound unless the uncertainty in the EoS lowers quite significantly~\cite{Yazadjiev:2018xxk}.

Another example of modified gravity theory where neutron stars can be analyzed is Hybrid metric-Palatini gravity with $f(R)$~\cite{Danila:2016lqx}.

The available energy that can be emitted as gravitational radiation can be very different from GR in $f(R)$ (and part of the radiation can be emitted in the scalar mode). To our knowledge however,  realistic waveforms to compare with aLIGO have not yet been constructed.

Additional very well known tests coming from the propagation of gravitational waves from their source to Earth include constraints on $f(R)$ or scalar-tensor theories of gravity from the modified dispersion relations, constraints on the number of space-time dimensions, or on whether grav. waves have additional polarizations; but these are of no concern to us as they do not involve the close proximity of the neutron star, of interest for hadron physics.

\subsubsection{Postnewtonian expansion}

\begin{figure} \begin{center} 
\includegraphics[width=10cm]{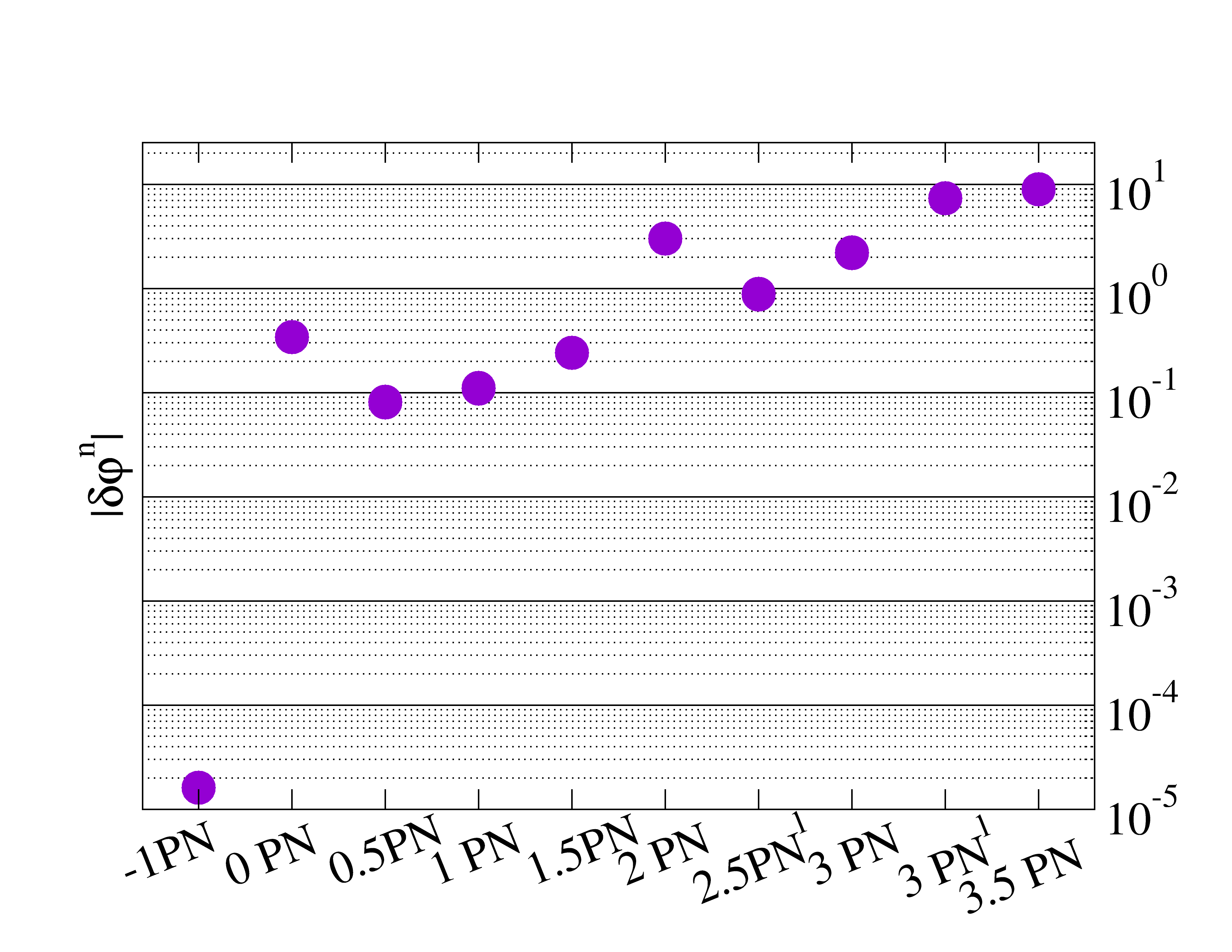}
\caption{\label{fig:postNewtonian}
Upper bounds (at 90\% confidence) on the deviations of the postNewtonian coefficients
parametrizing separations from General Relativity from analysis of GW170817~\cite{Abbott:2018lct}.
}
\end{center}
\end{figure}

In the philosophy of Effective Field Theories, one can, instead of model theories different from General Relativity, constrain the variations of the Lagrangian around the actual theory by constructing the possible operators allowed by symmetry and assign them a counting to decide which ones to keep and which ones to discard at a given order. A popular approach is to use a postnewtonian expansion around the Newtonian prediction for any observable, expanding in powers of $v/c$ (or, equivalently, field intensity). General relativity then predicts a specific combination of the postnewtonian coefficients $\varphi_n$ parametrizing those deviations to $O(v/c)^n$, and it is the task of observation to constrain variations $\delta\varphi_n$ respect to those values.
Figure~\ref{fig:postNewtonian} shows precisely such bounds on the deviations of postNewtonian coefficients from their General Relativity values imposed on modifications of GR by study~\cite{Abbott:2018lct} of the GW170817 neutron star merger event.
The bounds reported are competitive with the earlier ones from the Black Hole-BH merger GW150914~\cite{TheLIGOScientific:2016src} and other events, even combined. For example the bound on the coefficient "1PN" (first postnewtonian order) has gone down from $\sim 1000$ before aLIGO, to $\sim 1$ after the black hole mergers from 2015, to around $0.1$ with the neutron star merger of 2017. 
The coefficients that have not improved are those of order 0PN or less, that actually are larger at smaller velocities, $\phi_{-n} (v/c)^{-n}$. Neutron stars are just not sensitive to modifications of General Relativity of interest for large distances, including late--time cosmology.
The General Relativity value $\delta\hat{\varphi}_n=0$ is within the 90\% credibility value for all but the orders 3 and 3.5.
 
To produce those bounds, the collaboration generated numerical waveforms varied by replacing each of the postNewtonian coefficients $\varphi$ in General Relativity to $(\hat{\varphi}_n+
\delta\hat{\varphi}_n)$, one at a time, while allowing for the masses, spins and external parameters of the NS-NS system to vary in a multiparameter space, to best fit the experimental signal.

An important point and a thrust of our own research program is that the community constraining 
General Relativity are feeding into their simulations of neutron stars and binary mergers 
Equations of State and other microscopic properties in whose construction General Relativity has played a role (for example, if an EoS is forced to yield neutron star masses within astrophysical constraints, GR has been used to cut off parts of the hadron parameter space). Thus, testing theories of gravity requires feedback from hadron physics that is free of circular reasoning and where no astrophysical input related in any way to GR has been assumed~\footnote{Take~\cite{Carson:2019rjx} as an example: some EoS--quasiinsensitive relations are proposed to test GR, but the EoS used to establish the relations already assume GR in their using the 2$M_\odot$ neutron stars and GW170817.}.
We have setup a website providing hadron physics EoS that use the best available information in the literature from earthly laboratories and theory alone~\cite{Oter:2019kig}, in {\tt http://teorica.fis.ucm.es/nEoS/}\ .

In concluding this subsection, we would like the reader to consider that the EoS is rather well constrained at the scale of nuclei, and the extrapolation (in density) to the neutron star interiors extends a factor of 1.5-5. On the other hand, outside the star (where binary pulsar measurements constrain GR directly), $g\sim 300$m/s$^2$; inside white dwarfs, $g\sim 10^6$m/s$^2$; and inside the neutron star itself, $g\sim 10^{12}$m/s$^2$. That is, in proceeding to a neutron star interior, gravity needs to be extrapolated several orders of magnitude respect to observations in other systems.
Ideally, in the not too distant future, perhaps both EoS and GR can simultaneously be constrained.
But for now it seems as sensible, if not more, to extrapolate hadron physics over less than an order of magnitude to test GR, as extrapolating gravity over 10 and 6 orders of magnitude, respectively, to test QCD; only upcoming Black Hole tests at high field will level the field.

\subsection{How many events should we expect?} \label{subsec:howmany}

Before the detection of GW170817 it was considered possible that a double neutron-star merger would be found within the first three aLIGO runs~\cite{Baiotti:2016qnr}. At the time of Confinement XII~\cite{Llanes-Estrada:2016oom}, none had been detected to a distance of 70 Megaparsec (100 MPc in the case of merging NS-BH). 
Various studies were predicting between 0.2 and 200 NS-NS merger detections per year of aLIGO operation.
Finally, the second run of aLIGO, jointly with VIRGO, produced precisely one undoubted detection, that of GW170817  (out of some 8 certain GW events of astrophysical origin in that same run; the third run has already broadcast several additional alerts).

Once the associated electromagnetic signal (saliently the $\gamma$-ray burst) has been well understood, similar ones are been searched in gamma-ray databases; for example~\cite{Troja:2018ybt}, GRB150101B does look like the later blue kilonova event (but at a cosmological distance larger than 600 Mpc, the gravitational wave signal would be too small to detect). This event happened in a young galaxy with mean stellar age of $2^{+6}_{-1}$ Gyr.

To predict how many events are to be collected, simple algebra
\begin{equation}
N_{\rm merger} = T_{\rm operation}\times d_{\rm reach}^3 \times {\mathcal R}
\end{equation}
indicates that the key quantity is $\mathcal R$, the rate of detection per unit volume. In this formula, the reach $d_{\rm reach}$ of the aLIGO third run O3, simultaneous to observations by Virgo and GEO600, is expected to be 50\% larger than in the previous run, up to short of 170 Mpc, yielding a volume of 0.02 GPc$^3$.

Estimating the rate $\mathcal R$ requires knowledge of the number of neutron stars; the proportion of them in binary systems; and how tight these systems are so that merger happens in a Hubble time. The uncertainties are above an order of magnitude: the collapse leading to a neutron star formation is generically asymmetric, launching the core with a ``natal kick'' that can eject it from a binary system. And if the Roche lobe transfering mass between both stars (which is not corotating with them) is not ejected, a clear NS-NS binary system may not form.

Additional considerations that impact whether these events may be detectable as GW-events and well identified are the spatial distribution of the binary systems; the probability of the beam to be pointed towards Earth; the luminosity distribution of the mergers; any decrease of that luminosity due, for example, to Doppler smearing; and the luminosity of the host galaxies (for the EM counterpart to be detected and the source localized).

Now, our galaxy should contain $O(10^8-10^9)$ pulsars  (yielding a cosmological number density of perhaps 0.055/MPc$^3$),  of which 2300 are already known; some 18 are binary systems and up to a dozen are said to have orbital periods of order hours (so even if isolated they might merge in the lifetime of the universe). A recent account~\cite{Pol:2018shd} reduces the useful sample to 8.
It could help to know how many binary collisions there were in the Milky Way in, {\it e.g.}, the last million years. This could be found~\cite{Wu:2019xrq} by searching for sources with identified $^{126}$Sn, a heavy isotope with adequate lifetime of $\sim 10^5$ yr and near the nuclear r--process peak that is active in these mergers.

The binary systems relative distance distribution is described by a power law $dN/da\propto a^{-\beta}$~\cite{Chruslinska:2018ylm} and the distribution of times to merger follows as $dN/dt_{\rm merger}\propto t_{\rm merger}^{-(\beta+3)/4}$, the later exponent typically being -1 to -1.5

\begin{figure}
\begin{center}
\includegraphics[width=8cm]{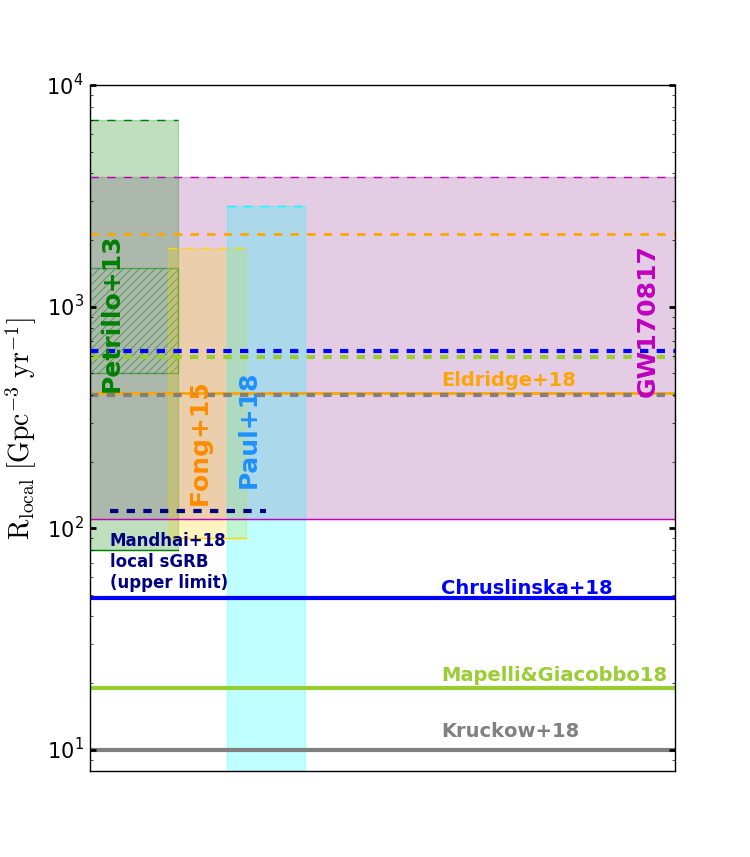}
\caption{\label{fig:numberofevents}
Expected binary neutron-star merger rate (number of events per unit time -year- and unit volume -cubic Gigaparsec-) updating the data from~\cite{Chruslinska:2018ylm} with the latest estimates from Gravitational Wave observations~\cite{LIGOScientific:2018mvr} and from short $\gamma$-ray bursts~\cite{Paul:2017rof,Mandhai:2018cdl}. 
\emph{Figure credit: personal courtesy of Martyna Chruslinska, unpublished.}}
\end{center}
\end{figure}

As an example, the detection rate per unit volume found for LIGO in~\cite{Pol:2018shd}  (at 90\% level)
and their update of earlier work by others~\cite{Kim:2013tca} is
\begin{equation}
{\mathcal R} = 0.18^{+0.13}_{-0.06}\times \left( \frac{1}{100 \rm Mpc}  \right)^3 yr^{-1}
\end{equation}
\begin{equation}
{\mathcal R} = 0.09^{+0.12}_{-0.06}\times \left( \frac{1}{100 \rm Mpc}  \right)^3 yr^{-1}\ .
\end{equation}
For the one-year observing time of run O3, these numbers imply that the detector needs to reach full sensitivity and reach out to 170 MPc to have a fair chance of finding an additional NS-NS merger.
In fact, looking retrospectively, with $d_{\rm reach}\sim 70$ Mpc, detecting GW170817 in the second run O2 possibly was a matter of good luck.
This is seen in figure~\ref{fig:numberofevents} where good modern calculations (for example, the blue line marked ``Chruslinska 18'') suggest that GW17817 would be predicted to be quite an unlikely detection: the uncertainty--in--$\mathcal R$ band suggested by the fact that it was actually detected lies quite higher than the theoretical calculation.

An additional surprise (see the second of~\cite{Chruslinska:2018ylm}) is that the event, statistically, should have taken place in a younger galaxy because the host NGC 4993 saw the peak of star formation 11Gyr ago and a steady decline since, with no star formation in the last 2Gyr in the region where the GW170817 event blasted. The type of models that produce binary systems with the needed characteristics in these environments seems to be somewhat extreme in that the neutron stars are barely kicked off upon being formed (so they end up in a bound state) and so on, but such models generate rates for the Milky Way that are at odds with observation (they would predict 1 event/1000 years, with the estimated rate from observation being smaller than 1/5000 years). Substantial reduction in model uncertainties is needed before the ``surprise'' can be put on solid statistical footing.

Conversely, current evolutionary models that fit the Milky Way population are in tension with the LIGO/Virgo finding, expected to be detectable only every 50-500 years. Again, the systematics need to be more thoroughly explored.

The merger rate was also predicted in a totally different way by noticing that Europium is produced by the r-process (now known to take place in binary merger kilonovae) in amounts of $O(10^{-5}M_\odot)$ per merger~\cite{Vangioni:2015ofa}, yielding an estimated detection rate of 2.5-11 yr$^{-1}$.

Mergers are not the only conceived sources of gravitational waves. For example,
 the LIGO-Virgo collaboration has carried out a study of 222 pulsars spinning faster than 10 Hz,
searching for gravitational wave emission~\cite{Authors:2019ztc} that has been found for none.
A direct bound on the quadrupole moment (and ellipticity) of those pulsars follows, that for young pulsars is tighter than those coming from spin-down as seen in the radio pulse frequency.
Searches for other long-duration signals have come back empty handed for the time being~\cite{Abbott:2019heg}.

It is also expected that the GW detector network could pick up the quadrupole part of the signal emitted by a core collapse supernova if it exploded in our galaxy or out to the Magellanic clouds. This should happen a couple of times per century, so there is a chance that we receive a picture of nuclear matter being quickly compressed and heated to a protoneutron star or to a hypermassive star on its way to becoming a black hole (see~\cite{Nakazato:2019ojk} and references therein).

Finally, the rate of neutron star-black hole mergers was predicted to be larger than that of binary neutron star systems, with some sources suggesting even a LIGO detection rate of some 10 NS-BH mergers per year~\cite{Abadie:2010cf} which is now somewhat discredited, and we could expect an associated $\gamma$-ray burst. So there is a working chance that the third run of aLIGO/Virgo will uncover one such event. 
As an example, in~\cite{Baibhav:2019gxm}, the number of events estimated for the third run is 0.11-3.4 for double neutron star mergers, and a larger 0.46-3.9 for neutron star--black hole ones.

A separate issue is the positive identification of the GW alert as a binary neutron-star merger candidate.
Here are of importance the groups searching for optical-counterpart kilonova events, {\it e. g.} the DLT40 survey~\cite{Yang:2019dws}. These groups are preparing databases of galaxies where the GW alert can trigger in the third run of aLIGO. DLT40 had a sample of 2200 luminous galaxies matching the GW170817 trigger. Only 23 of them where in the sky patch marked by the gravitational wave detectors, among which only one showed an optical transient and was thus identified as the origin of the gravitational signal. It seems to us that, since the reach of optical detectors is much larger than the $O(65)$Mpc of the GW detectors, the optical kilonova-GW merger association will be almost lossless, given similar or smaller sky patches to be searched.

\begin{figure}
\begin{center}
\includegraphics[width=11cm]{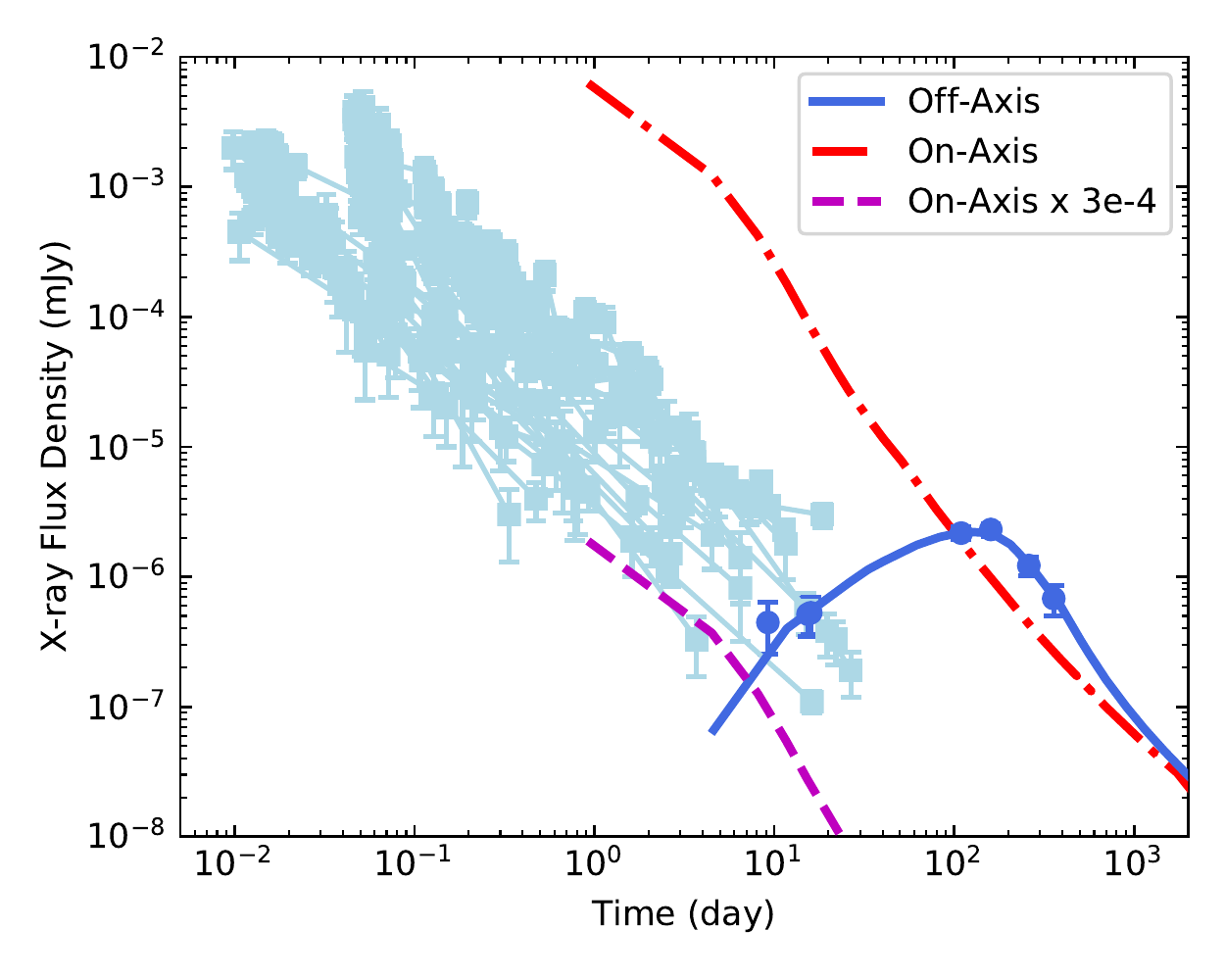}
\caption{\label{GWwasindeedGRB}
The X-ray flux following GW170817 (bottom right corner, solid line and data) seems to grow in a time span in which the bulk of known Short Gamma Ray Bursts (group of softly shaded squares with error bars across the figure) decrease in intensity. This has cast doubt on whether GW170817 was indeed a sGRB. 
However, two corrections to the new event's data bring it in line with the data base. The first (dash-dotted line at the top, red online) is to notice that we saw the new event off-axis so that the initial emission did not point our way until the jet cone opened enough. This corrects the slope but the resulting intensity would be much larger than typical sGRBs. The second correction (yielding the dashed line at the bottom) comes simply from the small distance to the new GW170817, hosted by the nearby galaxy NGC4993 at only about 41 MPc: if it was as far as the typical, cosmological sGRB, it would look fainter. After both corrections, the association of the gravitational wave event and a short gamma ray burst looks much more convincing. 
\emph{Figure courtesy of Yiyang Wu, based on~\cite{Wu:2019rla}.}
}\end{center}
\end{figure}

One alternative, indirect path for calculating the number of expected events is the association between GW and Short Gamma Ray Bursts (SGRBs). These have been copiously detected with $\gamma$-ray detectors and their red shifts have systematically been measured. From their population~\cite{Paul:2017rof}, a stringent lower estimate of 1.87 events/year for the aLIGO-VIRGO network has been put forward.
It is true that the association of GW170817 and a SGRB has been disputed because of the different intensity and distribution of the received radiation; but a convincing recent analysis~\cite{Wu:2019rla} accounting for the proximity of this event respect to other SGRBs, and the emission angle of the burst, accounts for the data, concluding that cosmological SGRBs likely are neutron star mergers. This association would bypass many of the uncertainties associated with stellar populations; because of its importance, we present the recent evidence~\cite{Wu:2019rla} in figure~\ref{GWwasindeedGRB}.

\newpage
\section{Static observables in a neutron star} \label{sec:static}

\subsection{Tolman-Oppenheimer-Volkoff equations and mass-radius diagram} \label{subsec:TOV}

\subsubsection{The TOV system of equations.}
The basic equations for hydrostatic equilibrium were laid out in 1939 by Oppenheimer and Tolman and by Volkoff;
as this has been astrophysics textbook material for half a century, this paragraph will be very terse.
In equilibrium, the pressure needs to compensate the weight of the upper layers. For a static, spherical body, the increase in pressure with depth in General Relativity is
\begin{eqnarray}
\label{basicTOV}
   \textcolor{blue}{\frac{dP}{dr}} &=& \textcolor{blue}{- \frac{G_N}{r^2}}
\frac{(\textcolor{blue}{\epsilon(r)}+P(r))(\textcolor{blue}{M(r)}+4\pi r^3P(r))}{1-\frac{2G_NM(r)}{r}}
    \\   
\color{black}\frac{dM}{dr} &=& 4\pi r^2 \color{black} \epsilon \ .
\end{eqnarray}
We have highlighted blue online the Newtonian equation. The relativistic extension includes the Schwarzschild factor of the metric, weighs the energy density $\epsilon$ instead of the rest mass density $\rho$, and includes the gravitational strength caused by the pressure $P(r)$.

As a first order differential system for two variables with a complicated right hand side, it is best solved by an initial-value algorithm such as a 4th order Runge-Kutta, integrating outwards from the star's centre (accredited colleagues can obtain a sample integrator from the authors). The star radius $R$ is determined by the condition $P(R)=0$ and at that point, the value of $m(R)$ determines the star's mass. This integrated quantity of energy-matter $M_{GR} =\int_{0}^R\,4\pi r^2 \epsilon(r) {\rm d}r$ coincides with the $M$ seen in the Schwarzschild metric outside the star and with the Newtonian mass $M$ obtained by matching to a Newtonian $M/r$ potential as seen by an observer at infinity, so it can be justly called the NS mass (in modified gravity this will no more be the case and one should distinguish ``baryonic mass'' or ``quantity of matter'', this $M$, from the gravitational mass).

This equation needs to be supplemented with the barotropic Equation of State $P=P(\epsilon)$ that describes the nuclear matter (generically, any additional dependence on the entropy is discarded for cold, stratified stars). This equation of state is one of the major threads of this review, and section~\ref{sec:EoS} is dedicated to its discussion.

\subsubsection{The $M(R)$ diagram.}\label{subsecMofR}

\begin{figure}
\begin{minipage}{0.45\textwidth}
\begin{center}
\includegraphics[width=8cm]{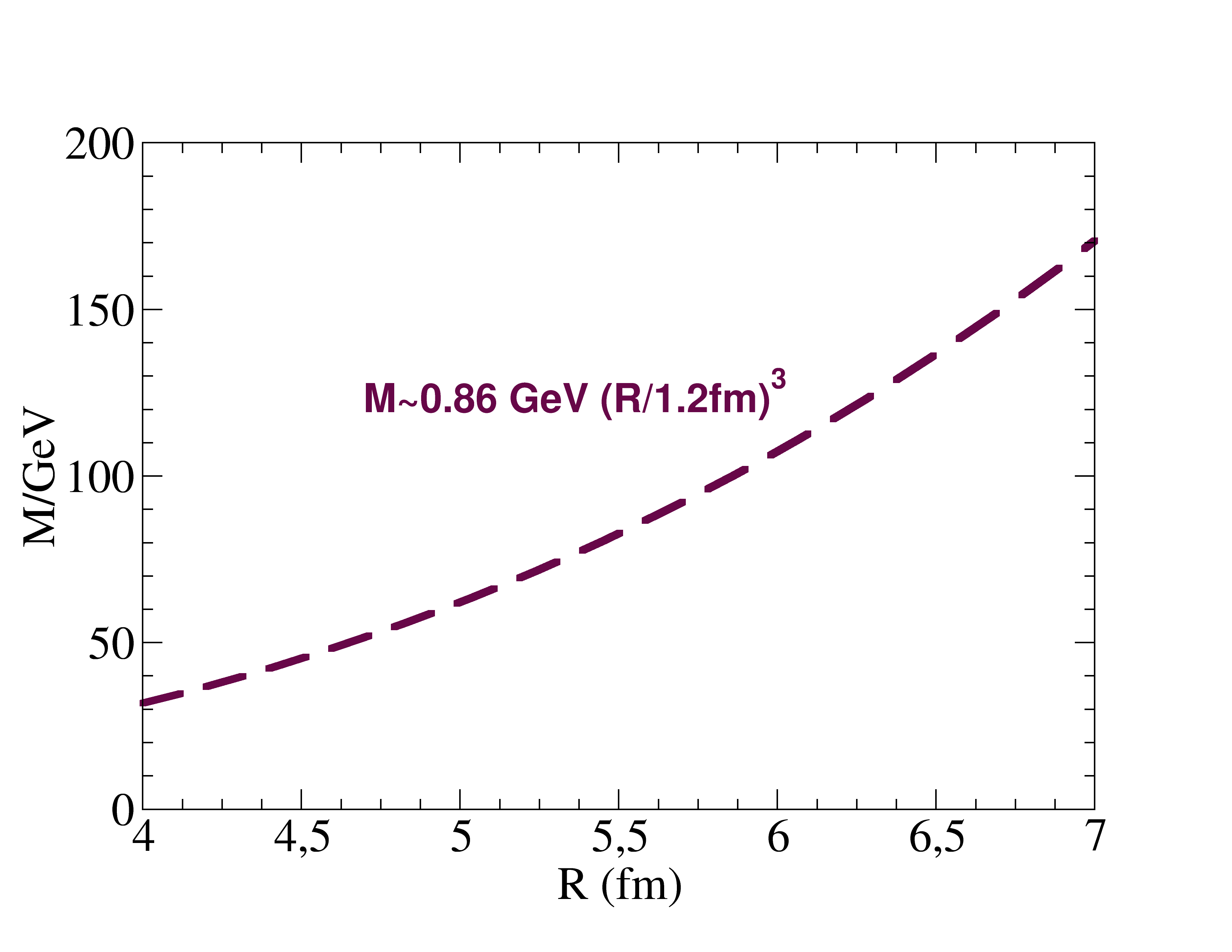}
\includegraphics[width=8cm]{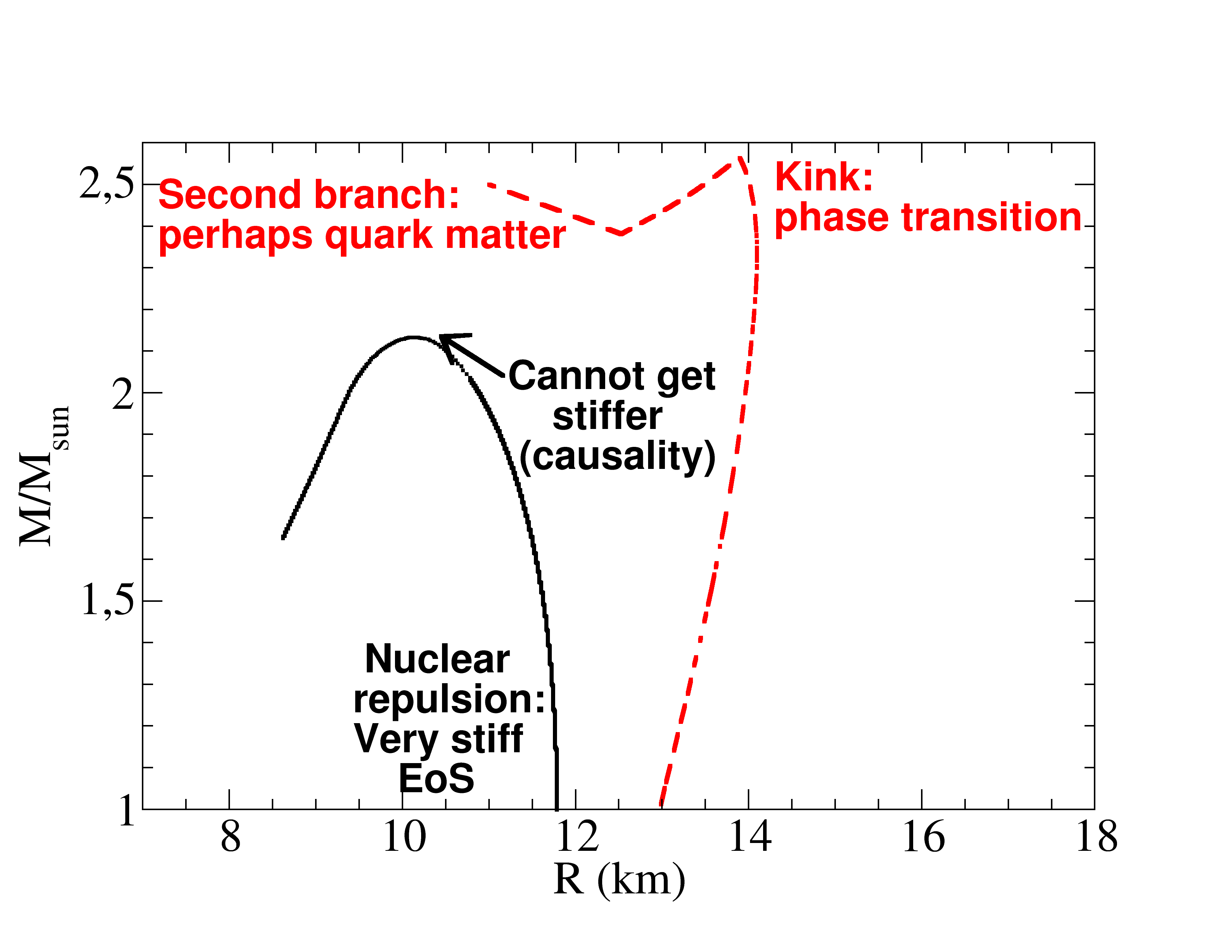}
\end{center}
\end{minipage} \hspace{1cm}
\begin{minipage}{0.45\textwidth}
\begin{center}
\includegraphics[width=8cm]{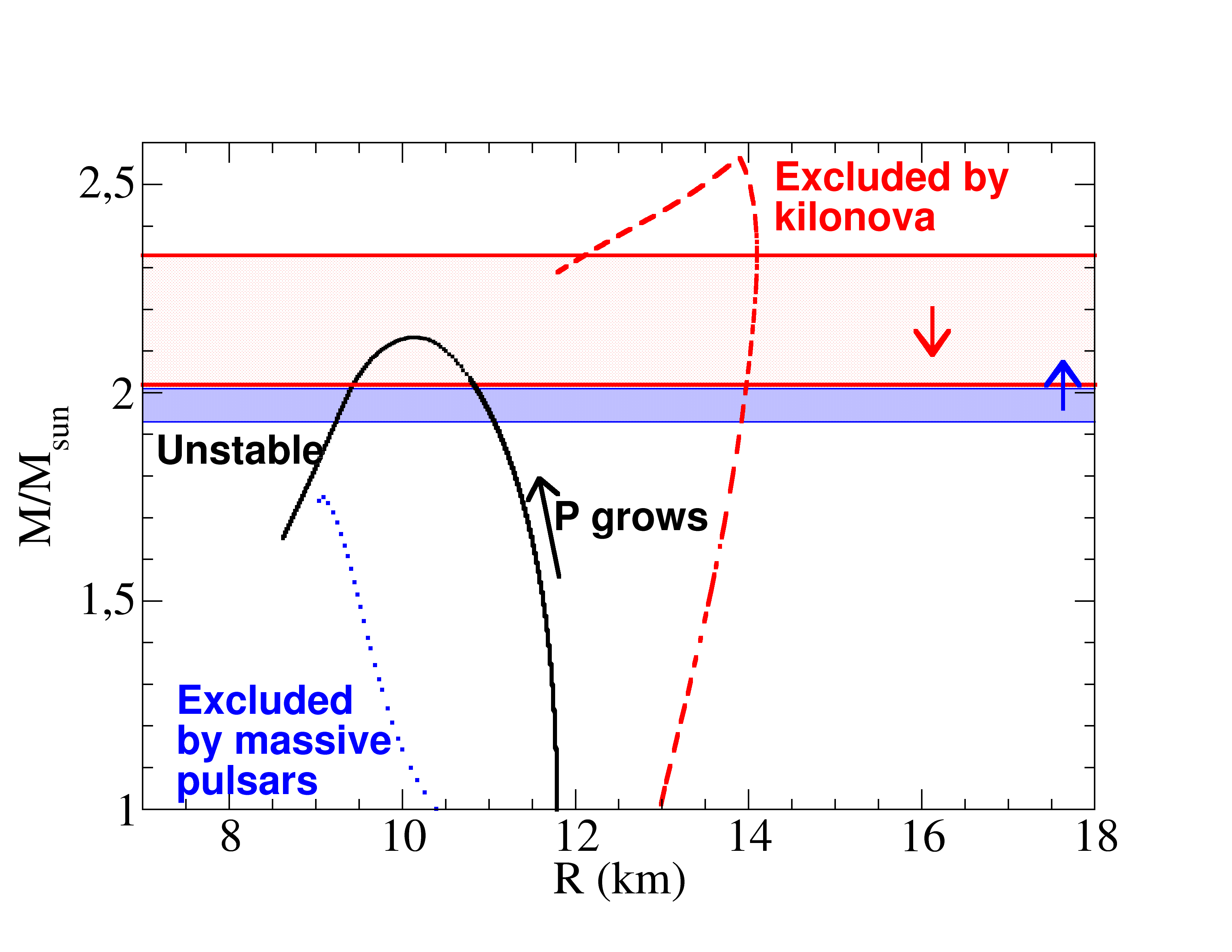}
\caption{\label{fig:MR} Top left: the radius of a nucleus grows with its mass as $M^{1/3}$ due to the stiffness of nuclear matter. Top right: Mass/radius diagram for a neutron star annotated with astrophysics comments. Bottom: Mass/radius diagram for a neutron star annotated with nuclear physics comments.
}
\end{center}
\end{minipage}
\end{figure}

Integrating the system of equations (\ref{basicTOV})  for different values of the starting central pressure $P(0)$ produces a family of stars that are plotted in the traditional $M(R)$ diagram of figure~\ref{fig:MR}.

The basics of this diagram is annotated in the figure. First, let us observe that ordinary nuclei increase in size with mass (top left plot), $M\simeq (M_N-B/A)\left(\frac{R}{R_0}\right)^3$. This is the telltale of the strong incompressibility of nuclear matter: the equation of state of nuclear matter is ``stiff'', so that the mass scales with the volume.

The top right plot in the figure shows three calculations of $M(R)$ with various equations of state from~\cite{Oter:2019kig} and~\cite{Dobado:2011gd} (the details are unimportant now). The true curve $M(R)$ is one of the traditional major goals of investigation in the field. Unlike in nuclei, adding mass makes the star smaller due to the force of gravity. 
As we will see in the next subsection~\ref{subsec:NSmasses}, whatever the true curve, it must reach a mass high enough to pass the narrow band (blue online) since two very massive pulsars take those values. And it should not exceed the broader band (red online) as the kilonova afterglow following GW170817 suggests that heavier neutron stars are not possible.

Moving from right to left, the pressure increases. When the slope changes sign and becomes positive, the star becomes unstable to perturbations (no such configuration can survive). This is because adding a small amount to $M$ should move the star upwards in the graph; but because of the inverted slope, this lowers the central pressure, and thus the additional weight of that extra mass cannot be supported and the star collapses.

The bottom plot is annotated with comments of interest for hadron physics. First, note that the curves are very vertical. This is because the EoS is very stiff: it takes a huge amount of pressure (and thus, weight) to compress nuclear matter slightly, due to a combination of Fermi repulsion upon squeezing neutrons, and internucleon repulsion. 

But there is a limit to the incompressibility of matter given by causality (see section~\ref{sec:EoS} below).
This means that the pressure cannot continue growing as fast as necessary to compensate the gravitational crunch of the additional mass. At that point the curve bends and stars cannot be any heavier. 

The verticality of the curve (radius insensitive to the mass) can be exploited to identify whether a merger is caused by a pair of neutron stars or a neutron star and a black hole~\cite{Chen:2019aiw}, given that $R\sim \rm constant$ for neutron stars except at the largest masses, while $R\propto M$ for a black hole, and the radius of the merging objects could be, in the future, estimated from the Gravitational Wave data. (Currently the NS-BH or NS-NS discrimination is made by the masses of the merging objects, as it is believed that no black holes populate the 1-2$M_\odot$ region whereas neutron stars do not populate 3$M_\odot$ and above; but this method provides a separate check that can work without population assumptions.)

In the upper curve we have highlighted one more feature of interest for section~\ref{sec:Phases}. When certain phase transitions that are strongly first order happen in nuclear matter, the $M(R)$ curve presents a sudden kink. In fact, a second branch of stable stars might be possible if the curve bends upwards again, a subject under investigation.

\subsubsection{The inverse problem}
That theorist's way of proceeding, to first approximate the Equation of state $P(\epsilon)$ to later compute the stellar structure and thus obtain the mass-radius diagram $M(R)$, amounts to a mapping between two function spaces over positive real numbers,  with $P,M\in \mathcal{F}(\mathbb{R}^+\to \mathbb{R}^+)$.
Obtaining the mass-radius diagram is to find the map $P(\epsilon) \mapsto M(R)$.

But if one knew from astronomical observations the shape of the $M(R)$ function, 
reconstructing from it the $P(\epsilon)$ EoS would amount to solving an inverse problem~\cite{Lindblom1992}. Several numerical works have focused on the quality of such reconstruction.
Of course, General Relativity is assumed to be valid throughout the star for such methods to be valid.

This task cannot, at the present time, be carried out:  we have numerous values of $M$ for various neutron stars but $R$ is not well measured. However, the situation may soon change as spelled out in~\ref{subsec:radii}. The question then becomes one of the precision of the measured $(R_i\pm \Delta R_i,M_i\pm \Delta M_i)$ pairs, that controls the width of the swath of compatible EoS. It is quite evident (but for a detailed numerical account, see~\cite{Weih:2019rzo}) that a good measurement of the radius of large mass neutron stars is most sensitive to the high--density EoS, that involves the most uncertain extrapolation from laboratory experiments as will be discussed below in section~\ref{sec:EoS}.

A similar inverse problem whose theoretical and numerical foundations are actively explored at this time is the reconstruction of the EoS from the (hypothetical) future measurement of the quasinormal modes of the neutron star: we comment on this timely topic in subsection~\ref{subsec:vibration} below.

\subsubsection{Modifications in the post-TOV formalism} 
In subsection~\ref{subsec:fR} we saw how a vibrant branch of research is assessing modifications of General Relativity; $f(R)$-based theories require modifications of the TOV equations that
we do not reproduce here, as they can be found in the literature~\cite{Resco:2016upv}. The metric has more independent degrees of freedom than in General Relativity, so the equations are more complex though easily manageable.

If the separations from GR are not too large, then the post-TOV formalism generalizes the postNewtonian formalism (necessary because of the not-small gravitational fields) to the static equilibrium equations~\cite{Glampedakis:2015sua,Glampedakis:2016pes,Will:1972zz,Shao:2019gjj}. The TOV system,
separating slightly from General Relativity by the small terms with coefficients $\delta_i$ (at first order) and $\pi_i$, $\mu_i$ (at second order) has been expressed as
\begin{eqnarray}
\color{black}\frac{dP}{dr} &=&  -G \frac{\epsilon + P}{r^2} \frac{m+4\pi r^3 p}{1-2Gm/r} 
\color{black}
- \frac{G m \rho}{r^2} ({\mathcal P_1+ P_2}) \\
\color{black}\frac{dm}{dr} &=& 4\pi r^2 \epsilon\color{black} + 4 \pi r^2\rho ({\mathcal M_1 + \mathcal M_2})\ .
\end{eqnarray}
There, $\rho$ is the baryonic rest mass density (as opposed to $\epsilon$, the total energy density) and the  basic TOV system in Eq.~(\ref{basicTOV}) can be identified; and the post-TOV correction terms are parametrized (with $\Delta=\frac{\epsilon-\rho}{\rho}$) as
\begin{eqnarray}
{\mathcal P_ 1} &=& \delta_1 \frac{Gm}{r} +\delta_2\frac{4\pi r^3 p}{m}\\
{\mathcal P_2} &=&  \pi_1 \frac{G^2 m^3}{r^5 \rho} + \pi_2 \frac{G^2m^2}{r^2}+\pi_3 Gr^2p +\pi_ 4 \frac{p\Delta}{\rho}\\
{\mathcal M_1} &=& \delta_3 \frac{Gm}{r} +\delta_4 \Delta \\
{\mathcal M_2} &=& \mu_1 \frac{G^2 m^3}{r^5\rho} + \mu_2 \frac{G^2m^2}{r^2}
+\mu_3 G r^2p +\mu_4 \frac{p\Delta}{\rho} + \mu_5 \frac{r}{Gm} \Delta^3 \ .
\end{eqnarray}

The $\mathcal M_i$ and $\mathcal P_i$ cannot be large from the success of solar system tests and 
outer-metric tests with pulsars. In particular, if as proposed by~\cite{Shao:2019gjj}, the first order coefficients are taken to vanish
$\mathcal M_1\simeq 0 \simeq P_1$ and a modification to GR appears through some of the $\mu_i$ appearing in $\mathcal M_2$, their presence
is in practice indistinguishable from a modification of the Equation of State. That is, with astrophysical data alone, we cannot guarantee that a presumed modification of the EoS is not mocking Modified Gravity. To lift the degeneracy, the community needs to keep improving the reliability of the {\it ab initio} equations of state.

The degeneracy is manifest from the strong Equivalence Principle and Einstein's equations,
\be
G_{\mu\nu} + \lambda g_{\mu\nu} = \frac{8\pi G}{c^4} T_{\mu\nu}\ ;
\ee
are any eventual disagreements between theory and observation to be assigned to the left side 
(gravity) or to the right side (hadrons)?

    \subsection{Neutron star masses}~\label{subsec:NSmasses}

Neutron star masses $M_{NS}$ (and $M_{\rm companion}$) can be accessed in binary systems by a study of their orbit.
This is achieved by means of the Keplerian orbital parameters~\cite{Vidana:2018lqp} via
\begin{equation}
\frac{(M_{\rm companion} \sin i)^3}{(M_{\rm NS}+M_{\rm companion})} = \frac{T v_i^2}{2\pi G}
\end{equation}
where the orbital period $T$, the projection of the velocity along the line of sight $v_i$, and the orbital inclination
$\sin i  =  \frac{Tv_i}{2\pi a_i}$ (in terms of the semimajor axis $a_i$) appear.
The formula is degenerate in that a pure extraction of $M_{\rm NS}$ is not possible.  
But it can be assisted by measurement of any two postKeplerian parameters (orbital corrections due to General Relativity).
A couple of salient ones are the orbital period decay $\dot{T}$ due to the emission of gravitational waves, and the advance of the periastron $\dot{\varphi}$, respectively
\begin{equation}
\dot{T} \propto \frac{M_{NS}M_{\rm companion}}{(M_{NS}+M_{\rm companion})^{1/3}}, \ \ \ \ \ \ \ \
\dot{\varphi}_{\rm min} \propto (M_{NS}+M_{\rm companion})^{2/3}\ .
\end{equation}
Because these relativistic corrections (and others) depend on different combinations of the neutron star and the companion masses, it becomes possible to reconstruct both from two such measurements.

An additional, very precise, general relativistic method is based on the Shapiro delay (of the companion's light crossing the gravitational field of the neutron star) 
$\Delta t_{\rm Shapiro}= -\frac{2G}{c^3}
{\rm ln} (1-\hat{\bf x}_{NS} \cdot   \hat{\bf x}_{\rm companion})\  M_{NS}$ 
in terms of the visuals from Earth to each of the two bodies, $\hat{\bf x}_i$.

At last, the chirp mass of Eq.~(\ref{chirpmass}) measured with gravitational waves, also allows an absolute mass scale calibration for the binary system in combination with another quantity. The measurement of a mass $M_{NS}$ therefore relies on a binary companion and is way more uncertain for isolated neutron stars.

\subsubsection{Maximum masses}
\label{subsec:maxM}

That there exists a maximum mass $M_{\rm max}$ for neutron stars has been known for decades. Adding material to the star increases the gravitational field pushing matter towards the center, and this needs to be compensated by additional pressure. Just as white dwarfs eventually exceed the ability to sustain their weight, neutron stars eventually collapse.

$M_{\max}$ can  be bound from below by directly finding neutron stars of increasing masses. The record mass claim for a neutron star~\cite{Cromartie:2019kug} stands at
$2.17^{+0.11}_{-0.10}M_\odot$ for
PSR J0740+6620, with runners up at
$(2.01\pm 0.04)M_\odot$ for PSR J0348 + 0432~\cite{Antoniadis:2013pzd}, and $(1.93\pm 0.02)M_\odot$ measured with the Shapiro delay method~\cite{Demorest:2010bx}. These stars are known to rotate slowly; therefore, the measurements can in practice be taken as a lower bound the maximum mass of a static star.
(In any case, section~\ref{subsec:rot}, the difference between $M_{\rm max}(\Omega)$ and $M_{\rm max}(\Omega=0)$ is known to be moderate, up to 20\% for the fastest possible rigid rotation.) 

This maximum mass should in principle be calculable in General Relativity if we knew the Equation of State of neutron matter with precision (section~\ref{sec:EoS}).
While this happens, one can use the low-density EoS constrained from nuclear data and extrapolate it to arbitrary density using the stiff-most EoS compatible with causality, namely $P=\epsilon$ ($v_{\rm sound}=c=1$).
An early computation along these lines~\cite{Rhoades:1974fn} yielded an upper bound on the maximum mass, $M_{\rm max}<3.2M_\odot$ in solar masses (trusting nuclear theory up to $n_{\rm match}=1.7n_0$; the limit decreases as $n^{-1/2}_{\rm match}$).
If the same reasoning is applied~\cite{Dobado:2011gd} to more modern chiral interactions in neutron matter up to a momentum scale $\mu\sim 500$ MeV, prolonged to higher densities with that stiffest EoS saturating causality, that maximum comes down to the 2.25 $M_\odot$ range. 

Extant microscopic models~\cite{Bhat:2018erd} yield masses between 2 and 2.5 $M_\odot$. These often have the problem that they are based on uncontrolled truncations of QCD, so that the resulting potentials are reliable for very low nucleon densities $n\simeq n_0$ where they are fit to data and any extrapolations to higher densities come without a counting that allows to assign a systematic uncertainty.

Therefore, a part of the community is pursuing an assessment with the order by order counting of ChPT. 
For example, Sammarruca and Millerson~\cite{Sammarruca:2018whh} use a higher order in the expansion with the counting in vacuo and find $M_{\rm max} \simeq 2.4 M_\odot$. However, the Darmstadt group~\cite{Kruger:2013kua} still found neutron stars all the way to 3$M_\odot$ with the same type of interactions. It is unclear to us how this disagreement comes about.

Another recent development is the understanding of the asymptotic phases of QCD at high density. The Helsinki group has calculated the EoS to second order in pQCD and applied it immediately to computations of the mass-radius diagram. They find that, interpolating the EoS between chiral perturbation theory at low density and their pQCD computations in the opposite limit, the mass distribution is populated all the way to about 2.75 $M_\odot$.

\begin{figure}
\begin{center}
\includegraphics[width=4in]{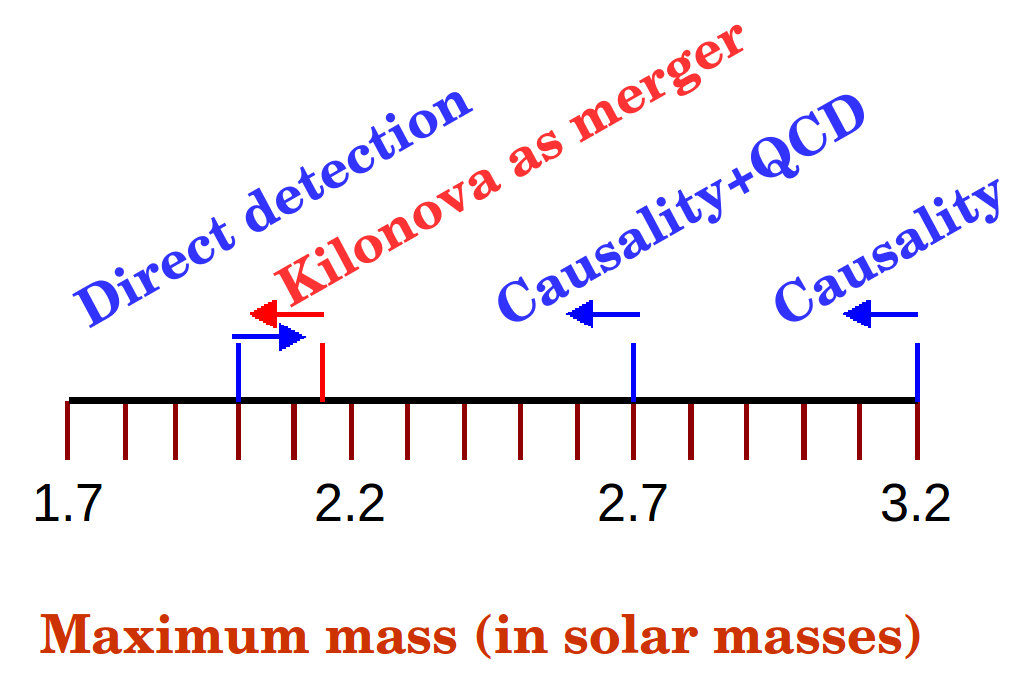}
\caption{\label{fig:maxmass}
Sketch of the determination of $M_{\rm max}$. Direct detection of 2$M_\odot$ pulsars constrains it from below; causality (eventually supplemented with pQCD) constrains it, conservatively, below 2.7 $M_\odot$. The novelty of the last works is the association of the electromagnetic kilonova following GW170817 to a NS-NS merger.}
\end{center}
\end{figure}

While theory comes to terms with what exactly is the prediction of General Relativity + QCD for $M_{\rm max}$, the phenomenology of GW170817 is already weighing in (see the sketch in figure~\ref{fig:maxmass}). 

The novelty~\cite{Margalit:2017dij,Ruiz:2017due,Rezzolla:2017aly}  following the discovery of the gravitational wave event GW170817 is the data suggesting that some 5\% of $M_\odot$ was expelled and seen as an electromagnetic kilonova afterglow.

This number, indirectly and with the input of numerical simulations, has been proposed to indeed constrain $M_{\rm max}$. To understand it, we need to mention the classification of objects that can be produced above $M_{\rm max}$ as results from those simulations.
\begin{itemize}
\item \emph{Direct collapse to a Black Hole} happens if $M_{\rm merged}>(1.3-1.6)M_{\rm max}$. The ejected mass expected is much smaller ($(10^{-4}-10^{-3})M_\odot$) than seen in the kilonova and this direct collapse is therefore disfavored for the GW170817 event.
\item \emph{A supra-massive neutron star (SMNS)} is supported by mostly rigid rotation and so its mass must be relatively close to $M_{\rm max}$, $M_{\rm merged}<1.2M_{\rm max}$ (see subsection~\ref{subsec:rot}). It can  survive several tens of seconds while spinning down by electromagnetic radiation. The Gamma-ray burst following GW170817 lasted at most 2 seconds, so this rigid rotation scenario is not likely either.
\item \emph{A hyper-massive neutron star (HMNS)} is supported by differential rotation $\Omega(r)$ and is intermediate in mass to the other two cases, probably $1.2-1.3 M_{\rm max}$.
The shear intrinsic to that differential rotation is damped quickly, and collapse ensues in 10-100ms after the merger. It is the best candidate to be assigned to the GW170817 remnant. 
\end{itemize} 

Thus, the merging mass known from the gravitational wave signal, $M_{\rm merger} = M_1+M_ 2 = 2.74^{+0.04}_{-0.01}M_\odot$ should be ascribed to this last HMNS object. Therefore, 
\begin{equation}\label{maxmassGW}
M_{\rm max} = \frac{2.74}{(1.2-1.3)} M_\odot\ \ \ \ \ \  \  M_{\rm max}\in (2.11,2.28)M_\odot\ .  
\end{equation}
In fact, \cite{Rezzolla:2017aly} quotes a maximum static star mass $2.16^{0.17}_{0.15}$ at $2\sigma$ (90\% confidence).
The upper limit still leaves room for discovering pulsars with masses slightly above the known 2$M_\odot$, but the lower limit would mean that the field would already be exhausted and no more record mass discoveries would be possible.

A small criticism that can be raised against this line of argument is that the proportionality between the mass sheds classifying the event as SMNS, HMNS or direct-BH  whether 1.2, 1.3, or 1.6, are extracted from numerical simulations that need as input the Equation of State, so they are contaminated by the uncertainty therein; especially because many of the EoS available have been constrained by maximum masses extracted from other sources, so there is some circularity in reasoning. The difficulty is aminorated by inspection of the so called ``quasi-universal relations'' that try to relate quantities independently of the EoS. 

In section~\ref{sec:EoS} we will address the EoS in detail.

The upper bound on $M_{max}$ has also been claimed~\cite{Shibata:2019ctb} to be weaker, at $2.3M_\odot$, in a reanalysis of the merger afterglow lit by lanthanides. With the somewhat uncontrolled uncertainties and hypothesis inherent to this type of analysis, we remain agnostic as to the precise value of the upper bound, beyond that it does exist.

\subsubsection{Mass distribution}
\begin{figure}
\begin{center}
\includegraphics[width=5in]{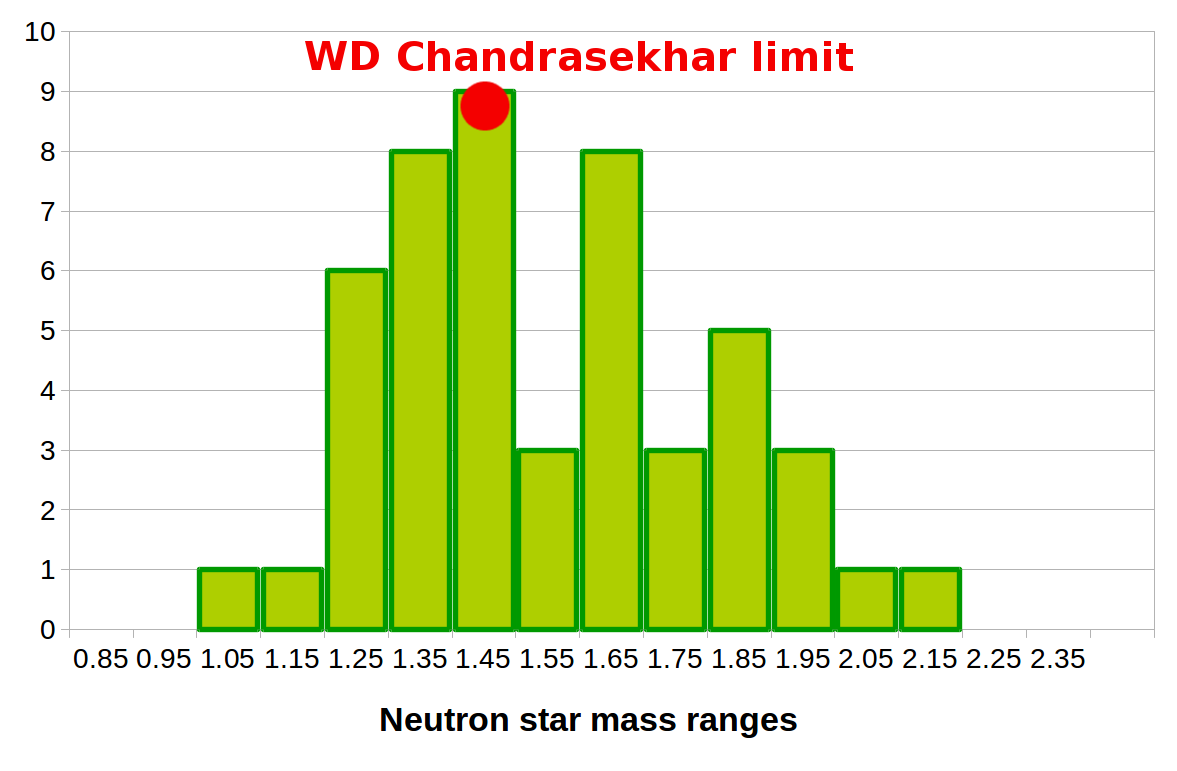}
\caption{\label{fig:massdistribution}
Mass distribution of neutron stars.
There is a large clustering around and below 1.4 $M_\odot$ which coincides with the Chandrasekhar limit of a white dwarf (some 1.44 $M_\odot$) and a maximum mass just above 2 $M_\odot$. Data as of 2016 from the compilation in~\cite{Ozel:2016oaf} (the highest point corresponds to the recently reported J0740+6620 at 2.17 $M_\odot$~\cite{Cromartie:2019kug}).}
\end{center}
\end{figure}

Now that we have a clear idea of what the maximum neutron star mass can be, let us have a quick glance at the rest of the distribution, shown in the histogram in figure~\ref{fig:massdistribution} as of 2016 (in the last couple of years the last point at highest mass has fluctuated).
As seen in the mass-radius diagram (figure~\ref{fig:MR}) and from TOV calculations, the theoretical lower limit to a neutron star mass is quite open: one can have neutron stars with a very small fraction of $M_\odot$. That the histogram collecting the observations ends on the left is possibly due to the difficulty of forming such stars: a collapsing white dwarf exceeding the Chandrasekhar limit (marked in the figure) would have to shed a very large amount of mass to become so light, but avoid being completely torn apart without a runaway thermonuclear explosion. Unbinding so much matter in an accretion--induced collapse (with typical ejection of 2-5\% of the total mass~\cite{Sharon:2019mji})  seems energetically difficult: much of the gravitational energy is employed in overcoming nuclear pressure and Fermi repulsion. In any case, no pulsars are known with masses below about 1$M_\odot$. 

Thus, it is not unnatural that the distribution has a peak at that Chandrasekhar mass limit around 1.4$M_\odot$,  marked in the figure..
Neutron stars that are much heavier tell either of accretion from or merger with a binary companion; or from fall--back during the supernova explosion.

As for the distribution of masses of binary neutron star systems, of interest for the aLIGO-Virgo program, a recent study~\cite{Farrow:2019xnc} has reestablished that binary systems are rather symmetric: the mass ratio $M_-/M_+$ of the lightest to the heaviest companion is, at 99\% confidence level, larger than 0.69. This is consistent with the one event known, GW170817, where that ratio is in the interval (0.73, 0.86).

\subsubsection{Maximum mass beyond General Relativity}

To understand how the maximum mass of a neutron star can exceed the maximum in General Relativity,
we can first think that weakening gravity allows to jam more matter in the neutron star without forcing it to collapse. thus, $M_{NS}^{\rm max}$ can exceed 3$M_\odot$ depending on the size of the modifications of GR.

In modified theories of gravity, the mass of the neutron star does not coincide anymore with $\int \varepsilon(r)$ over the star. The correct definition is to obtain it from matching to the Newtonian potential at infinite distance from the star; then the mass can actually receive contributions from the gravitational field outside the star. To calculate,  use can be made of the Schwarzschild parametrization of $A(r)$ and $B(r)$ in the metric 
\begin{equation} \label{metric}
ds^{2}=B(r)\,dt^{2}-A(r)\,dr^{2}-r^{2}(d\theta^{2}+\sin^{2}{\theta}\,d\phi^2)
\end{equation}
for all $r>R_{NS}$ but allow the mass therein to be $r$-dependent, resulting in a function $M(r)$. 
The Runge-Kutta numerical integration extends much farther away than the 10 km of the star's edge until the asymptotic regime is recovered.

The static, spherically symmetric solutions in $f(R)$ theories can be tagged with the quantity of matter up to the star's edge, $M_{\rm NS}= M(R_{\rm NS})$. However this tag is different from the label that we would assign them from Newton's potential at infinite distance, $\Phi\to -GM_\infty/r$. Both quantities are equal in GR, $M_\infty=M_{NS}$; not so in modified theories.

A further note of interest is that, if beyond-GR theories allow for NS to have masses above the maximum  (2-3)$M_\odot$, the earlier BH-BH identification of several aLIGO GW signals becomes less firm, since it is based in the blief that $M_{NS}<M_{NS}^{\rm max}\sim (2-3)M_\odot$.

\subsection{Constraints on neutron star radii}\label{subsec:radii}

We dedicate these subsection to some comments on the size of a spherical neutron star. 
An important motivation for nuclear and particle physicists is that a determination of radius and mass of the same object would seriously constrain the EoS from the $M(R)$ diagram within GR. But precise estimates of NS radii are very difficult and more model dependent than those of masses, as they are indirect observations affected by large uncertainties ({\it e.g.}, composition of the NS atmosphere, distance to the source, magnetic field, accretion). 
Until very recently, the radii of neutron stars were only known with huge uncertainties, but this is now changing rapidly.

Currently, radii are constrained from several sources: 
(a) Quiescent X-ray transients in low-mass X-ray emitters, (b) X-ray bursts for rotation-powered millisecond pulsars (RPMSPs) where radii can be determined from the shape of the X-ray pulses, and newly (c) gravitational waves.

Several astrophysical analyses seem to favour small values, mostly in the range 9 $-$ 13 km \cite{Verbiest:2008gy,Bogdanov:2012md,Guillot:2013wu,Steiner:2012xt,Poutanen:2014xqa,Ozel:2015fia}.
A particular stir was caused by the quiescent X-ray method~\cite{Guillot:2014lla} that produced 
$R_{\rm NS} = 9.4\pm 1.2$ km, a number uncomfortably low for standard Neutron Star theory 
(rather predicting 11-13 km). 

The method employs the X-ray luminosity, apparent from Earth, of a neutron star softly heated by accretion of surrounding material, related to the luminosity at the source by
\begin{equation} \label{radius}
\color{black} L_{\rm apparent} \color{black} d^2 = L_{\rm EM} R_{\rm NS}^2\ .
\end{equation}
The right hand side of this equation contains two unknowns, the radius  (that we seek) and the actual star luminosity (this includes an implicit dependence on the radius because of the General Relativistic red shift of the radiation emitted, see Eq.~(\ref{surfaceredshift} below).
The distance $d$ in the left side is inferred from other measurements, such as knowledge of the cloud or stellar group where the X-ray emitter is.
To obtain $R_{\rm NS}$, we need one additional equation. This can be obtained from a thermal fit of the X-ray spectrum. With the extracted temperature one can use Stefan's law for the absolute luminosity, 
\begin{equation}
L_{\rm EM} = \sigma (\pi R_{\rm NS}^2) T^4
\end{equation}
and substituting in Eq.~(\ref{radius}), $R_{\rm NS}$ can then be extracted. A way to give it, accounting for the general relativistic redshift of the radiation leaving the source, is the algebraic equation for $R_{NS}$~\cite{Vidana:2018lqp}
\begin{equation}
R_{NS} = \sqrt{1-\frac{2GM}{c^2 R_{NS}}} \sqrt{\frac{L_{\rm EM}}{4\pi \sigma T^4}}\ .
\end{equation}

The measurement's~\cite{Guillot:2014lla} disagreement with theory prompted assessments of the systematic uncertainties, especially absorption and reradiation by the stellar atmosphere, and questioning the poor fits to a thermal spectrum.

\begin{figure}
\begin{center}
\includegraphics[width=5in]{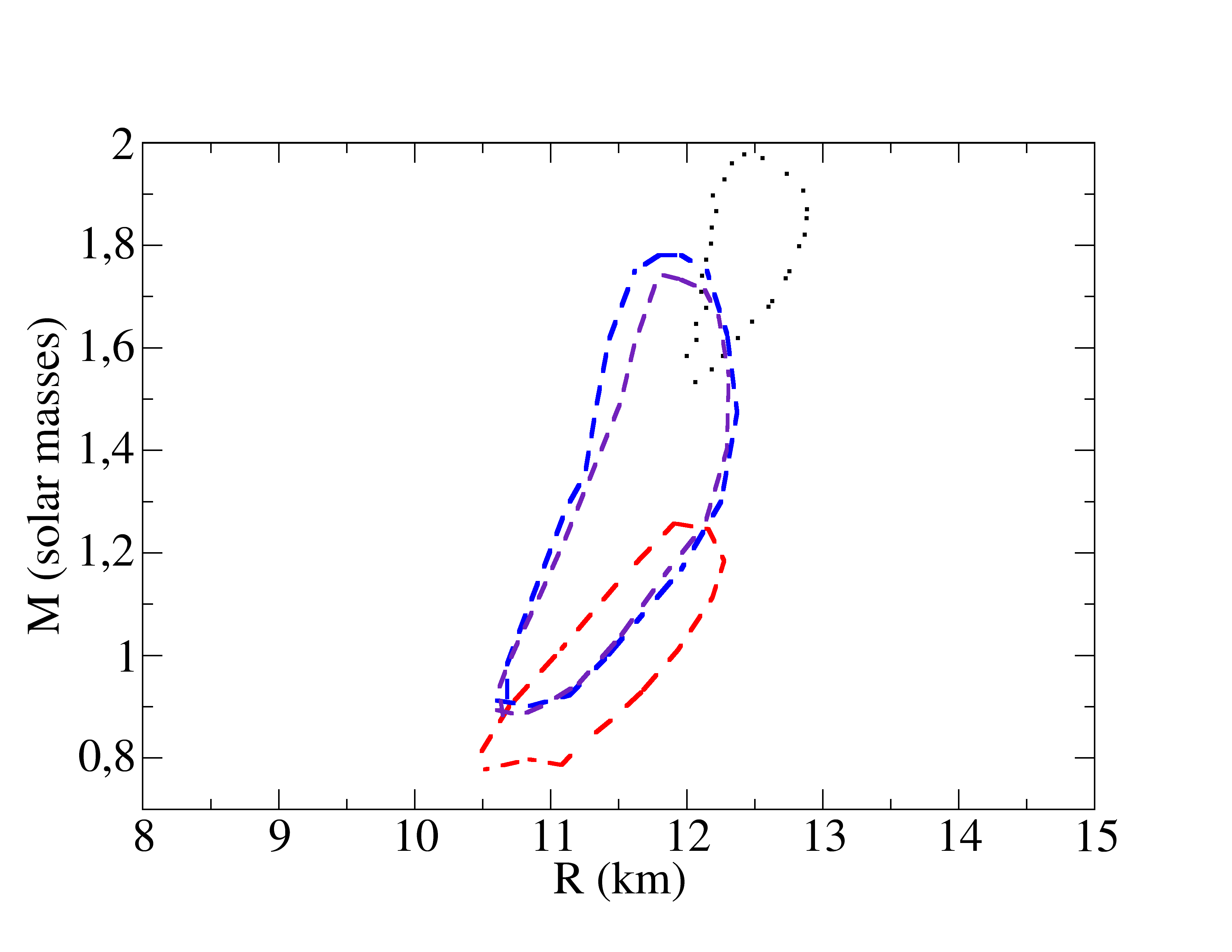}
\caption{\label{fig:radii}  Contours showing the radius of several observed pulsars in the $M(R)$ diagram. Dotted line: analysis with model D of~\cite{Nattila:2017wtj}. Dashed lines: the three pulsars analyzed with model C from~\cite{Nattila:2015jra}. }
\end{center}
\end{figure}

The second method uses thermonuclear-burst sources (believed to quickly process chunks of infalling matter) as opposed to quiescent stars, and yielded $R_{\rm NS}\in (\sim 10.4,\sim 12.9)$ km, more in line with theory expectations~\cite{Steiner:2012xt}.
In figure~\ref{fig:radii} we show four 68\% contours in the $M(R)$ diagram from~\cite{Nattila:2017wtj,Nattila:2015jra} as an example of the precision achieved in the last few years.

Such joint analysis of mass and radius for the same star using Bayesian inference are now becoming standard~\cite{Raaijmakers:2018bln} and the last developments suggest that one obtains different results if only the Eq. of State is chosen as a prior, or whether the choice of priors is made in a joint space of masses and radii that include the exterior Schwarzschild distribution.

In the immediate future, the NICER mission (``Neutron star Interior Composition Explorer'' at the International Space Station) will simultaneously pursue soft X-ray timing and spectroscopy.  It aims to reaching better than 10\% precision in determining $R$, even down to 130 eV at 6 keV (uncertainty that the hypothetical TES telescope might bring down to 2-3 eV)~\cite{Hey:2019hof}.  It is designed to enable rotation-resolved spectroscopy of the thermal and non-thermal (burst-like) emissions for 0.2-12 keV X-rays. But the theory of the continuous-spectrum pulses is affected by severe modelling uncertainties~\cite{Watts:2019lbs}.

The relevant discovery here would be that of resolved discrete spectral lines: their Doppler shift respect to the laboratory would reveal the rotation velocity, that together with the angular frequency (known from either the pulsar period or the burst oscillations) would immediately lead to the star radius. To date, the X-ray spectrum has not been resolved into individual lines.

Among other methods to measure the radii~\cite{Vidana:2018lqp}, one is of particular interest for particle physicists: the positron annihilation $e^- e^+\to \gamma\gamma$ that produces the characteristic 511 KeV photons can be used to measure the redshift
\begin{equation} \label{surfaceredshift}
z=\frac{1}{\sqrt{1-\frac{2GM_{NS}}{c^2 R_{NS}}}} -1
\end{equation}
that is sensitive to the compactness $M_{NS}/R_{NS}$. This is not yet precise enough as the spread of measured redshifts $z\in(0.2,0.5)$ is not competitive with other determinations.

Finally, the detection of gravitational waves from merging compact stars, GW10817, has provided important new insights into the NS radius~\cite{Bauswein:2017vtn} from multimessenger analysis, or by means of the measurement of tidal deformabilities~\cite{De:2018uhw} in a binary system that we discuss in the next subsection~\ref{subsec:tidal}. 

Here, GW measurements of the tidal deformability (discussed in subsec.~\ref{subsec:tidal}) or the maximum spinning frequency of a light enough merger end product with discernible quadrupole (figure~\ref{fig:GWpulse}) could bring a new measurement with totally different systematics. 

Several meaningful constraints on the radius of neutron stars from GW170817 are collected in the appendix.

\subsection{Tidal deformability} \label{subsec:tidal}

\subsubsection{Extraction from binary mergers}

Gravitational wave detection happening at a retarded time $t$ ($t'=t-|{\bf x}-{\bf y}|/c$) 
is sensitive to the neutron star energy-stress tensor at $t'$, in linear approximation
\begin{equation}
\bar{h}_{\alpha\beta}(t,{\bf x}) = \frac{4G}{c^4} \int d^3 {\bf y} 
\frac{{\color{black} T_{\alpha\beta}(t',{\bf y})}}{|{\bf x}-{\bf y}|}\ .
\end{equation}
The information about hadron physics is contained in ${\color{black} T_{\alpha\beta}(t',{\bf y})}$.
Most of the work is based on an isotropic ideal fluid with 
\begin{equation}\label{stressT}
T^{00}=\epsilon\ ,\ \  T^{ij}=P\delta^{ij}\, \ \  T^{0i}=0\ ,
\end{equation}
with pressure given as a function of the energy density by the EoS $P(\epsilon)$ (see section~\ref{sec:EoS}).

Unfortunately, the GW-caused strain at the detector is extremely weak, and the luminosity
(extracted  from data as shown in figure~\ref{fig:GWpulse}), for such weak radiation,
is almost entirely obtained from the lowest multipole, via Einstein's second quadrupole formula, 
\begin{equation} \label{luminosity}
L = 5 \frac{G}{c^5} \color{black} \langle \dddot Q_{ij} \dddot Q_{ij}\rangle \color{black}
\end{equation}
(analogous to the usual dipole radiation in electrodynamics,
$ L_{EM} = \frac{2}{3} \frac{K}{c^3} \langle \ddot D_i \ddot D_i \rangle$).

Thus, the information we receive about $T$ at the neutron star tells about  the quadrupole $Q$.
Isolated stars minimize free energy by adopting a spherical shape. The quadrupole comes about because the binary system has cylindrical, not spherical symmetry, and it varies while the stars orbit, changing the orientation of the symmetry axis linking them.  
But when at close quarters, the tidal field of each star induces a quadrupole in the other, which in linear response reads
\be
Q_{ij}^{\rm 1-star}= -M^5{\color{black} \Lambda} E_{ij} = - \lambda  E_{ij}  \ .
\ee
Substitution in Eq.~(\ref{luminosity}) shows that the gravitational wave signal is sensitive to the \emph{tidal deformability} $\Lambda$, defined as the coefficient of proportionality, of both stars, though in the combination of Eq.~(\ref{joint:tidal}). 

Even before the contact between the two inspiralling neutron stars, the tides excited due to their
finite size detract energy from the orbit, so that the inspiral  takes less time.
But in an NS-NS merger, \emph{both} stars are distorted, so that what the GW signal actually allows is the extraction of the binary tidal polarizability parameter $\widetilde{\Lambda}$, defined as a mass-weighted average of the individual $\Lambda_{1,2}$ :
\begin{equation} \label{joint:tidal}
\widetilde{\Lambda} =\frac{16}{3}\left[\frac{(m_1+12m_2)m_1^{4}\Lambda_1}{(m_1+m_2)^{5}} + \frac{(m_2+12m_1)m_2^{4}\Lambda_2}{(m_1+m_2)^{5}}\right] \ .
\end{equation}

The strongest  parametric dependence of the tidal deformability on the star's mass and radius can be extracted to read
\begin{equation}\label{Love}
\Lambda = \frac{2}{3} k_2 \frac{R^5}{M^5}\ .
\end{equation}
In geometrodynamic units, of course, $[R]=[M]$ (appendix~\ref{app:units}) so that the tidal deformability is dimensionless. 
The information about neutron matter is then carried out by two dependences: that in the compactness $R/M$ which has been exhaustively studied for decades, and that in the second Love number $k_2$ that has more recently been addressed. After extracting that fifth power of the compactness, the remaining Love number $k_2$is only weakly dependent on it (see figure 1 of~\cite{Hinderer:2007mb}).

\begin{figure}
\begin{center}
\includegraphics[width=8cm]{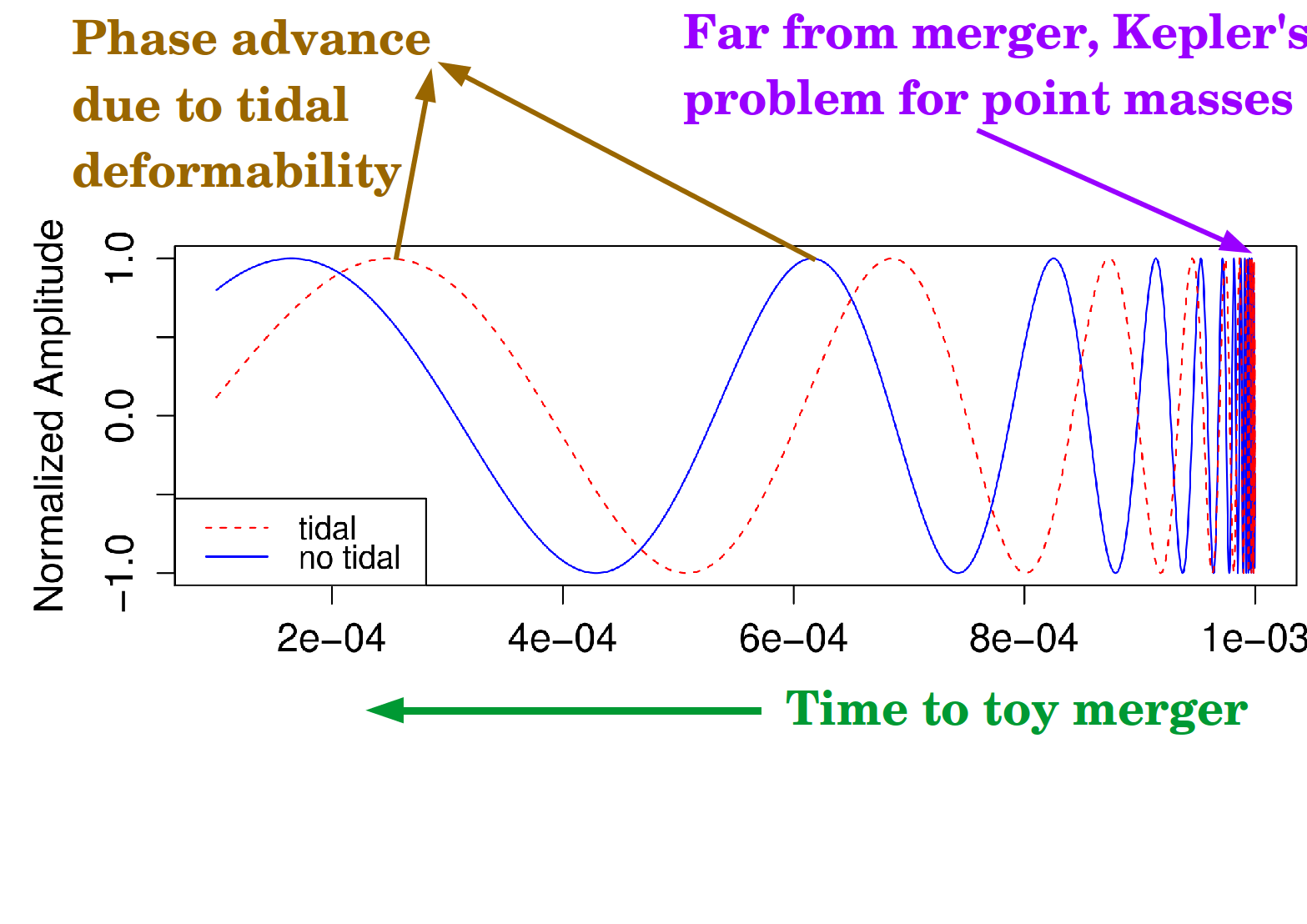}
\caption{\label{fig:tidal} Time advances leftwards to a binary merger; the blue line is a simple sinusoidal wave (gravitational radiation not taken into account, so the phase should not advance). However, the broken line advances because of the tidal deformability.}
\end{center}
\end{figure}

Because of that power of $M^{-5}$, $\widetilde{\Lambda}$ appears as a postNewtonian fifth-order correction to the wavefront phase: the phase advances faster due to the accelerated merger (see the cartoon in  figure~\ref{fig:tidal}). That high order makes it small and therefore difficult to extract from GW observations. Further,  its presence manifests itself only during the last few orbits and it is correlated with the other star parameters being extracted so that a precise value is not yet at hand  ~(\cite{Flanagan:2007ix,Damour:2009dn}). 

The stationary phase approximation ({\it e.g.}~\cite{Choi:2018zbi}) can be written as function of the frequency $f$, $\hat{h}(f)=\frac{A(\{\theta_i\}}{f^{7/6}} e^{i\psi(f)}$ with $A$ carrying the dependence on the parameters such as the chirp mass $\mathcal M$, and the phase $\psi$ expressed in the postNewtonian expansion, which in General Relativity yields
\begin{equation} \label{phase}
\psi(f) = 2\pi f t_{\rm coalescence} -(\phi_{\rm coalescence} +\frac{\pi}{4}) 
+ \frac{3M^2}{128m_1m_2v^5}\left(\Psi^{\rm pointlike}_{3.5\rm PN} + \color{black} \Psi^{\rm Tidal}\color{black}\right)
\end{equation}
(the coalescence time and phase can be chosen arbitrarily).
Equation~(\ref{phase}) shows the addition of the accumulated tidal phase to that from the orbiting stars taken as if they were structureless objects.
That tidal phase is 
\begin{equation}
\Psi^{\rm tidal}(f) = - \frac{39 v^{10}}{2} \widetilde{\Lambda} 
\end{equation}
in terms of the joint $\widetilde{\Lambda}$ defined in Eq.~(\ref{joint:tidal}).
The correction to 6th postNewtonian order is known~\cite{Choi:2018zbi} and only needed if $\widetilde\Lambda$ could be measured to percent precision.

 Before discovery, the tidal deformability $\Lambda$ was estimated to be constrainable in order of magnitude from a single observed NS-BH merger~\cite{Kumar:2016zlj} or to $O(10\%)$ from 25-50 observations combined, by studying  the ratio of  gravitational wave signals $h_{NS-BH}/h_{BH-BH}$, whose magnitude and phase can be simulated.

The initial work of the aLIGO-Virgo collaboration constrained it by  $\widetilde{\Lambda} \leq 800$, 
but this analysis did not assume both objects to have the same EoS. 
Some reanalysis have since improved the situation, with the aLIGO-Virgo collaboration reporting (for GW170817~\cite{Abbott:2018exr}) limits of 70 $\leq \widetilde{\Lambda} \leq$ 720  at 90$\%$ confidence. Assuming that both merging objects were NS governed by the same EoS, that band contracts to 70 $\leq \widetilde{\Lambda} \leq$ 580  with 90$\%$ of confidence ~\cite{Abbott:2018exr}. 

The tidal deformability for a canonical $1.4M_\odot$ NS has been subsequently obtained by extrapolation (from the measured masses, employing a family of parametrized EoS and somewhat EoS--insensitive relations)~\cite{Abbott:2018exr} as $\Lambda_{1,4}=190^{+390}_{-120}$, also within the aLIGO-Virgo collaboration. Nevertheless, numerous authors have revised these claims applying numerous techniques (Bayesian analysis, combination with multimessenger-electromagnetic data, theory extrapolations to different NS masses...) so that we have seen fit to collect several contemporary results in a table for the reader's convenience, see appendix~\ref{app:tablas}.

A very interesting correlation between radius and deformability of the binary system $\tilde{\Lambda}$ has been put forward, profiting from the steep $\Lambda\propto R^5$ dependence~\cite{De:2018uhw}:

\begin{equation}
R_{1.4M_\odot} = (11.2\pm 0.2) \frac{\mathcal{M}_{\rm chirp}}{M_\odot} \left(\frac{\tilde{\Lambda}}{800}\right)^{1/6}{\rm km}
\end{equation}
that allows for quick estimates (though the numerical coefficient has been given also~\cite{Zhao:2018nyf} as $11.5\pm 0.3$ and it does have some dependence on the EoS).

The correlation between the uncertainties in the measurement of the tidal deformability and the neutron star radius were explored early on~\cite{Hotokezaka:2016bzh}, finding that  $\delta \Lambda$  depends on $\delta R$ as
\be
\frac{\delta R /0.91{\rm km}}{R/13{\rm km}}  =  \frac{\delta \Lambda/400}{\Lambda/1000} \ .
\ee

Finally, we should also mention that the gravitation community is trying to obtain 
hadron--physics free relations among their observables, ``universal relations''~\cite{Yagi:2016qmr}, to bypass the difficulties inherent to strongly coupled QCD.

\subsubsection{Computation with a TOV solver}

The theory necessary to compute the tidal deformability simultaneously with the star's mass and radius for a static relativistic star has been very competently laid out in~\cite{Hinderer:2009ca} (see also~\cite{Hinderer:2007mb,Flanagan:2007ix}).

The static Schwarzschild metric gets distorted by the quadrupolar correction to
\begin{eqnarray}
ds^2 &=& - e^{2\Phi(r)} \left[1 + H(r)
Y_{20}(\theta,\varphi)\right]dt^2  + e^{2\Lambda(r)} \left[1 - H(r)
Y_{20}(\theta,\varphi)\right]dr^2 \nonumber \\
& & + r^2 \left[1-K(r) Y_{20}(\theta,\varphi)\right] \left( d\theta^2+ \sin^2\theta
d\varphi^2 \right)\ , \nonumber\\ 
\end{eqnarray}
and this leads to a system of extended TOV equations that modify Eq.~(\ref{basicTOV}) requiring the simultaneous solution of two differential relations for $H(r)$ and $dH(r)/dr\equiv \beta(r)$ (so the system remains first order), that with $f=\frac{d\epsilon}{dP}=\frac{1}{c_s^2(\epsilon)}$ becomes
\begin{eqnarray} \label{computeTidal}
\frac{dH}{dr}&=& \beta\\
\frac{d\beta}{dr}&=& 2 \left(1 - 2\frac{m_r}{r}\right)^{-1} H\left\{-2\pi
  \left[5\epsilon+9 P+f(\epsilon+P)\right] +\frac{3}{r^2}+2\left(1 - 2\frac{m_r}{r}\right)^{-1}
  \left(\frac{m_r}{r^2}+4\pi r P\right)^2\right\}\nonumber\\
&&+\frac{2\beta}{r}\left(1 -
  2\frac{m_r}{r}\right)^{-1}\left\{-1+\frac{m_r}{r}+2\pi r^2
  (\epsilon-P)\right\}\ .\nonumber
\end{eqnarray}
At the center of the star one sets $H(r)=a_0r^2$ and $\beta(r)=2a_0r$ (the $a_0$ deformation constant cancels upon constructing $\Lambda$). As for Eq.~(\ref{basicTOV}), the edge of the star is reached when $P(R)=0$ at which point the auxiliary quantity $y=\frac{ R\beta(R)}{H(R)}$ gives the Love number
\begin{eqnarray}
\color{black} k_2\color{black} &=& \frac{8C^5}{5}(1-2C)^2[2+2C(y-1)-y] \times\bigg\{2C[6-3y+3C(5y-8)]\nonumber\\
      & & +4C^3[13-11y+C(3y-2)+2C^2(1+y)] +3(1-2C)^2[2-y+2C(y-1)] \ln(1-2C)\bigg\}^{-1},\nonumber\\
\label{Love2}
\end{eqnarray}
with $C\equiv M/R$ being the NS compactness. Eq.~(\ref{Love}) then provides the deformability. We have solved this system ourselves and there are no particular numerical difficulties in handling it, the 4th order Runge-Kutta algorithm is up to the task. 

Among the many studies that have computed the tidal polarizability, that will reappear later in the article let us for now mention the one (see for example Piekarewicz and Fattoyev~\cite{Piekarewicz:2008nh}) tight relation
$\Lambda\left(\frac{M}{R}\right)$ between the polarizability and the compactness $M/R=R_{Schwarzschild}/(2R)$. This is quite independent of the nuclear EoS so it can be used as a prediction of General Relativity to be contrasted with data if both quantities can be simultaneously measured, similarly to the known I-Love-Q relations between the moment of inertia and the quadrupole moment, found by Yagi and Yunes~\cite{Yagi:2013bca}. To obtain information about the underlying hadron matter, additional correlations are needed.

\newpage
\section{The equation of state of cold hadron matter} \label{sec:EoS}
The sheer size of neutron stars make necessary a statistical treatment of the many-neutron system. The main point of contact between macroscopic astrophysical studies and microscopic nuclear and particle physics work is the Equation of State to which we dedicate this section. Here, ``cold hadron matter'' is used as opposed to the hadron matter created in high--energy heavy--ion collisions, that can reach up to 500 MeV in temperature at the LHC. Whereas most of this
section is dedicated to near $T=0$ matter, subsection~\ref{subsec:finiteT} is dedicated to small but finite Temperature $T<100$ MeV.

\subsection{General principles}

An equation of state (EoS) relates thermodynamic variables describing the state of matter under given physical conditions. For the TOV equations~(\ref{basicTOV}),  the required relation is one between the pressure and the energy density, $p(\epsilon)$. A second independent variable is necessary to completely describe a mechanical gas (for example, the baroclinic EoS in the atmosphere depends also on the entropy density or temperature, so that isobaric surfaces do not coincide with isodensity ones). In section~\ref{subsec:finiteT} below we will add temperature as the additional second variable, an incipient subfield. Most of the time, we will restrict ourselves to EoS at $T=0$.

Hadron matter in neutron stars is often considered as a fluid (though crystalline phases have been proposed~\cite{Canuto:1974gi,LlanesEstrada:2011jd}); explicit gravitational effects do not have to be included as the EoS is a local thermodynamic equilibrium property of matter, in a region small enough that space-time can be taken as Minkowskian.
Intensive thermodynamic variables such as temperature $T$, pressure $P$, or chemical potentials $\mu_i$ are well defined and take constant values in local equilibrium. An appropriate thermodynamic potential (that attains a minimum in the ground state) is the Helmholtz free energy $F(T,N_i,V)$, depending on the particle numbers $N_i$ occupying a volume $V$; these are traded in the thermodynamic limit for the particle number densities $n_i =N_i/V$. 
Nuclear reactions can alter the $n_i$ proportions: in equilibrium, the matter is characterised by a perhaps smaller number $N_{cons}$ of independent conserved charges. Then, the individual densities $n_i$ are connected by the conditions of statistical equilibrium $\mu_i = \partial F/ \partial N_i$ \cite{Oertel:2016bki}.

Simulations demonstrate quick thermal and mechanical equilibrium establishing $T$ and $p$. Chemical equilibrium, granting the use of an EoS, is only justified if the timescales of the chemical reactions are much shorter than those of the system’s hydrodynamic evolution. This can cause a distinction between the treatment of nuclear matter in static stars respect to collapsing systems (supernovae or mergers).
Typically it is assumed that a temperature on the order of 0.5 MeV and above is sufficient to reach the so-called nuclear statistical  equilibrium (NSE) \cite{Iliadis:2007ili} 

\subsubsection{Chemical equilibrium}

A typical set of conserved charges are the total baryon $N_B$, (electric) charge $N_Q$, electronic lepton $N_{L(e)}$, and strangeness $N_S$ numbers.  
Correspondingly,  the chemical potential of each particle carrying those net numbers is given by
\begin{equation}\label{particlepotential}
 \mu_{i} = B_{i}\mu_{B} + Q_{i}\mu_{q} + L_{i}^{(e)}\mu_{le} + S_{i}\mu_s \ .
\end{equation}
For a composite  nucleus (or other aggregate) with $N_a$ neutrons and $Z_a$ protons, this becomes
\begin{equation}\label{nucleipotential}
 \mu_a= (N_a+Z_a)\mu_B + Z_a\mu_q \equiv N_a\mu_n  + Z_a\mu_p
\end{equation}
where $\mu_n$($\mu_p$) is the chemical potential of neutrons (protons). Conditions on (electric) charge neutrality and weak equilibrium can further reduce the number of independent particle numbers or chemical potentials.

Weak interactions are not always in equilibrium. Specifically, in core collapse supernovae (CCSNe), the electron capture reaction $p + e^− \rightarrow n + \nu_e$ is not equilibrated for  baryon number densities below $n_B \approx 10^{- 3} - 10^{- 4}$ fm$^{- 3}$ (or equivalently,
for mass-energy densities below $\epsilon_B \approx$ 10$^{11}$ g cm$^{- 3}$), since the relevant timescales can exceed the dynamical timescale of the astrophysical object of interest. In the first stages of CCSNe, neutrinos are not in equilibrium and are not included in the EoS, but treated with transport schemes.   They determine the electron number densities, which remain a degree of freedom of the EoS. 
If density is high--enough,  strangeness--changing weak interactions activate, with estimated chemical timescales at  $\sim 10^{- 6}$ s or below  \cite{Oertel:2016bki}. Therefore, for all purposes in this article, strangeness--changing weak equilibrium is a good approximation, {\it i.e.}, $\mu_s$ = 0; strangeness is not conserved and is not taken as an independent thermodynamic variable.
In later stages of CCSNe, and in (proto-) neutron stars ((P)NSs), neutrinos become trapped and $\beta-$ equilibrium is achieved. They can thus be included in the EoS by the neutrino fraction 
$Y_{\nu_e} = n_{\nu_e}/n_B$ or the lepton fraction $Y_{L_e} = Y_{\nu_e} + Y_e$ with the electron fraction $Y_e = n_e/n_B$.

At a later cooling stage of the neutron star, neutrinos become untrapped, i.e., their mean free path becomes comparable to the system size and $\beta$-equilibrium without neutrinos is established. This condition can be expressed by setting the electronic lepton chemical potential $\mu_{le}$ to zero in Eq.~(\ref{particlepotential}) as for cold NSs. Together with charge neutrality, this implies that $n_e$ or $Y_e$ are fixed by $n_B$ and are no longer free variables of the EoS.

Assuming lepton flavor conversion via neutrino oscillations to be negligible, the heavy flavor lepton numbers are conserved independently of the electronic lepton number. 

Charge neutrality is a second important condition.
To all microscopic purposes, a neutron star is infinitely large. Charge neutrality is necessary to avoid instabilities due to  strong electric fields. It can be locally formulated as 
$n_Q = \sum\nolimits_{i}Q_in_i = 0$. Thus $n_Q$ is not an independent thermodynamic degree of freedom and it is convenient to introduce the hadronic charge density $n_q = \sum\nolimits_{i}' Q_i n_i$ summing over all hadrons (and/or quarks, if present). If electrons are the only leptonic component, this implies $n_q = n_e$.

Inhomogeneous matter as present in the star's crust can suffer from locally imbalanced charge distributions with neutrality only global. The resulting competition between nuclear surface and Coulomb energies causes the formation of clusters developing into ``pasta phases''~\cite{Ravenhall:1972zz,Hashimoto:1984hsy,Williams:1985koo}.

\color{black}
\subsubsection{Thermodynamic consistency and stability}

A valid EoS must satisfy the thermodynamic conditions of consistency (that follow automatically if it is derived from an energy potential, so that modelling the underlying theory is a preferred strategy) and stability (positive sound speed squared $c_s^2>0$). Let us examine them in turn.

From the internal energy per unit volume  $e$, expressed as a function of entropy $s$, and the likewise specific volume $v$: 
 
\begin{equation}
 du =Tds -Pdv 
\end{equation} 

The thermodynamic definitions of pressure P and temperature T are:
\begin{equation}\label{Energy}
 P =-\left.\frac{\partial u}{\partial v}\right\vert_s,
 T =\left.\frac{\partial u}{\partial s}\right\vert_v, 
\end{equation}
and Schwarz's lemma implies the thermodynamic consistency condition
\begin{equation}\label{crossder}
 \left.\frac{\partial P}{\partial s}\right\vert_v=
 -\left.\frac{\partial T}{\partial v}\right\vert_s\ , 
 \end{equation}
a differential equation that a valid equation of state must satisfy.

With temperature and density being independent variables, and with $v=1/n$, the first of the conditions (\eqref{Energy}) takes a curious form (that can be used to constrain simple EoS models such as Eq.~(\ref{thermalpressure}) below),
\begin{equation}\label{firstconstrain}
 P= T\frac{\partial P}{\partial T} +n^2\frac{\partial u}{\partial n} \ .
 \end{equation}
 
 The really important point is that  thermodynamic stability requires that the Hessian of $u$ be jointly convex in $v$ and $s$, which leads to the conditions:

\begin{equation}\label{convexHessiant}
 \left.\frac{\partial^2 u}{\partial s^2}\right\vert_v \geq 0, 
 \left.\frac{\partial^2 u}{\partial v^2}\right\vert_s \geq 0,
 \left.\frac{\partial^2 u}{\partial s^2}\right\vert_v\left.\frac{\partial^2 u}{\partial v^2}\right\vert_s \geq\left(\frac{\partial^2 u}{\partial s \partial v}\right)^2
 \end{equation} 
 
Taking as independent variables temperature and density,  these are satisfied if 
\begin{equation}\label{secondconstraint}
 \frac{\partial u}{\partial T} \geq 0, \  \frac{\partial P}{\partial n} \geq 0,
 \end{equation}
(note that $\partial u/\partial T= \partial u/\partial s \partial s/\partial T$ and likewise $\partial/\partial v = -n^2\partial n)$.
The equations in~(\ref{secondconstraint}) are the monotony conditions; pressure needs to increase with the number density and energy with the temperature. Stable equilibrium entails satisfaction of these relations.

\subsection{From laboratory observables to the crust EoS}\label{subsec:lab}

Terrestrial nuclear laboratories can constrain the Equation of state at low density, near its value in conventional stable nuclei. It is customary to parametrize the binding energy per nucleon $E$ in nuclear matter with a few bulk coefficients in a Taylor expansion around small density~\cite{Piekarewicz:2008nh}. Some recent work has focused on constraining those coefficients from gravitational--wave data~\cite{Carson:2019xxz,Li:2019sxd}
which may be a workable strategy if the merger of two very light neutron stars is found (for heavier stars, these low--density parameters are subsumed into the larger EoS problem treated throughout this section).

\subsubsection{Saturation density and compressibility}

For the convenience of the particle physics reader we will briefly recall the reference point taken for studies of neutron (and generically, nuclear) matter.
The basic idea is that nuclear radii for large nuclei roughly scale as $R=R_0 A^{\frac{1}{3}}$ (that would mean that the nuclear matter would be incompressible). This entails that 
$n|_{\rm nuclei}=A/V\simeq n_0$, a saturation number density. Thus,
\begin{equation}\label{satdensity}
n_0=\frac{3}{4\pi R_0^3}. 
\end{equation}
(In reality, nuclear radii have pronounced jumps easily interpreted in the shell model as subshell closure shifts, see {\it e.g.}~\cite{Angeli:2013epw} for a compilation of data.)
In practice, Eq.~(\ref{satdensity}) is used as it stands with $R_0\in(1.1,1.2)$ fm, yielding a reference value which is broadly used to discuss neutron matter (though it does not play a specific role in neutron stars, where higher densities are quickly reached), $n_0\simeq 0.16$fm$^3$.

To increase the density of nuclear matter beyond the saturation one, energy must be provided (in the case of neutron stars, by the gravitational binding potential).
The standard parametrization of this energetic cost is by convention written in terms of the binding energy per nucleon, $E/A$ (we will immediately drop the $A$), for symmetric nuclear matter ($Z=A/2$),
\begin{equation}
E(n) = \epsilon_0 + \frac{K}{18}\left(1-\frac{n}{n_0}\right)^2 + \dots 
\end{equation}
where the binding energy at saturation is $E_{sat} \equiv \epsilon_0= -16.0 \pm 1.0$ MeV at $n_0 = (0.16 \pm  0.01)$ fm$^{-3}$
from fits to nuclear binding energies.

The $K$  coefficient is the incompressibility of symmetric nuclear matter~\cite{Piekarewicz:2008nh}. 
Taking $n=1/v$ in terms of the specific volume per nucleon (that matches the choice of binding energy per nucleon), using the chain rule $d/dv = -n^2d/dn$, and taking the nucleon mass as constant, so that $u=m_N-E(n)$, we have, for symmetric nuclear matter
\begin{equation}
P = \frac{du}{dv} = \frac{K}{9}\frac{n^2}{n_0^2}(n-n_0)\ .
\end{equation}
Thus, increasing the number density $n$ increases the energy density $u$ and  the pressure, complying with Eq.~(\ref{secondconstraint}) and $K$ is the coefficient controlling the rate of change. Determining the nuclear compressibility from experimental data, basically the giant monopole (scalar) resonances assisted by theory to provide the uncertainties has been the subject of a recent review~\cite{Garg:2018uam} in this journal and the interested reader can refer to it. From $^{208}Pb$ data, $K\simeq (240\pm 20)$ MeV, about $10\%$ higher that it was estimated three decades ago, but within the old uncertainty band.

\begin{figure}
\begin{center}
\includegraphics[width=0.75\columnwidth]{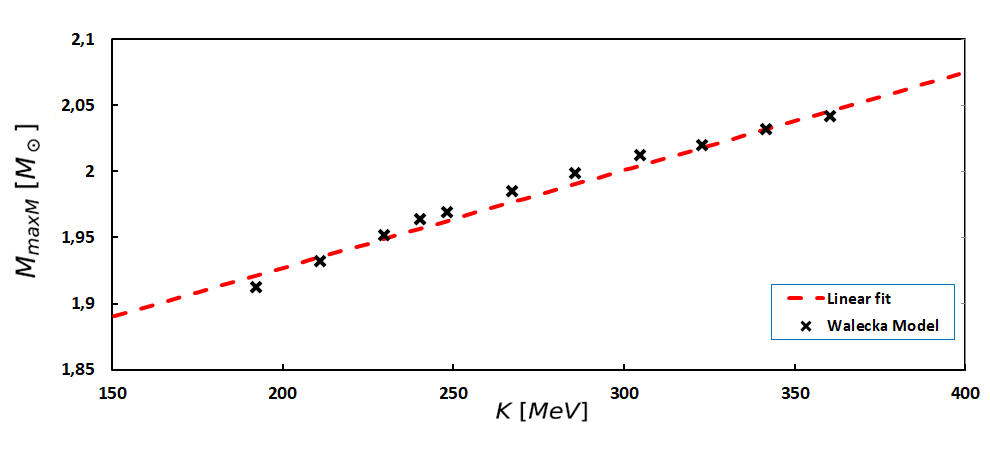}
\end{center}
\caption{
The maximum neutron star mass, $M_{Max}$  as function of the compression modulus, K. The dashed line is the best linear fit which is plotted $M_{Max} = 1.779 +0.0076 $; the black symbols mean the maximum mass $M_{Max}$ by using Walecka model with a bosonic interaction term  for the $\sigma$ field $\sigma^3+\sigma^4$ (From the data of \cite{Posfay:2019nto}).
\label{fig:massversusK}
}
\end{figure}

The relation between maximum star mass and compressibility is shown in Figure \ref{fig:massversusK}. This relation has been obtained using the Walecka model with $\sigma$ (scalar) and $\omega$ (vector) fields and bosonic interaction term $U_{34}= \sigma^3 + \sigma^4$,  and with the Landau mass fixed for the value at nuclear saturation $m_L$ = 0.83 $m_N$ (MeV) ($m_N$ is the nucleon mass) from the data of~\cite{Posfay:2019nto}.

\subsubsection{Symmetry energy}

Because neutron matter is very far from being isospin symmetric, laboratory measurements (in which the number of protons is nonnegligible) need to be extrapolated to larger asymmetries to be of interest for the physics of neutron stars. The asymmetry parameter that quantifies the separation from $n_p=n_n$ is $\alpha:= \frac{n_n-n_p}{n_n+n_p}$ in terms of the number densities for protons and neutrons.

The energetic cost of a separation from isospin-symmetry characterized by $\alpha$ is then the Symmetry Energy, 
\begin{equation}
E_{\rm sym} (n=n_n+n_p) = \frac{1}{2} \left.\frac{\partial^2 E(n,\alpha)}{\partial \alpha^2}\right\vert_{\alpha=0}
\end{equation}
that can be determined experimentally from the binding energy per nucleon $E(n,\alpha)$. For example~\cite{Russotto:2016ucm}, the ASY-EoS experiment at GSI,  colliding $^{197}_{79} Au$ isotopes at 400 MeV/nucleon (this provides a high density environment, in the range of 1-2 times the nuclear saturation density $n_0$)
supplemented by UrQMD simulations, has provided a handy parametrization of the symmetry energy
\begin{equation}
E_{\rm sym}({\rm MeV}) = 22 \left( \frac{n}{n_0}\right)^\gamma + 12 \left(\frac{n}{n_0}\right)^{2/3}\ ;
\end{equation}
the extracted $\gamma$ coefficient is, at 1$\sigma$, in the interval $(0.49,0.87)$ which translates into a symmetry energy at $2n_0$ of $50-60$ MeV. 

The symmetry energy from low-density measurements has been shown not to correlate strongly with either of the star radius or the tidal polarizability~\cite{Tsang:2018kqj}; that calculation is based on a large swath of (Skyrme) Equations of State, using constraints on the $\alpha$ to exclude parts of them. Still, other properties such as the proton fraction (and hence, the electron density, and in consequence the transport coefficients) should be affected by the symmetry energy, see subsection~\ref{subsec:transport}.

Putting together the expansion in $x\equiv\frac{n-n_0}{3n_0}$ beyond the symmetric saturation point from the preceding paragraph and the asymmetry parameter $\alpha$, and keeping the cubic order in terms of the skewnesses $Q$, we have
\begin{equation}\label{expandingenergy} 
E(n,0)=\epsilon_0 +\frac{1}{2}K_{\rm sat} x^2+\frac{1}{6}Q_{\rm sat} x^3+ ..
\end{equation}
and the symmetry energy can likewise be expanded around $n_0$
 \begin{equation}\label{symmetry}
 E_{sym}(n)= S_0+ L\ x + \frac{1}{2}K_{\rm sym}x^2+\frac{1}{6}Q_{\rm sym}x^3+ ...
\end{equation} 
New parameters are the coefficient $S_0$ that determines the increase in the energy per nucleon due to a small asymmetry $\alpha$ at number density $n_0$, the isovector incompressibility $K_{\rm sym}$ that gives the curvature of $E(n, \alpha)$ at $n=n_0$,  the slope of the symmetry energy per nucleon,  $L$, and the skewness $Q$. The set ($S_0,L, K_{\rm sym},Q_{\rm sym}$) contains the parameters needed for pure neutron matter, with much larger uncertainties than those for symmetric matter in Eq.~(\ref{expandingenergy}). They can be extracted from theoretical calculations of the binding energy per nucleon by taking appropriate derivatives.
We collect some of them in table~\ref{tab:nucl.properties}.

\begin{table}
\caption{\label{tab:nucl.properties} We collect several results on the symmetry energy $S_0$ and its slope parameter $L$ at saturation from various experimental input.}

\begin{center}
\begin{tabular}{|c|c|cc|}\hline
 S$_0$ (MeV)   & L (MeV)  &Method & Reference \\ \hline
  (29.0 - 32.7)      &  (40.5 - 61.9)& Nucl. masses+dipole resonances+ neutro skin thickness    &  \cite{Lattimer:2012xj} \\
 (29.0 - 32.7)       &    (44 - 66)     & Nucl. masses+dipole resonances+ neutro skin thickness &  \cite{Lattimer:2014sga}
\\
  (33.5 - 36.4) & (70 - 101) & isobaric analog. states+ skin thickness meas. & \cite{Danielewicz:2016bgb} 
\\ 
  (28.5 - 34.9)      &  (30.6 - 86.08) &normal and radioactive nuclear beams                   &
 \cite{Oertel:2016bki}
\\ 
  (28 - 35)          & (30 - 86) &UG and nuclear binding-energy constraints                    &\cite{Kolomeitsev:2016sjl}
\\
\hline
\end{tabular}
\end{center}
\end{table}
These nuclear matter parameters are strongly correlated. Their density dependence and these correlations have been the object of numerous recent studies. Once determined or modelled, one can address the stellar crust equation of state, composed largely of nuclei~\cite{Piekarewicz:2018sgy} or aggregates thereof. 

\begin{figure}
\begin{center}
\includegraphics[width=13cm]{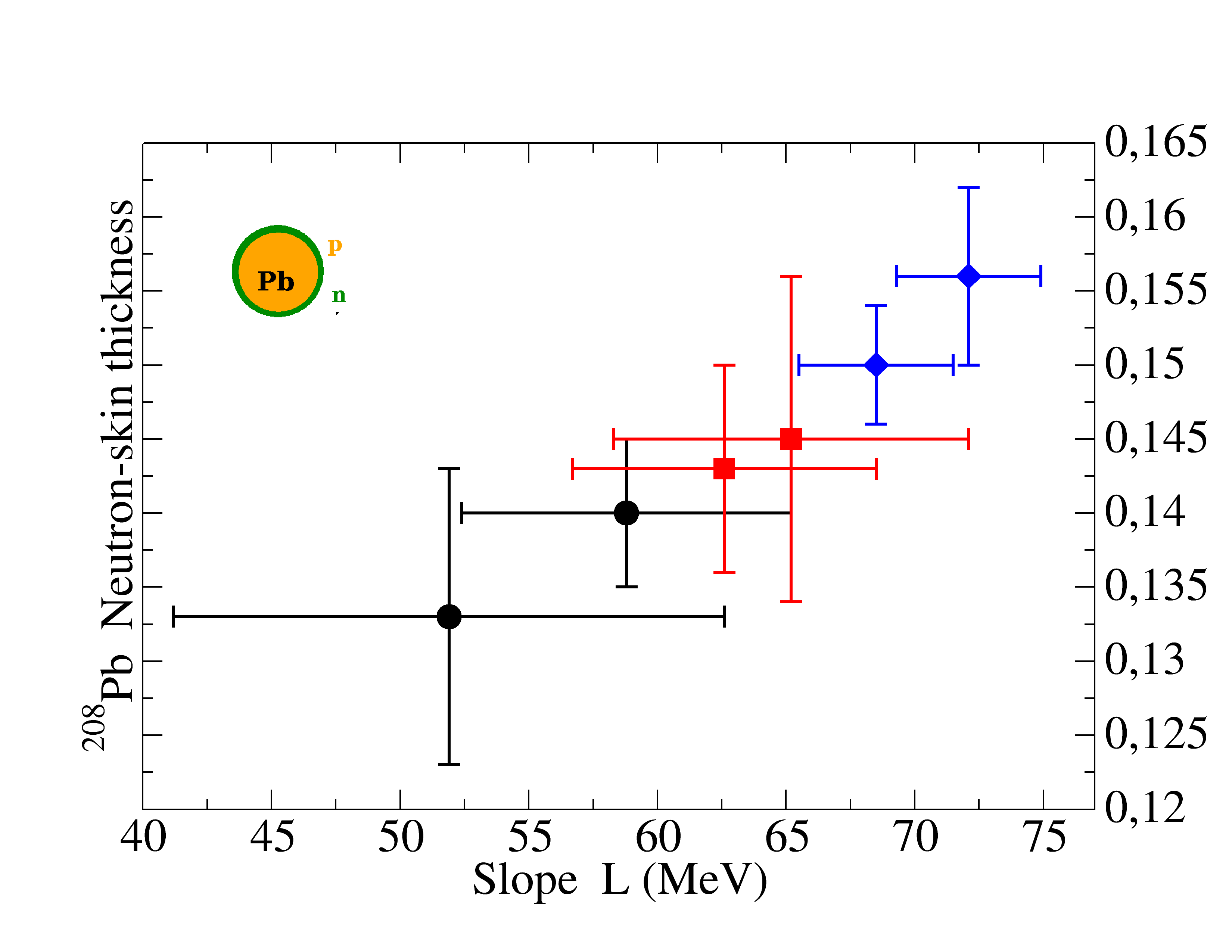}\\
\includegraphics[width=13cm]{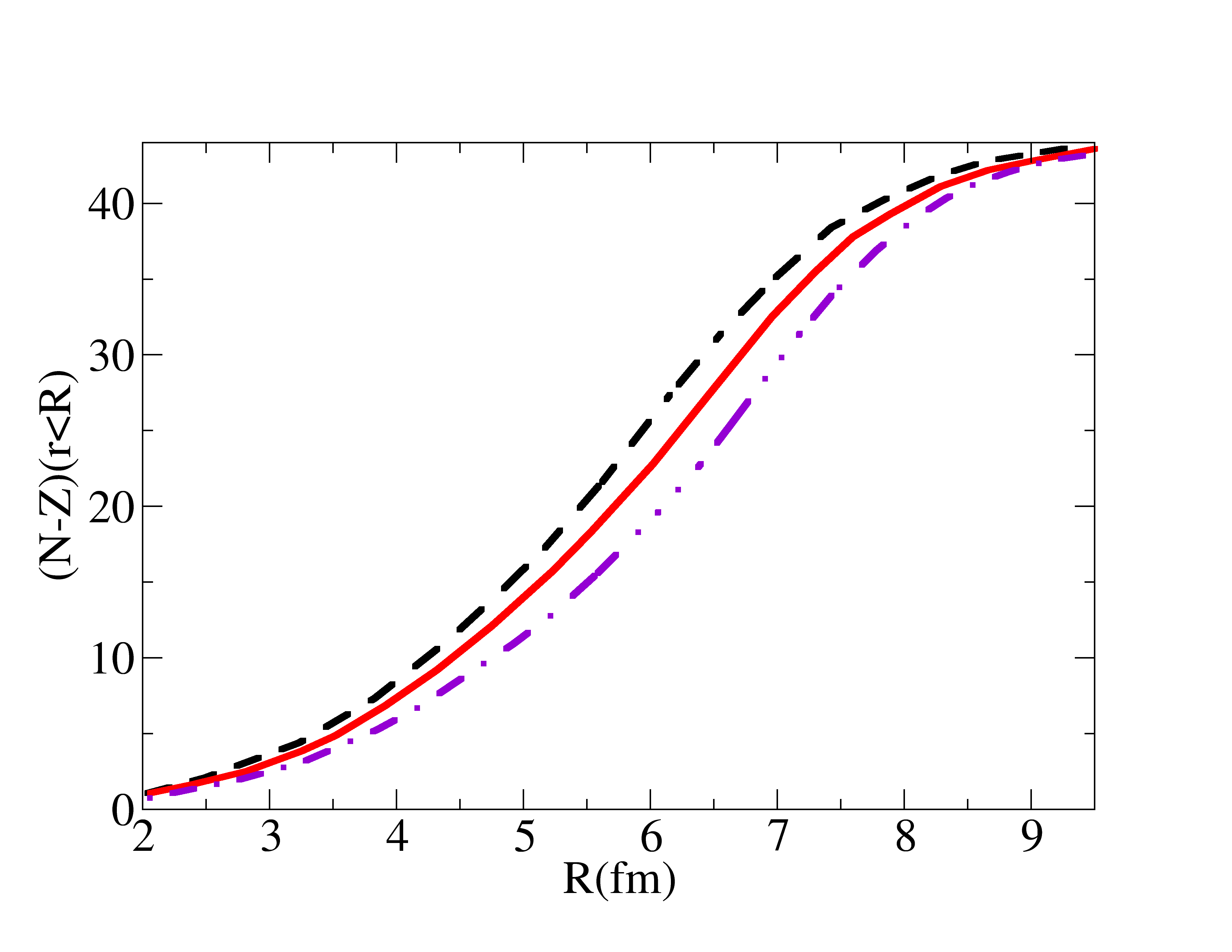}
\caption{\label{fig:skinthickness}
Top: the neutron skin thickness (difference in radii between the neutron and proton distributions in a nucleus, sketched at the top) is correlated with the slope parameter of the symmetry energy, $L$ in Eq.~(\ref{symmetry}). Computation with chiral interactions from~\cite{Millerson:2019jkg}.
Circles (black online): N$^2$LO. Squares (red online): N$^3$LO.
Diamonds (blue online): N$^4$LO. For each pair, the lowest and highest points have cutoffs $\Lambda=450$, and 500 MeV, respectively. Bottom plot: cumulative distribution of the excess of neutrons over protons from $r=0$ to 10 fm in a $^{126}_{82}Pb$ nucleus~\cite{Piekarewicz:2019ahf}. A larger slope $L$ of the symmetry energy pushes the neutrons towards the nuclear edge (to the right in the plot). From left to right, the lines correspond to 48, 63 and 135 MeV respectively.}
\end{center}
\end{figure}

The slope $L$ is correlated with and can eventually be extracted from laboratory data of the the neutron--skin thickness~\cite{Millerson:2019jkg}. This is depicted in figure~\ref{fig:skinthickness}
The upper plot in the figure shows how the correlation is rather good, though the convergence of the chiral series for these quantities is not convincing as the three orders of perturbation theory shown are drifting right and upwards.
The lower plot reproduces a calculation from old--fashioned nuclear potentials~\cite{Piekarewicz:2019ahf} that shows, for $^{126}_{82}Pb$, how the 44 excess neutrons are distributed as a function of the symmetry energy slope $L$. Thus, this parameter is quite directly accessible to laboratory experiments such as PREX.

As for the new observables, the tidal deformability, relevant for gravitational waves, is largely independent of the nuclear crust~\cite{Piekarewicz:2018sgy} (this is supported also in~\cite{Kalaitzis:2019dqc}). They model the inner crust EoS with $P = A+B\epsilon^\gamma$ and obtain the values of the polarizability shown in table~\ref{tab:Piekarewicz} for different indices $\gamma$ (the liquid core is modelled once more by a Walecka--type model in which nucleons exchange $\sigma$--scalar and $V$--vector mesons).

\begin{table}
\begin{center}
\caption{\label{tab:Piekarewicz}  The tidal polarizability (last column) seems to be quite  independent of the crust EoS and size: its dependence on the crustal radius apparently cancels out with 
that of the Love number~\cite{Piekarewicz:2018sgy}.}
\begin{tabular}{|ccccc|}\hline
$\gamma$  &  $R_{\rm crust}$(km) & $R$(km) & $k_2$ & $\Lambda$ \\ \hline
1          &  1.40                & 13.25   & 0.087 & 623.7 \\
4/3        &  0.98                & 12.83   & 0.102 & 623.1 \\
2          &  0.75                & 12.60   & 0.111 & 623.2 \\ \hline
\end{tabular}
\end{center}
\end{table}
The last column is indeed showing a very small dependence on the crust: though it contributes a sizeable amount to the size of the star, its weight is not enough to make a substantial contribution to the polarizability. 
In any case, since the nuclear crust is important for many observables, a good starting point to deploy an EoS in that part of the neutron star is~\cite{Sharma:2015bna}.

\subsection{Model--independent information from Hadron physics and QCD} \label{subsec:limits}

The structure of the inner crust is riddled with model dependencies, so many studies eschew computing it altogether and use the few solid laboratory-data based numbers described in subsection~\ref{subsec:lab} to use as starting point for the hadron-matter equations of state based on modern Chiral Effective Theories. With all their limitations, these have the advantage that they have a systematic counting that allows to assign order by order uncertainties. These equations are the state of the art in the lower density part of the star's core. The intermediate and high density parts within the star are not accessible to first-principles theory, but in the extreme high--density limit, we know that the asymptotic phase of QCD matter is the Color-Flavor locked phase (CFL), and many computations can be carried out within it, thus providing a safe upper range where modelling or interpolation of the intermediate density region can be anchored.

\subsubsection{Low-density limit}

To describe the EoS of neutron star matter at lower densities, a starting point is pure neutron matter (PNM), an idealized, infinite system consisting solely of neutrons, much easier to compute with than a less isospin asymmetric system. Computing the EoS of PNM requires first a nuclear Hamiltonian, and once specified, an appropriate many-body method~\cite{GrossBoelting:1998jg,Gandolfi:2006vp} with small systematic bias.

The interaction Hamiltonian can be constructed within the systematic framework of Chiral EFT. It has been successfully deployed to calculate nuclei and nuclear matter \cite{Hebeler:2015hla}.
In general, the nuclear Hamiltonian contains two-body (NN), three-body (3N), and higher many-body (AN) forces,
\begin{equation}
\label{ManybodyHamiltonian}
H = T + V_{NN} + V_{3N} +\dots 
\end{equation}
and an advantage from Chiral EFT is that the order in which they appear is specified  by its systematic expansion of the nuclear potential in powers of momenta. 

The expansion ceases to converge around a breakdown scale, $\Lambda_b$, which
indicates the need of additional higher--momentum physical degrees of freedom. $\Lambda_b$ is expected to be of the order of 450-600 MeV, limiting the applicability of chiral EFT for only the lower densities encountered in neutron stars not too far from the saturation density (that lies at Fermi momentum around $k_F^{\rm sat}\simeq 300$MeV~\cite{Holt:2019oys}):
the theoretical uncertainty band grows rapidly with the density beyond $n_0$. 

\begin{figure} \begin{center}
\includegraphics[width=0.4\textwidth]{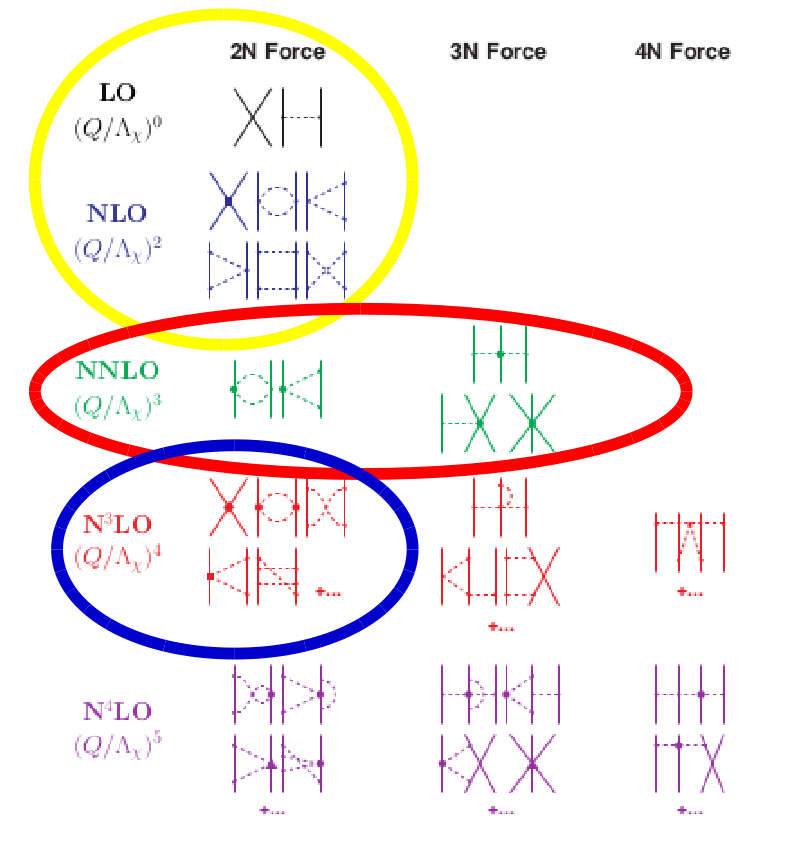}\ \ \ \ 
\includegraphics[width=0.4\textwidth]{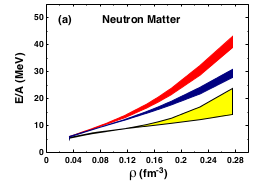}
\caption{\emph{Left}: the nucleon-nucleon interaction in Chiral EFT is computed order by order in a low-momentum, small pion-mass expansion. It controls  the order in which potential terms and relativistic, many-body corrections are to be used; it also allows for a systematic estimate of their values and uncertainties. An important point is that 3-body forces first enter at NNLO. Currently, the $NN$ interactions is being worked to sixth order~\cite{Entem:2015xwa} but computations of interest for neutron stars to such high accuracy will have to wait.
\emph{Right}: PNM energy per nucleon  at increasing order in Chiral EFT. The shaded lines, colored online, 
match the highlighted diagrams in the left plot. The width of the bands is obtained by varying the cutoff in loop momenta in the interval 450-600 MeV.   
\emph{Reprinted, with permission, from~\cite{Sammarruca:2016ajl} under Creative Commons License.}\vspace{0.5cm}\label{fig:counting}}
\end{center} 
\end{figure}

Still, several reliable calculations have been performed up to twice that saturation density, beyond which uncertainties were estimated by analysing the order-by-order convergence in the chiral expansion and the many-body perturbation theory \cite{Sammarruca:2014zia,Hu:2016nkw,Holt:2016pjb,Drischler:2016djf}. They were performed by extending the calculations of PNM with developed local chiral N$^2$LO and N$^3$LO interactions, including two- and three--body forces (see figure~\ref{fig:counting}).

From the data of this and other contemporary works just cited, the EoS $P(\epsilon)$ can be obtained from a derivative of the energy per nucleon,
\begin{eqnarray} \label{nonrelEoS}
\epsilon= n\left(M_N+\frac{E}{A}\right)\\ 
P= n^2\,\frac{d(E/A) }{d n}\ .
\end{eqnarray}

One caveat on this type of calculations is that they employ the counting of ChPT developed in a vacuum, and the new scale brought about by the baryon chemical potential $\mu$ or, equivalently, by the Fermi momentum $k_F$, is not quite treated on the same  footing, but typically a many body method is deployed to solve the in-medium problem. It would be cleaner to develop a counting in which the interactions in the medium are also chirally expanded and quantities that cannot be expanded are exactly resummed by means of Dyson-like equations. Exactly such a study has been carried out, though for a low order in the expansion, in~\cite{Lacour:2009ej,Oller:2019ssq}.

\subsubsection{Effect of three-body forces}

\begin{figure}
\begin{center}
\includegraphics[width=3.5in]{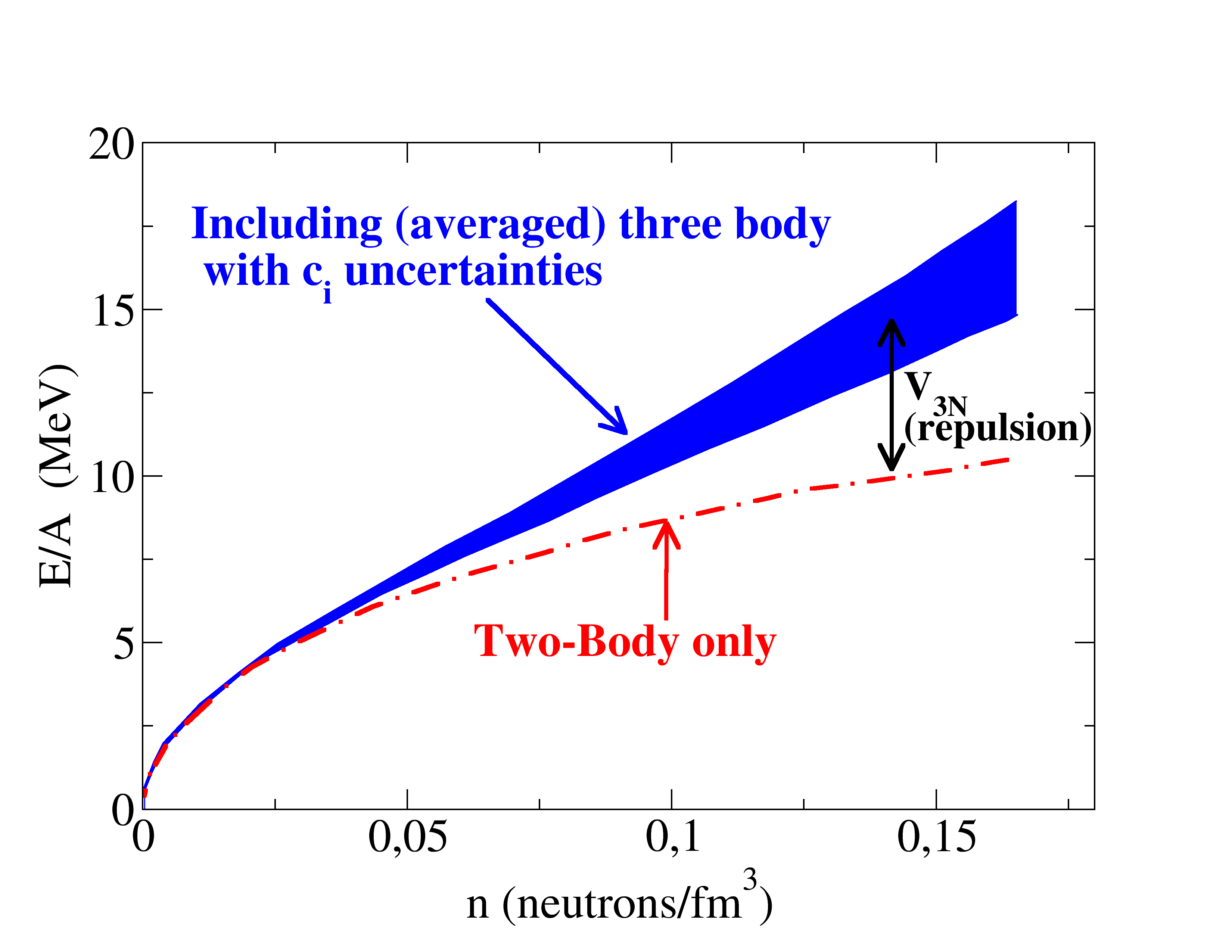}
\caption{\label{fig:3body}
The effect of the three-body forces, averaged over one of the nucleons
that is summed over the Fermi sea, is to increase the repulsion of 
the $NN(N)$ forces, and thus, to increase the energy per nucleon of the 
neutron star matter.}
\end{center}
\end{figure}

Three-body forces appear from $N^2LO$ in the chiral expansion and are necessary in nuclear spectroscopy, but pose a challenge to computation, as the resummations yielding the matter's energy per nucleon (and thus, from Eq.~(\ref{nonrelEoS}), the EoS) become multivariate integral equations. A standard approximation is to average (trace over spin and integrate over momentum) one of the
three nucleons according to 
\begin{equation}
\label{3Nforce}
\overline{V}_{3N}(1,2) = {\rm Tr_3}\int^{k_F} \frac{d{\bf k}_3}{(2\pi)^3} V_{3N}(1,2,3) 
\end{equation}
to obtain a modification of the conventional nucleon-nucleon potential.
The effect of this (approximate) inclusion of the three-body potential is shown in figure~\ref{fig:3body} (with computer data taken from \cite{Hebeler:2009iv})

The effect of the three-body force is clearly repulsive, increasing the energy per nucleon and thus stiffening the EoS. A quark-model argument of why contact terms should provide some repulsion is that the Pauli principle disfavors the proximity of the $(6d,3u)$ quark structure necessary to seed three neutrons, but that the long-range (pion-exchange induced) part of the 3N force turns out to also be repulsive is an original contribution of those investigations. 
Recently, complete formulae for the density-dependent contribution to the NN interaction due to the long-range part of the $N^3LO$ (subleading) three-body potentials have appeared~\cite{Kaiser:2019yrc}; this opens the way to an upcoming inclusion of a second order of the three-body force in neutron matter computation, which will allow for convergence checks.

\subsubsection{High density limit and pQCD}

At asymptotically high density, we can trust hadron matter to enter a regime tractable with pQCD. Actually, the asymptotic phase is CFL, but the difference in what concerns the EoS is minor. This is because the Cooper pairing in that phase distorts the Fermi sphere, of radius $k_F\sim O(400)$ MeV by an amount $2\Delta\sim O(20)$ MeV proportional to the pairing gap. 
Indeed, the EoS of quark matter in the perturbative regime could be parametrized as
a power expansion in the chemical potential,
\begin{equation}
P(\mu) = a_4 \mu^4 - a_2 m_s^2 \mu^2 + b_2 \Delta^2 \mu^2 + B +\dots
\end{equation}
At large density or chemical potential, the first term is the one discussed in the rest of this paragraph, Eq.~(\ref{pQCDPressure}). The second trades two powers of the chemical potential for two powers of the strange quark mass, which is much smaller. Only the third term sees the influence of the Cooper pairing. The fourth one, equivalent to a bag constant/vacuum energy, might even be sizeable in comparison.

Thus, we concentrate on the largest $\mu^4$ term, that has been studied in~\cite{Kurkela:2009gj,Kurkela:2014vha} and a new, partial  N$^3$LO calculation has been presented in~\cite{Gorda:2018gpy}.
We find useful a handy parametrization~\cite{Kurkela:2014vha}  of the field--theory calculation.
The pressure (in GeV$^4$) is given as a function of the chemical potential for baryons $\mu_B$ in GeV (it is straightforward to convert between $\mu_B$ and $n$)
 and a parameter $X$ proportional to the renormalization scale ($X\equiv 3 \bar\Lambda/\mu_B$) 
\begin{equation} \label{pQCDPressure}
P_{\rm pQCD} = \left[ \frac{3}{4\pi^2} \left(\frac{\mu_B}{3}\right)^4\right] \left( 0.9008 - 
\frac{0.5034 X^{-0.3553}}{\mu_B - 1.452 X^{-0.9101}}
\right)\ .
\end{equation}
This expression includes the free gas in the first factor between brackets,  and $O(\alpha_s^2)$ corrections in the second factor. The match to the pQCD regime is typically performed for values $\mu_B \approx 2.6- 2.8 GeV$.

At values of $\epsilon$ below this matching point, QCD is still strongly coupled (non-perturbative) and we are unable to predict whether deconfined quark matter exists or not at appropriate neutron star, intermediate densities. Statements in that regime are generally model dependent:
many phases with different degrees of freedom have been proposed. 

But instead of needing to extrapolate from the low--density EoS alone, the high--density limit of pQCD allows to pin any model EoS at the upper extreme of the density range, thus diminishing the uncertainty.

\subsubsection{Interpolations at intermediate density}
This is the most important range to determine many basic properties of NS, such as $M_{\rm max}$. However, it is also the region where a consistent expansion of QCD is not workable.
Extrapolation schemes are constrained by nuclear physics inputs at low densities and observational constraints, with a very popular choice~\cite{Read:2008iy,Sammarruca:2018whh}  being piece--wise polytropes $P\propto \epsilon^\gamma$.
The EoS is then much more constrained if the high--density information is included~\cite{Annala:2019puf}; a band of EoS compatible with all available information is then produced. 
This is a very recommendable reference to consult when an EoS through the neutron star with theory--controlled uncertainty band is needed.

Of course, if one intends to put bounds on departures from General Relativity, the astrophysical constraints cannot be imposed on the EoS  because they already assume GR 
(in the use of Eq.~(\ref{basicTOV}) to calculate an observational mass-radius diagram, or  of the GR gravitational wave solutions, for example).

Therefore, we have developed a set of EoS recently reported in~\cite{Oter:2019kig}
to sample EoS at intermediate densities.
We sample the uncertainty bands of Chiral Perturbation Theory (at low density), the band of perturbative QCD (at high density) and interpolate between both of them at intermediate density.
Only basic theoretical properties have been used: these are the conditions of causality and monotony encoded in
\begin{equation}\label{constraint}
0\leq \frac{dP}{d\varepsilon} \leq 1\ .
\end{equation}

For the low--density end, we  collected calculations of the neutron matter EoS from microscopic nuclear forces at different orders in Chiral effective field theory  from several standing works ~\cite{Sammarruca:2014zia,Hu:2016nkw,Holt:2016pjb,Drischler:2016djf,Hebeler:2015wxa}. 
Those sets extend only up to nucleon-number density  n of order of  1-2 $n_0(\epsilon\sim 250 $MeV/fm$^3$) (only~\cite{Hu:2016nkw}  extends the computation up to  $n= 2.5 n_0$ ($\epsilon <$ 380 MeV/fm$^3$ for N3LO), so that the nucleons can still be reasonably taken as not too relativistic.
The cutoff and order of perturbation theory is sampled.

For high--energy density, the NNLO pQCD result of~\cite{Kurkela:2009gj} is employed at baryon chemical potential above $\mu_{\rm match} \sim 2.6$ to 2.8\, GeV by using Eq.  \eqref{pQCDPressure} and taking the renormalization scale parameter  X = 1 for the lower band an X = 4 for the upper band.
 
\begin{figure}
\centerline{\includegraphics[width=4in]{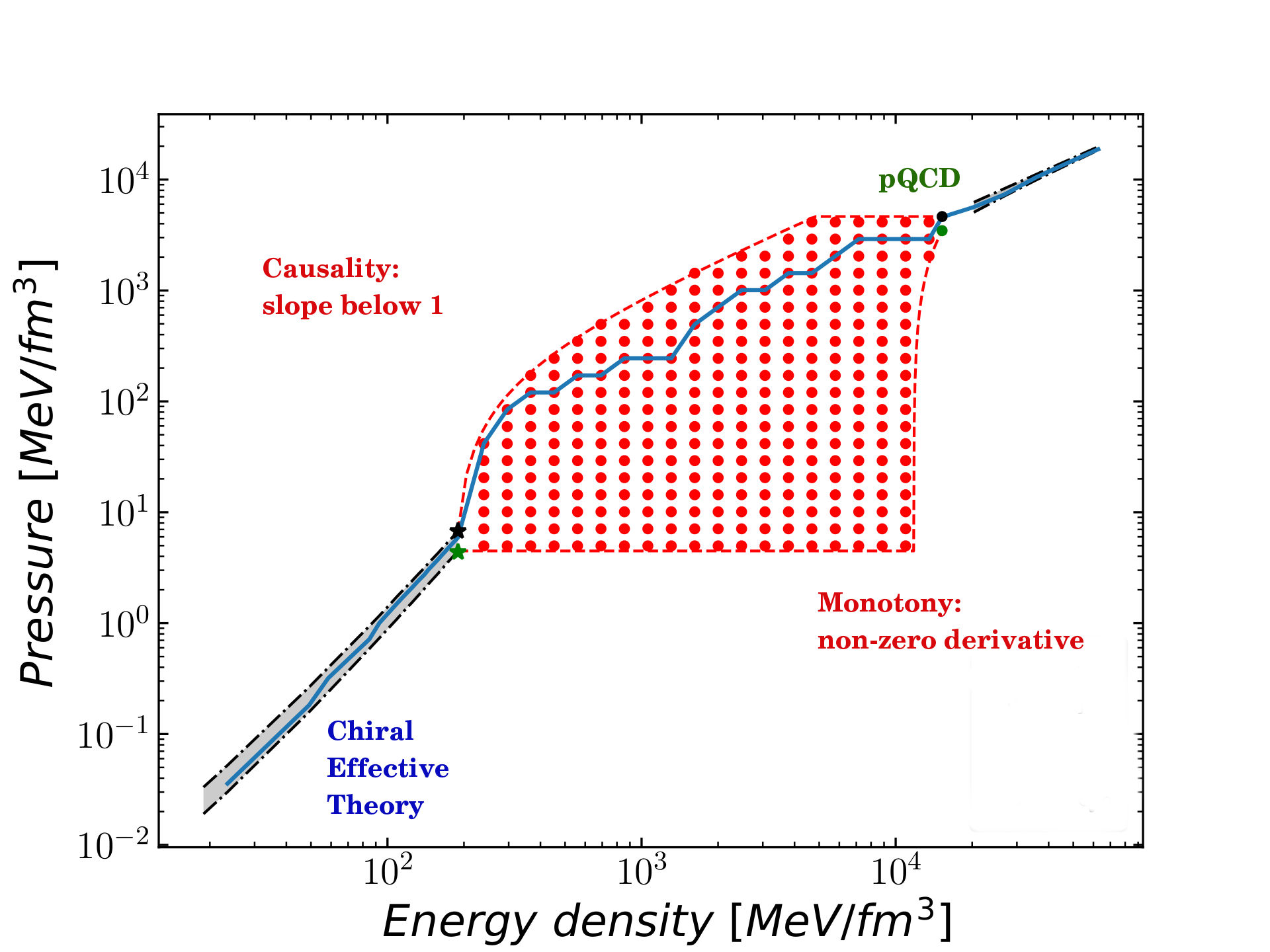}}
\caption{Scheme of the interpolation  through intermediate densities 
between a chiral computation~\cite{Drischler:2016djf}  and the high-density pQCD physics at $\mu_{\rm match} \sim 2.6$ GeV.  An example EoS is shown (blue solid line).  Because the grid points maintain a finite spacing, strong phase transitions (horizontal lines) are allowed. \label{fig:grid}
}
\end{figure}

 The boundary region resulting between chiral perturbation theory and high-density physics treated with pQCD is shown in figure~\ref{fig:grid}. The approach features a simple numerical interpolation scheme on a square (logarithmic) grid in the $(\varepsilon,P)$ energy-density/pressure two dimensional space. To each energy density point of the grid we associate a set of pressure points inside the boundary of the region highlighted in the figure. These points are used to generate the EoS candidates.     
 The construction of  the EoS requires first to fix the low density and high density limits of the intermediate band. At the low density end this is read off from the data $(\rho,E/A)$ in the references mentioned above. At the high-density end it is computed from the parametrization of~\cite{Kurkela:2014vha} given in Eq.~(\ref{pQCDPressure}).

The EoS construction is carried out in three steps.
\begin{enumerate}
\item  At low densities, by randomly selecting for each $\epsilon$ point a  value of $P$ between the upper and lower limits of the allowed pressure band corresponding to each set. 
Because $P$ grows very slowly within the low--$\epsilon$ band, touching the bound $c_s>c$ is extremely unlikely and thus not a concern, so that  only the monotony condition is explicitly enforced. 
\item Through the intermediate density area:
i) first, appropriate matching points with the low- and high- density regions are prepared.
ii) A logarithmic grid in the ($\epsilon$, P) space (red points in figure~\ref{fig:grid}) is prepared. 
iii)finally, starting from a randomly chosen starting point at the bottom end, 
a forward random walk by choosing points of the grid is effected, with the constraints of monotony and causality imposed in a Von Neumann rejection scheme.
\item At high densities, we again randomly selected $P$  between the upper and lower limits of the  pQCD band for each energy density point. Once more, only the monotony condition is needed. Here, c$_s^{2}$ approaches 1/3 from below and again, exceeding the speed of light is not a concern.  
\end{enumerate}

\begin{figure} \begin{center}    \includegraphics[width=3.2in]{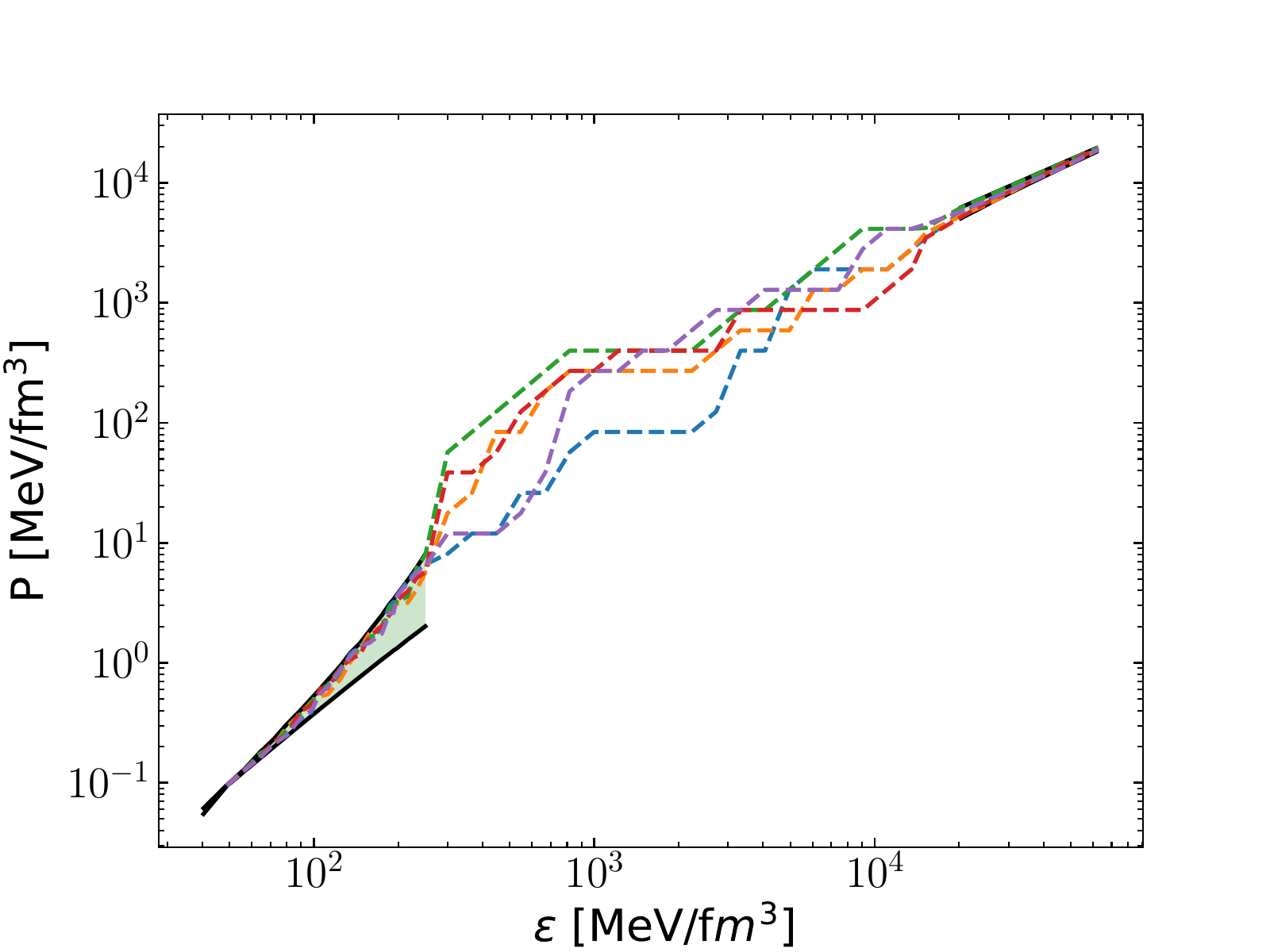} \ \
\includegraphics[width=3.2in]{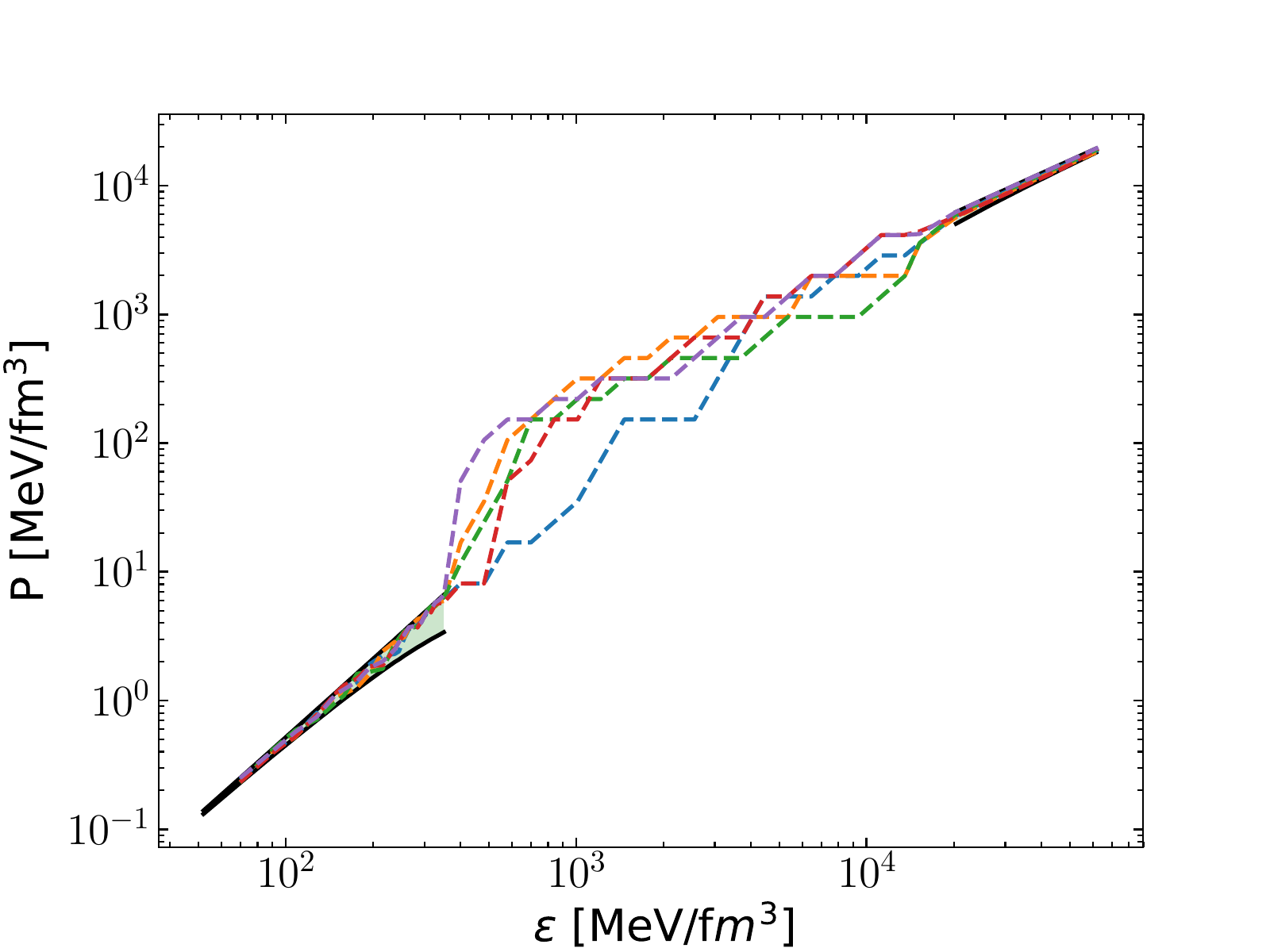} \\
   \includegraphics[width=3.2in]{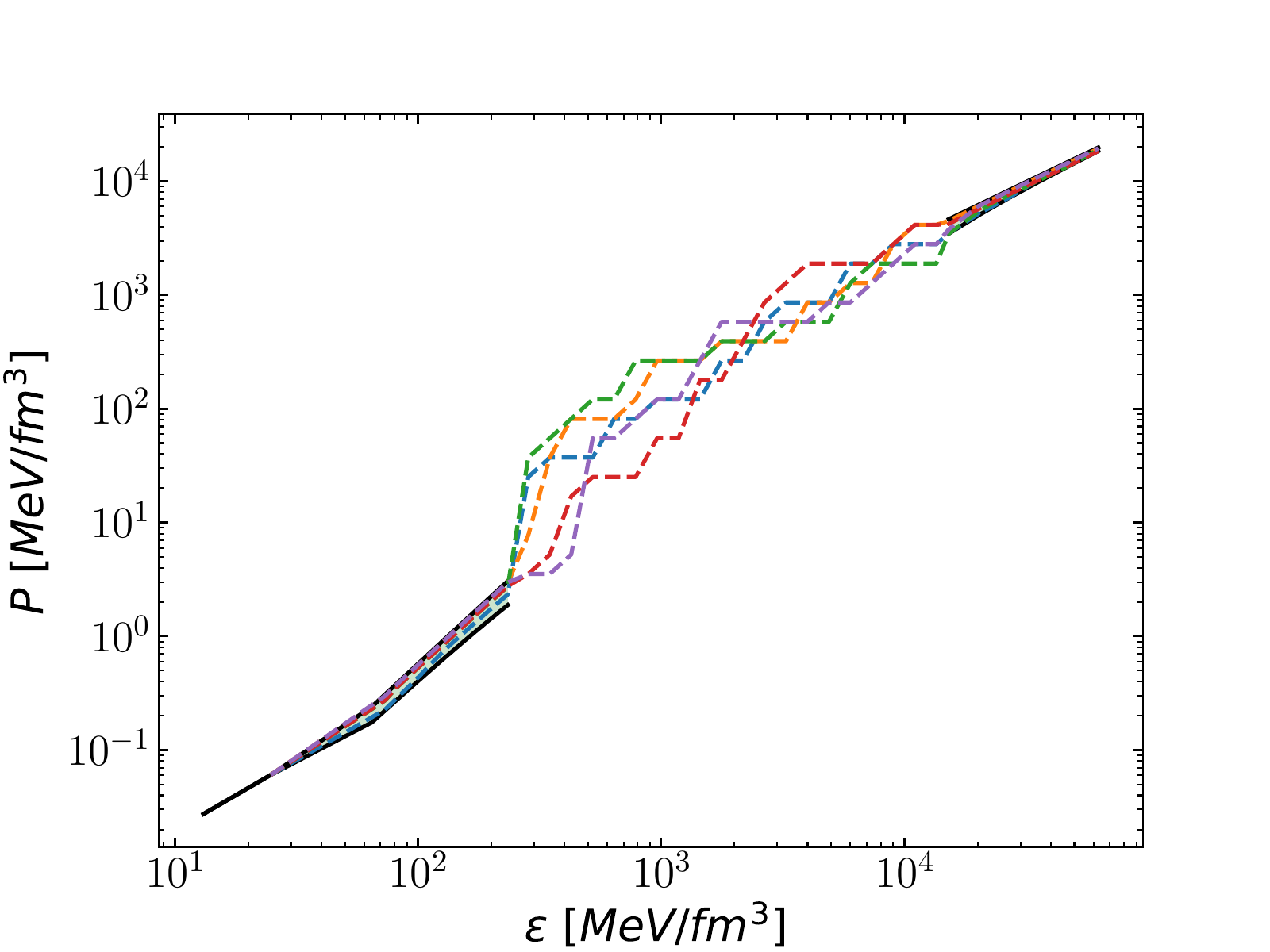}\ \ 
   \includegraphics[width=3.2in]{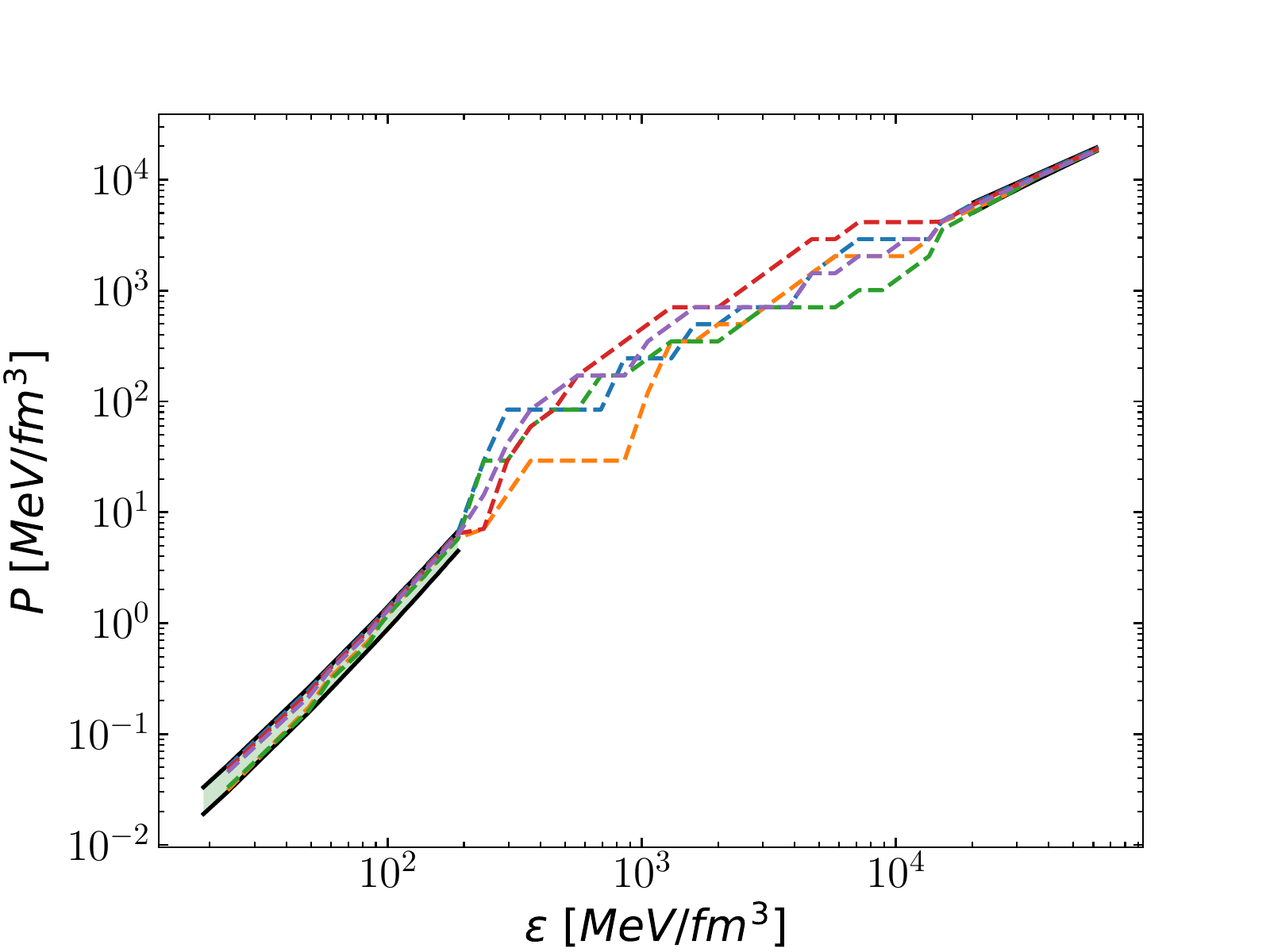} \\
  \caption{Several example EoS satisfying only constraints from hadron physics. The high--density band is the pQCD constraint from Kurkela \emph{et al.}~\cite{Kurkela:2014vha} matched at $\mu_{\rm match}$=2.6 GeV. At intermediate densities only monotony and causality are imposed.The low-density band is constrained by NLO chiral potentials. Clockwise from the top left, from~\cite{Sammarruca:2014zia}  with cutoff $\Lambda=450$ and 600 MeV; from~\cite{Holt:2016pjb} with cutoff at 450 and 500 MeV; from~\cite{Hu:2016nkw} at N$^3$LO with R=0.9 fm and R=1.0 fm;
  and from~\cite{Drischler:2016djf}. For the sensitivity to the choice of $\mu_{\rm match}$ see the analogous plots with $\mu_{\rm match}=2.8$ GeV in~\cite{Oter:2019kig}.
 \label{fig:EoSsamples}
} \end{center}
\end{figure}

The code, tables with 1000 EoS each and smaller samples of the EoS obtained are assembled at the already mentioned website
{\tt http://teorica.fis.ucm.es/nEoS/}. A few samples are shown in figure~\ref{fig:EoSsamples}.

The resulting tidal deformability and mass-radius diagrams are as discussed in section~\ref{sec:static}.

\newpage
\subsection{EoS at finite (but small) temperature} \label{subsec:finiteT}

\begin{figure}
\begin{center}
\includegraphics[width=0.5\columnwidth]{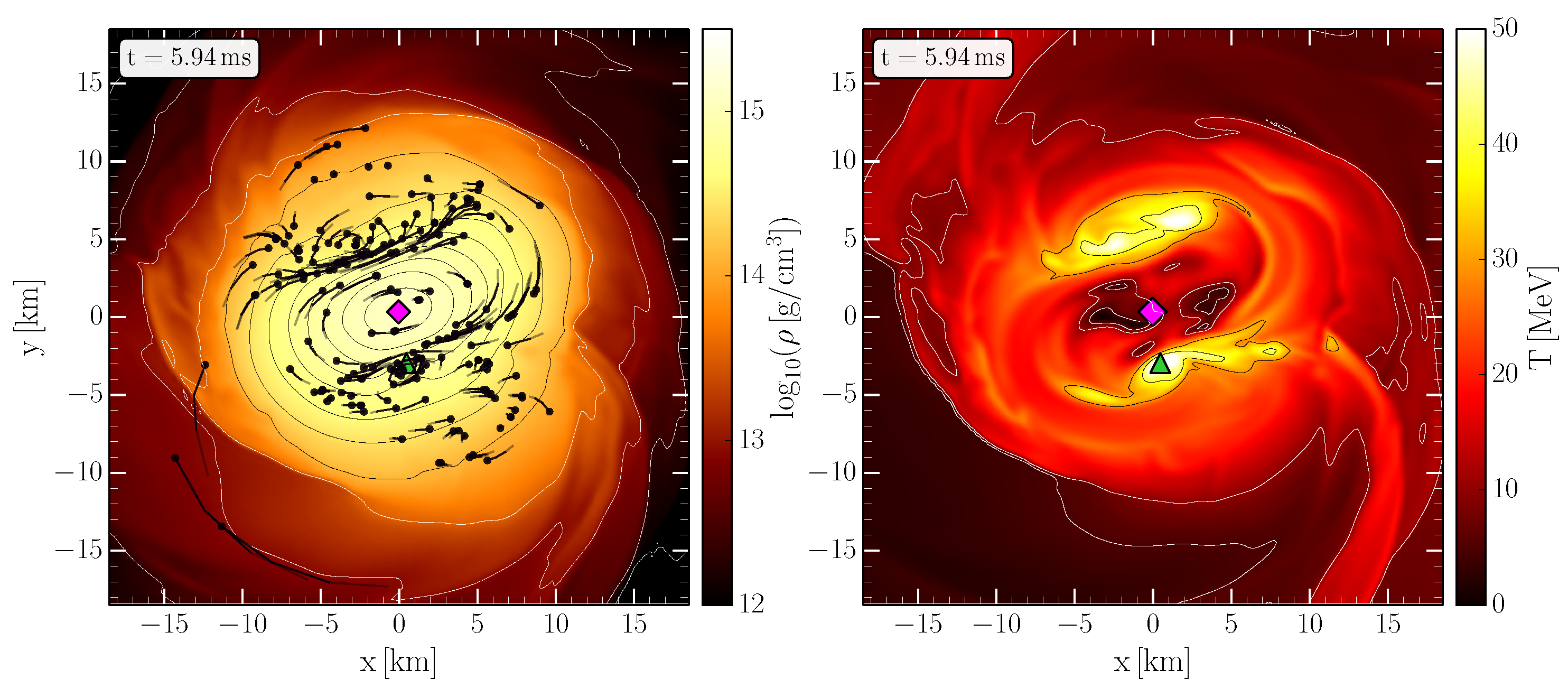}
\caption{\label{fig:FiniteT1} Snapshot of the density profile (left) and the temperature profile (right) about 6ms after the merger of two neutron stars, in a simulation reported in~\cite{Hanauske:2019qgs}. The temperature typically reaches several tens of MeV, here up to 50 MeV. This is large enough that one can question the adequacy of the EoS computed in cold nuclear matter at $T=0$ to describe the merger dynamics.
\emph{Reproduced from~\cite{Hanauske:2019qgs} under the terms of the} \href{https://creativecommons.org/licenses/by/4.0/}{\tt Creative Commons 4.0 License}.
}
\end{center}
\end{figure}

Sufficiently old Neutron Stars have settled in a configuration of high density and low temperature, ``cold nuclear matter'', so that calculations of the EoS are traditionally carried at $T=0$.
However, a binary star mergers frees large amounts of gravitational energy that heats up the neutron matter up to 50 or more MeV (see figure~\ref{fig:FiniteT1}). 
A recent detailed study of the pressure and temperature conditions after the merger~\cite{Perego:2019adq} shows temperatures between 10 and 100 MeV; demonstrates that trapped neutrinos decrease the pressure at most 5\% (because their appearance as an additional degree of freedom softens the EoS as usual) and the temperature at most 10\%, so they are not very crucial; and suggests that at the higher temperatures one could even have thermal pions, which is very interesting because the NN interaction potential would then be modified. 

Knowing the finite temperature EoS is thus necessary to calculate quantities with observational impact such as simulating the amount of ejected mass yielding the electromagnetic kilonova afterglow.

Current practice in the simulation of these events is to split the EoS in the sum of two parts, one of cold matter and one of finite temperature,
\begin{eqnarray} \label{thermalpressure}
P(T,\mu) \simeq P(T=0) + P_{\rm thermal}\\
\epsilon(T,\mu) \simeq \epsilon(T=0) + \epsilon_{\rm thermal}
\end{eqnarray}
with the thermal part taken from the ideal gas law, 
$P_{\rm thermal} = \frac{\rho}{M_N}k_B T$ and $\rho$ the rest mass density. Contributions due to relativistic particles $\propto T^4$ and thermal corrections due to the nuclear part of order $T^2$ can be found in a very recent paper~\cite{Hanauske:2019qgs}.  

The thermal pressure is also often expressed in terms of an adiabatic index $\Gamma$ and the thermal energy density,
\begin{equation} \label{Adiabindex}
P_{\rm thermal} =(\Gamma-1) \epsilon_{\rm thermal}
\end{equation}
where $\Gamma$ is taken, for many astrophysical applications, to be constant (a nonrelativistic ideal gas would have $\Gamma=5/3$ while the relativistic value is $\Gamma=4/3$~\cite{Zeldovich}). 

Such semi--empirical methods have also recently been improved greatly in~\cite{Raithel:2019gws} by splitting the energy--density interval carefully and considering what contribution is dominant in each one, but they clearly remain a very crude approximation that can be improved upon by field-theory based fundamental studies of nuclear matter. While chiral perturbation theory results are not yet at hand, there is at least one microscopic calculation using the Brueckner-Hartree-Fock method that we briefly describe~\cite{Lu:2019mza}.

The free energy density of the nucleonic part (dominant) is obtained from 
\begin{equation}\label{BruecknerHF}
f_N= \sum_{i=n,p} \left[
2\sum_k f_i(k)  \left( \frac{k_i^2}{2m_i} + \frac{1}{2} U_i(k)\right)-Ts_i \,
\right]
\end{equation}
where $f_i$ is the Fermi--Dirac free--gas number density $(exp((e_i-\mu_i)/T)+1)^{-1}$ and $s_i$ its associated entropy density for each of the proton and neutron components,
\begin{equation}
s_i = -2\sum_k \left(f_i(k)\ln f_i(k) + (1-f_i(k))\ln (1-f_i(k))\right)\ . \
\end{equation}
The potential is in turn obtained from a self--consistent equation that yields the in--medio $K$--matrix for nucleons 1 and 2
\begin{equation}
K(E) = V + V {\rm Re}\sum_{1,2} \frac{\arrowvert 12\rangle (1-f_1)(1-f_2) \langle 12 \arrowvert}{E-e_1-e_2+i\epsilon} K(E)
\end{equation}
from which it is obtained for nucleon 1 after tracing over 2 (mean--field approximation for 1),
\begin{equation}
U_1 = {\rm Re} \sum_2 f_2 \langle 12\arrowvert K(e_1+e_2) \arrowvert 12\rangle\ .
\end{equation}
The temperature dependence accompanies the density dependence through the Fermi--Dirac function in a natural way. (Thermodynamic consistency however requires that the free--gas guess for the chemical potential $\mu_i$ be recalculated after Eq.~(\ref{BruecknerHF}) to yield and updated chemical potential and particle density.)

\begin{figure}[h]
\begin{minipage}{0.45\textwidth}
\begin{center}
\includegraphics[width=8cm]{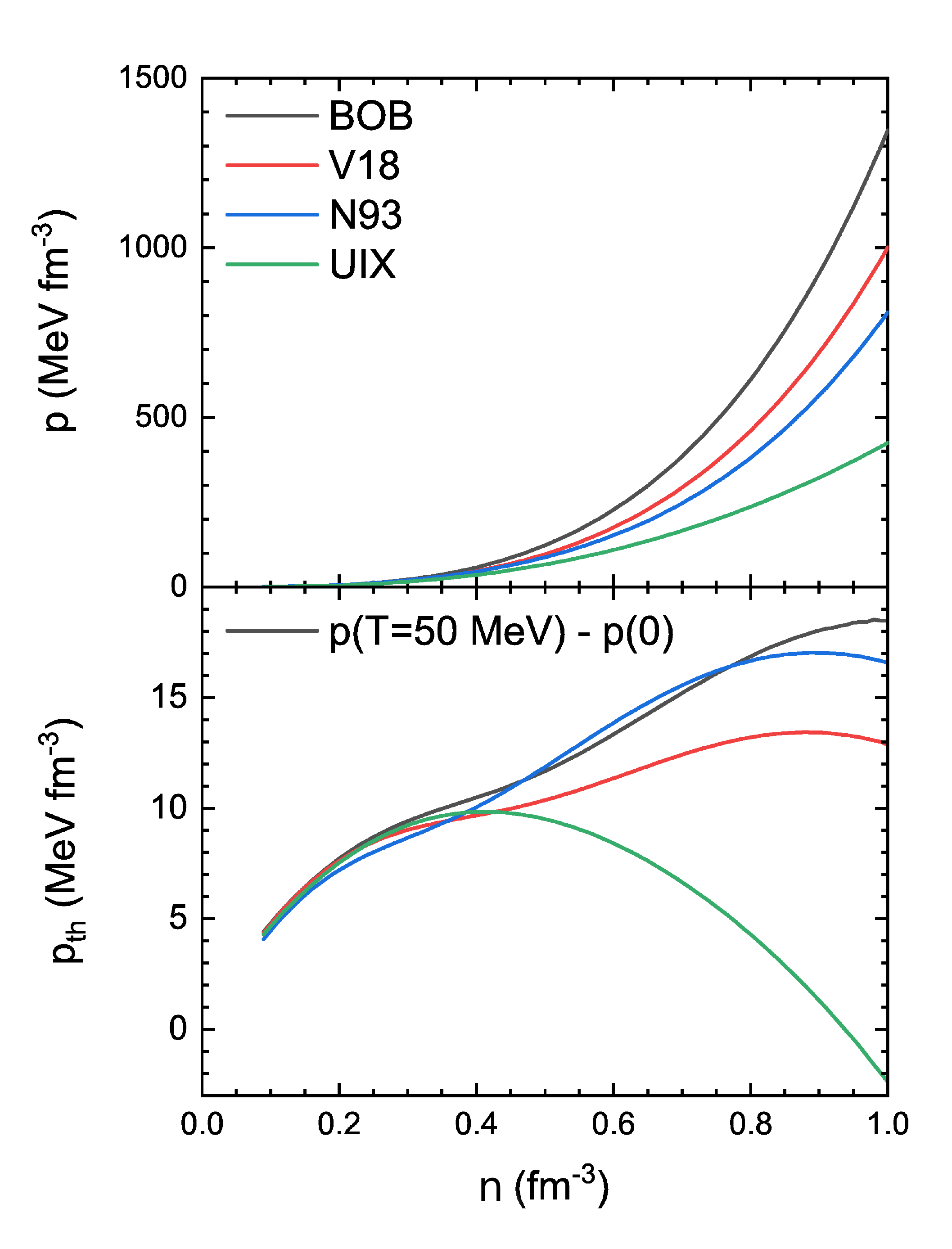}
\end{center}
\end{minipage}
\begin{minipage}{0.45\textwidth}
\begin{center}
\caption{\label{fig:thermalEoSBHF} Brueckner-Hartree-Fock estimate of the thermal contribution to the EoS $P(n)$ (instead of $P(\varepsilon$). Top plot: several model microscopic EoS at $T=0$. Bottom plot: the thermal corrections to those models at $T=50$ MeV are seen to be smaller by a large factor of order 10-100.
\emph{Adapted from~\cite{Lu:2019mza}, courtesy of Jiajing Lu.}}
\end{center}
\end{minipage}
\end{figure}

The method has been pursued with several microscopic potentials and the thermal corrections, shown in figure~\ref{fig:thermalEoSBHF}, are seen to be of order 1-5\% up to a temperature of 50 MeV: density is still the dominant feature in neutron star mergers.

Another interesting \emph{ab-initio} study~\cite{Carbone:2019pkr} employs the chiral interactions within a self-consistent Green's function method (based on the Schwinger-Dyson equation for the nonrelativistic nucleon propagator within ChPT), in a similar spirit to the study just described, and explores the values of the adiabatic index~$\Gamma$ in Eq.~(\ref{Adiabindex}). Far from being constant, $\Gamma$ varies from about 1.65 (near 5/3) at low density down to 1.25--1.45 depending on the level of approximation (and consistently with 4/3, though the computation is not relativistic) for $n=0.3\ {\rm fm}^{-3}\simeq 2n_0$.

\begin{figure}
\begin{minipage}{0.45\textwidth}
\begin{center}
\includegraphics[width=8cm]{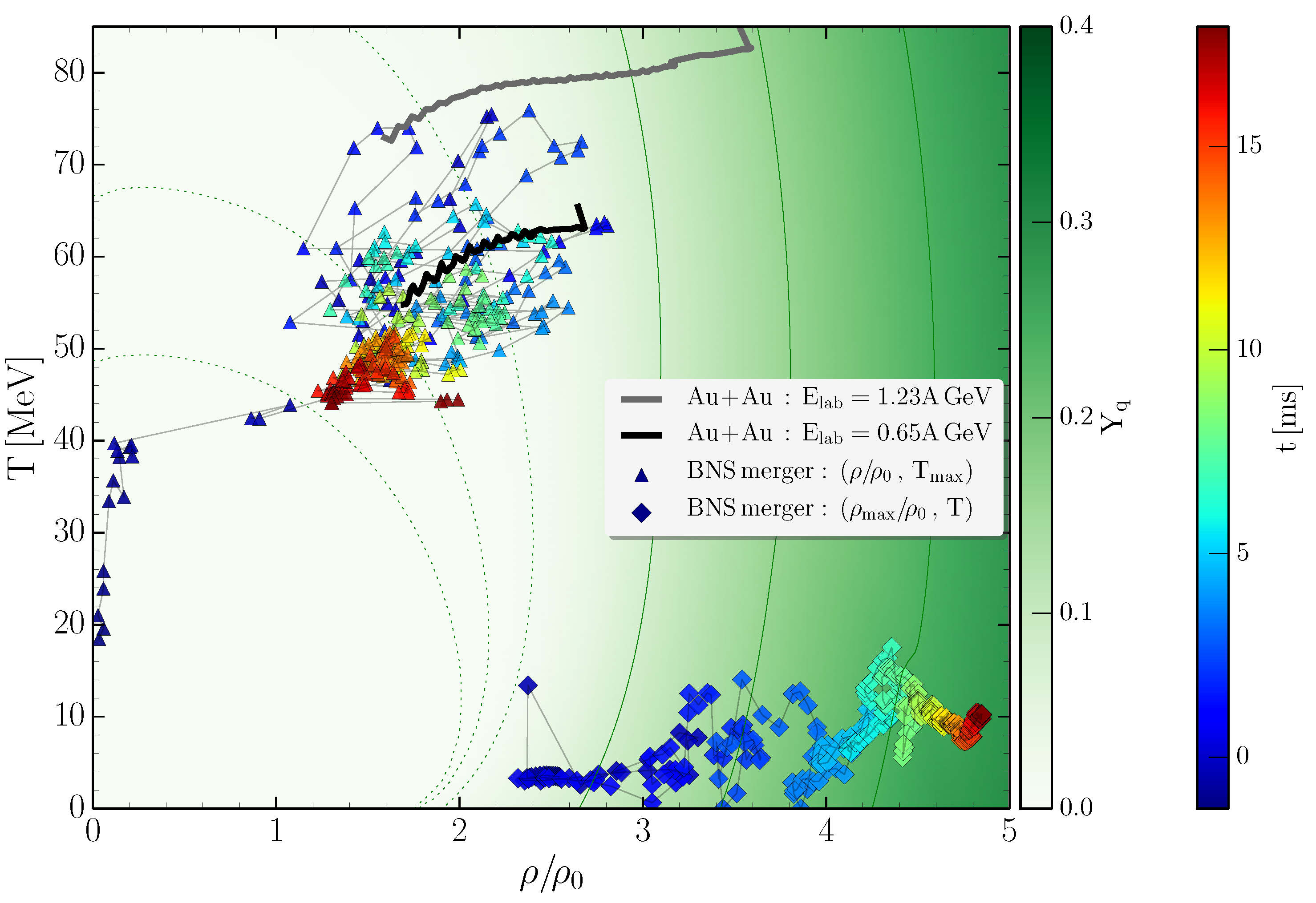}
\end{center}
\end{minipage}
\begin{minipage}{0.45\textwidth}
\begin{center}
\caption{\label{fig:FiniteT2}
Maximum value of the temperature (triangles) and rest-mass density (diamonds) at the equatorial plane inside the merger remains in a simulation as function of time after merger in ms.
Near the top are a gray and a black lines: they correspond to simulations of Au-Au ion collisions
with energies E=1.23 $A$ GeV and E=0.65 $A$ GeV. 
(The background color corresponds to a model-dependent assignment of quark fractions in the dense matter and is not discussed here). 
\emph{Reproduced from~\cite{Hanauske:2019qgs} under the} 
\href{https://creativecommons.org/licenses/by/4.0/}{Creative Commons 4.0 License}.
}
\end{center}
\end{minipage}
\end{figure}

Such a novel feature of neutron star mergers as probing the thermal EoS adds value to the heavy-ion collision experimental program; when two ions collide, the energy density takes values well in excess of that in nuclei at rest, but unavoidably, the nuclear matter is also heated. Heretofore this was an obstacle to really relate the EoS measured in the laboratory (of necessity at finite $T$) with the EoS used in isolated neutron stars (at $T=0$). But if merger data needs and provides feedback on the EoS at finite temperature, it will bridge the gap between the two fields. This is just what the simulation in figure~\ref{fig:FiniteT2} is showing.

Indeed, the simulation of the Gold-Gold ion collisions in the picture nicely overlaps with the region of interest for the maximum temperatures reached in the merger. Thus, a new promising avenue of research for nuclear matter has appeared. The EoS determined from the nuclear experiments can be used in astrophysical simulations, in which the temperature will be an internal field of the simulation hard to measure directly.
It is also worth observing that in the field of Heavy Ion Collisions, the EoS at sizeable temperature but negligible baryon density (and even perhaps small baryon density, through Taylor expansion) is amenable to first--principles calculations within lattice QCD. At smaller temperatures, a matching can be performed to chiral perturbation theory or to hadron--resonance simulations~\cite{Monnai:2019hkn}, providing one more solid theory anchor point for EoS studies.

One remaining obstacle is the different isospin composition of the nuclear matter in a binary merger (very asymmetric, with a very small proton fraction) and in ion collisions (with a significant proton fraction). An extrapolation in isospin will be necessary for a meaningful comparison, perhaps employing data from different nuclei.

As a last observation, let us note that the finite temperature range available in merger can provide sensitivity to various quark matter phases, particularly unpaired quark matter, that might not be obtainable in static stars (due to insufficient density or temperature). The next section~\ref{sec:Phases} describes some.

\newpage
\section{Phase transitions in hadron matter}\label{sec:Phases}

\subsection{Neutron Star Structure: a dozen phases!}

Different phases of matter are characterized by different symmetries, whether spatial --whose breaking leads to inhomogeneous condensates and crystals, or EM gauge invariance, yielding superconductors, or flavor/color internal quantum numbers~\cite{Alford:2019oge}. (A recent review~\cite{Orsaria:2019ftf} tries to address their possible influence in the emission of Gravitational Waves.)
Particularly, there is much variety of possible quark matter phases intermediate between hadrons and the asymptotic Color-Flavor Locked phase, due to the 9 fermion species (u,d,s flavors in three colors) that can form a variety of Cooper pairs.

\begin{figure}
\begin{center}
\includegraphics[width=0.75\columnwidth]{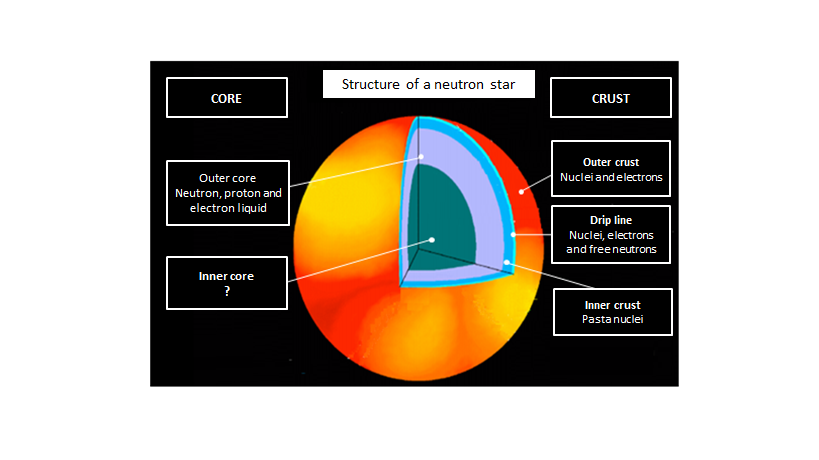}
\caption{
Schematic diagram (not to scale) of a NS interior, showing the outer crust, the inner crust (neutron drip regime with a sea of free neutrons), the region of non-spherical pasta nuclei in the inner crust; in the outer core a hadronic liquid, and perhaps a quark liquid in the inner core.
(Our much simplified rendering of the figure in~\cite{Piekarewicz:2009zc}).
\label{fig:structure}
}
\end{center}
\end{figure}

The presumed structure of a neutron star is shown in figure~\ref{fig:structure}. 
It can be divided into a crust (outer and inner, with certain confidence) and a core (where the number of layers with different phases of nuclear matter is not yet clear).

For  $\epsilon<10^4$ g/cm$^3$, one expects an atmosphere of partially ionized atoms and electrons.
At higher densities, the region composed of inhomogeneous nucleonic matter and unlocalized $e^-$ in $\beta$-equilibrium is called the crust. Beneath this outer crust with a plasma of nuclei and electrons as degrees of freedom, an inner crust with unbound neutrons likely exists. 
The outer crust surface layer at $\epsilon\sim 10^7$ g/cm$^3$ is made of $^{56}Fe$ and neighboring ions immersed in a sea of electrons. With increasing $\epsilon$ the matter becomes more and more neutron rich, evolving from $^{56}Fe$ (ideally) at the surface to $^{62}Ni$ and then $^{64}Ni$. Beyond, the nuclear shell model predicts a sequence of nuclei with an $N=50$ closed shell, spanning between $^{84}Se$ and $^{76}Ni$, followed by one with an $N=82$ closed shell, from $^{124}Mo$ to $^{118}Kr$.

Further, nuclei become more massive and neutron rich, reaching the neutron drip line (the last neutron is bound at zero energy) defining the boundary to the inner crust ($\epsilon \sim 4\cdot 10^{11}$ g/cm$^3\simeq 0.22$MeV/fm$^3$). There, the $\mu_n$ becomes too high for nuclei at the lattice sites to bind additional neutrons. A sea of free neutrons permeates that inner crust, and they are believed to be BCS-paired in a $^1S_0$ state. Then calculations suggest a sequence of ``pasta'' phases in which nuclei, 
equilibrating surface and Coulomb energies, become rod- and later sheet- shaped. Finally, the crust probably dissolves into a sea of free neutrons, with a smaller fraction of free protons and electrons \cite{Baym:2019yyo}. Representative calculations of the equation of state in the crust, particularly below the pasta region, include Refs. \cite{Douchin:2001sv,Baym:1971pw,Haensel:2002cia}, while details of the pasta phases can be found in Refs. \cite{Ravenhall:1983uh,Williams:1985prf,Oyamatsu:1993zz,Watanabe:2004tr}.
At about half the saturation density $n_0/2 = 0.08 $fm$^3$, that is, $\epsilon \approx 2.7\times 10^{14}$ g/cm$^3$, the nuclei completely melt and the NS core could begin. 

Up to that point, nuclear theory seems a reasonable guide. But important features of the hadronic liquid interior remain unsolved. First, nuclear many-body forces remain uncalculated, particularly with increasing density, though Chiral Effective Theory now provides a way to address them. In addition, computations of neutron--rich nuclear matter are lacking the accuracy  of those for pure neutron matter ($Z=0$) and symmetric nuclear matter ($Z=N$).

At even higher densities, above $\approx$ 2$n_0$, the composition of matter is basically unknown. 
Other particles than nucleons and electrons are expected to appear such as hyperon matter \cite{Chatterjee:2015pua} or quark matter (see e.g. \cite{Weber:2004kj,Alford:2006vz,Weissenborn:2011qu,Buballa:2014jta,Zacchi:2015lwa}. In the literature muons, pions, kaons and their condensates, hyperons, nuclear resonances and quarks have been considered \cite{Glendenning:1997wn}. 
Further various phases of pure quark matter~\cite{Haensel:1987fy} are under investigation, and even absolutely stable strange quark matter~\cite{Witten:1984wi} has been proposed in spite of the $m_s-m_d\sim 100$MeV mass difference. At not too high temperatures the appearance of clusters and crystalline structures is expected.

In summary, the outer regions of the NS, which are better understood, consist of a solid crust and a nuclear matter liquid just inside the crust, spanning baryon densities up to 1-2 $n_0$. The NS interior, for  baryon densities  $n \sim$ (2-10)$n_0$ is the most difficult to describe from first principles. Here, hyperons, quark matter in various phases (presumably unpaired at finite $T$, otherwise partly or totally condensed such as in the 2SC or CFL phases respectively) may exist, but $\mu<\mu_{pQCD}$. Only at much higher densities (n$_B >$ 10n$_0$, beyond those expected in neutron stars) can one apply perturbative QCD for dense matter directly.

Several insightful models (but with uncontrolled uncertainties) have been applied to describe quark matter in compact objects: the MIT bag model~\cite{Nicotra:2006eg}, the NJL model~\cite{Buballa:2008zza}, and the Dyson--Schwinger Equations~\cite{Muller:2013pya}. Alternatively, instead of attempting a Lagrangian-based description, one can try to parametrize our ignorance.
For example, a density-independent speed of sound with a jump to allow for a phase transition has been investigated in \cite{Chamel:2013wza,Zdunik:2012dj,Alford:2013aca}.

If the degrees of freedom at high $\epsilon$ are taken to be quarks, a hybrid EoS with phase transitions is used: hadronic EoS for the crust and the outer core (considering not only nucleons but also hyperons in the inner core) and a high-$\epsilon$ EoS describing quark matter. 

It is believed that quark matter, when it appears in neutron stars, is too soft to support high masses against gravitational collapse, and as a consequence quark cores in neutron stars must be small \cite{Baym:1976yu}.

Therefore, we can consider three types of Neutron Stars: purely nuclear matter stars (NS), hybrid hadron/quark matter star (HS) and Strange quark matter star (SQS).
In the first case (NS),  purely nucleonic EoS that has been adapted to the description of neutron stars whose cores are assumed to be non-exotic.

\subsection{Construction of an EoS with explicit phase transitions} \label{subsec:phases}

That body of theoretical work needs to be followed up by astrophysical observation. So a focus of the community is to construct observables that can be examined. Often here the intermediate steps in a calculation -such as the EoS itself- are simplified. Constructing a unified EoS from the surface to the center of a NS is a challenging task.

In addition, the stiffness or softness (larger or smaller sound speed) of the EoS is not a closed issue.

That  $M_{\rm max}\sim$ 2.0-2.2$M_\odot$ needs to be reached requires stiffness. But finding new degrees of freedom by definition relaxes the thermodynamic potential and softens the EoS.
For example, hyperon EoS are difficult to reconcile with a 2 $M_\odot$ NS. However, one expects that hyperons appear at $n \sim$ 2--3$n_0$.  Sometimes this is called
the ``hyperon puzzle'' in the literature \cite{Lonardoni:2014bwa}. A similar effect is observed with nuclear resonances \cite{Drago:2014oja} and meson condensates.

There are two credible solutions to the puzzle. 
One is that the chemical potential of the hyperons $\mu_\Lambda$ only becomes smaller than the neutron $\mu_n$ one at very large densities, perhaps too high to be relevant in neutron stars.

\begin{figure}
\begin{center}
\includegraphics[width=8cm]{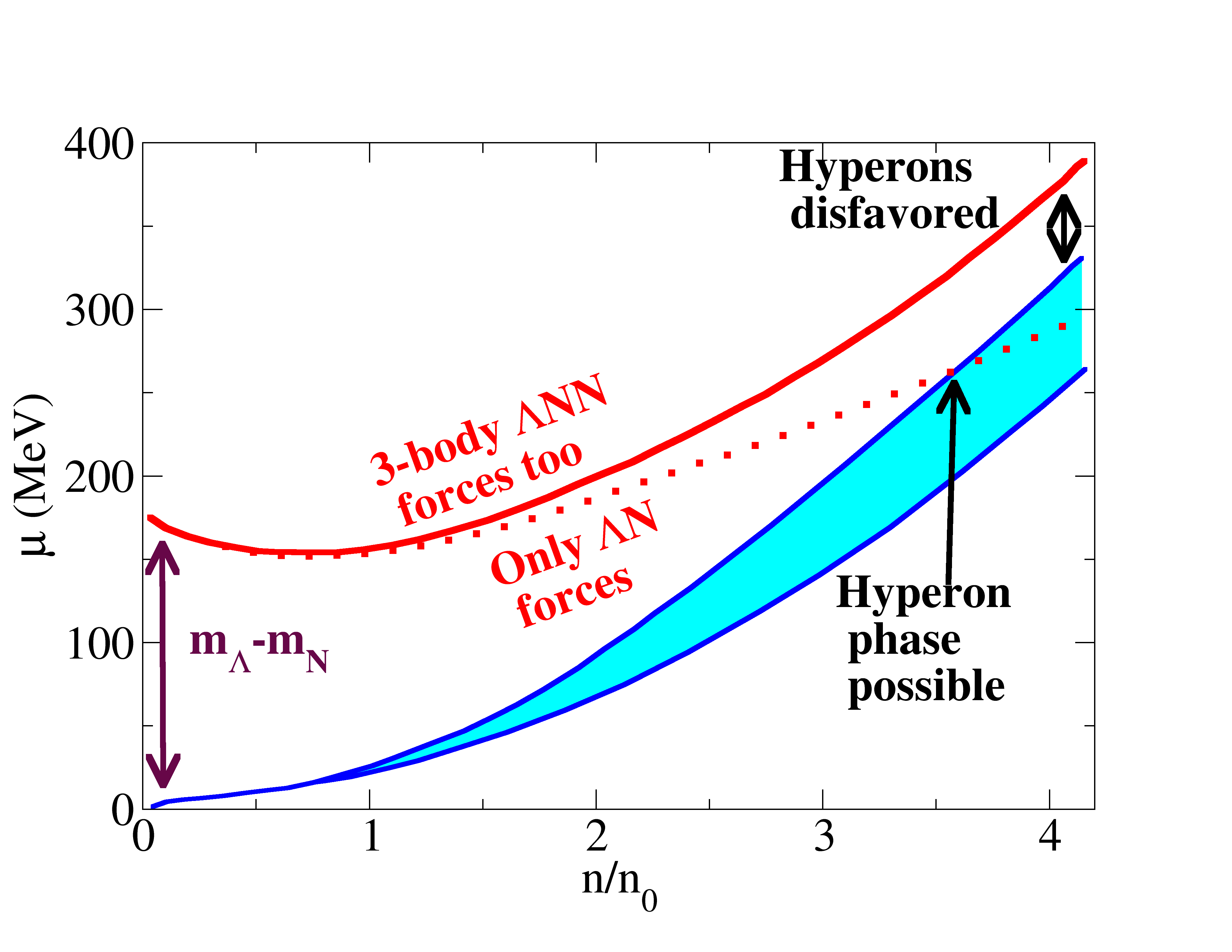}
\caption{\label{fig:Weise}
A possible solution of the hyperon puzzle: while two--body forces alone could make the production of hyperons energetically favorable (the dotted $\mu_\Lambda$ crosses into the $\mu_n$ band, and if $\mu_\Lambda<\mu_n$ additional baryons will be hyperons),
three-body forces from including $\langle V_{\Lambda NN}\rangle$ make the top line stay away from the band. Therefore hyperons are less favored and the EoS does not soften by the additional degrees of freedom.
Data from figure 7 of~\cite{Weise:2019mou}.}
\end{center}
\end{figure}

Earlier calculations, based on two-body $\Lambda N$ interactions constructed from $SU(3)$ symmetry and ChPT did not indicate that this would be the case, and indeed the corresponding $\mu_\Lambda(n/n_0)$ line in figure~\ref{fig:Weise} enters the band of credible $\mu_n$ indicating that it should be favorable to produce strangeness.
However, the top $\mu_\Lambda(n/n_0)$ line in the figure, computed including $\Lambda NN$ three--body forces, stays away and above  the $\mu_n$ band much longer~\cite{Weise:2019mou} (making the production of hyperons disfavored).

\begin{figure}
\begin{center}
\includegraphics[width=7cm]{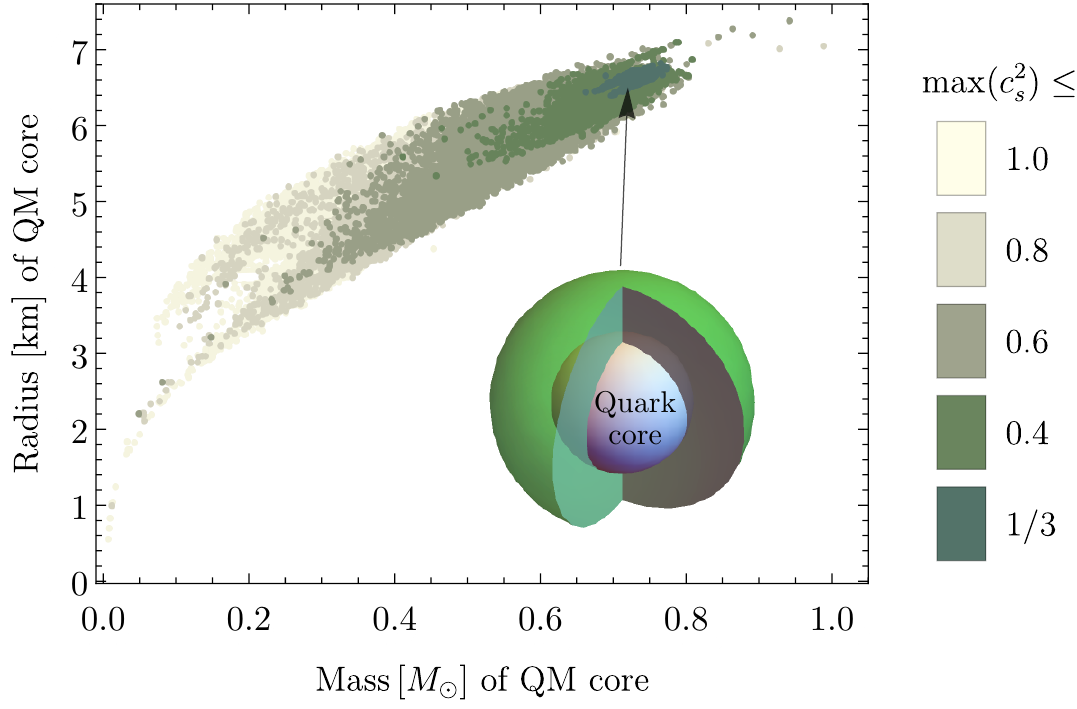}
\caption{\label{fig:core} Simulated  size and mass of a neutron star's quark core as function of the maximum value that the sound speed is allowed to take within the star. Even for EoS that would not exceed the conformal value $c_s^2=1/3$ there could be a window, on the top right, supporting a quark core. But smaller cores seem to require some significant repulsion exceeding that limit, which is natural in the hadron phase. 
\emph{Original figure courtesy of Tyler Gorda, based on~\cite{Annala:2019puf}.}}
\end{center}
\end{figure}

A second possibility is that the PNM phase before any exotic degrees of freedom appear is extremely stiff (very repulsive $NN$ interactions saturating causality), so that room for the phase transition's softening is made. Such later softening is even welcome to accommodate the small values of the tidal deformability deduced from the aLIGO data.
(A recent exception has been provided~\cite{Annala:2019puf}, in which a set of EoS that are not stiffer than the conformal limit, that is, $c_s^2<1/3$ for the entire density range can still produce a quark core and not disagree with any observables, as shown in figure~\ref{fig:core}.) 

Because of the difficulty in treating hadrons and quarks in an integrated framework, the conventional description of the onset of quark matter has been to construct  a thermodynamically consistent transition bridging between the (hadronic) nuclear and quark matter distinct phases.
Stars computed with such a "hybrid" equation of state - hybrid stars - consist typically of a small quark matter core surrounded by hadronic matter and we turn to them next.

Two possibilities for the transition are employing a strong first order hadron-quark phase transition (Maxwell construction) or a continuous crossover (Gibbs construction).

Other extrapolation schemes in use are constrained by nuclear physics input at low densities and observational constraints, such as, {\it e.g.},  a scheme  based on piecewise polytropes, a scheme based on parametrizing the speed of sound (the derivative of $P(\epsilon)$ instead of the function itself) or the scheme based on  phenomenological interpolation. 

In the next paragraphs we present a summary and commentary of some of the many recent publications related to constructing a hybrid EoS. The literature explores very thoroughly a first order interface (sharp interface) facing the uncertainty in estimating the values of the surface tension $\sigma$, but all options are under consideration.

\subsubsection{First order phase transition}

In the first order case, one simply determines at which critical $\mu_c$ the pressure of the two phases are equal to find the transition point (Gibbs condition for thermodynamic equilibrium):
\begin{equation}\label{Gibbscondition}
      T^H = T^Q = 0 \ , \ \ \ \ 
      \mu  = \mu_c \ , \ \ \ \ 
      P^H = P^Q = P_c\ .
\end{equation}

 By construction, this ``switches'' from a given hadronic EoS to a softer quark matter EoS. 
 Therefore, this procedure requires a nuclear EoS which is stiff enough to support at least a two solar mass neutron star, and a quark matter EoS which is softer but stiff enough to do the same.

Additionally, the long-range electromagnetic interaction between all constituents has to be considered, specifically requiring charge neutrality. Recently, surface effects and the range of the interactions have explicitly been taken into account. The controlling parameter is the surface tension between phases. Typically, for high surface tensions, the phases tend to approach a configuration which resembles the case of local charge neutrality  \cite{Heiselberg:1992dx,Maruyama:2007ss,Yasutake:2014oxa}. Infinite surface tension corresponds to the Maxwell first order construction and thus to $\Delta P = 0$, while a vanishing surface tension returns the crossover construction under global charge conservation.   

Whenever the surface tension is smaller than a critical value $\sigma_c$, a mixed phase could develop. The obvious example is the physics of pasta phases, a complicated problem taking into account the different sizes and shapes of possible structures as well as transitions between them. It has been dealt with in the literature within different methods and approximations \cite{Maruyama:2007ss,Yasutake:2014oxa,Watanabe:2003xu,Horowitz:2005zb}.
 
At the hadron-quark transition, despite the efforts made to constrain $\sigma$, it still remains uncertain. Recent works have placed this surface tension $\sigma_{\rm HQ} >$ 70 MeV /fm$^2$  \cite{Voskresensky:2002hu,Yasutake:2014oxa} which could imply a sharp interface between phases of pure hadronic matter and pure quark matter. In \cite{Xia:2019pnq} a value $\sigma_c = 79.1$ MeV/fm$^2$ has been reported.

\subsubsection{Crossover}
 
 An example $T=0$ quark-hadron crossover EoS is presented in~\cite{Baym:2019iky}. 
 This ``QHC19'' EoS describes hadronic matter with the EoS of Togashi {\it et al.}~\cite{Togashi:2017mjp,Akmal:1998cf}, and quark matter from a Nambu-Jona-Lasinio model with two phenomenological parameters, a short range universal repulsion between quarks, and an attractive diquark interaction, producing quark pairing. Between these two limits, a simple polynomial interpolation in $P$ versus $\mu$ is performed. 
 
 The improvement over an earlier QHC18 is the use, for all densities, of the extension of the old APR EoS~\cite{Akmal:1998cf} by  Togashi {\it et al.} \cite{Togashi:2017mjp}, who constructed a thermodynamically consistent EoS from the Yukawa--exchange nuclear potentials. 
 (To our knowledge the same construction has not yet been carried out for the more modern Chiral EFT potentials). 
 Furthermore the Togashi equation of state is readily generalized to finite $T$. This consistency is not easy to accomplish by joining equations of state separately constructed in different density regions.

Beyond some 2$n_0$, exactly how matter transitions from hadronic to quark degrees of freedom is still unknown. 
The high-density regime above some 4-7$n_0$ where quark degrees of freedom could be dominant is described first. At baryon densities relevant to the core of the neutron stars (from several to 10 $\times n_0$, still below pQCD), the baryon chemical potential $\mu_B$ is in the range 1.5-2 GeV, $\alpha_s$ is of order 1, and perturbation theory is not applicable.
Therefore the Nambu-Jona Lasinio (NJL) effective model (reviewed in Ref.\cite{Buballa:2003qv}) is adopted. This global--color contact model contains three quarks (u d s); a flavor-dependent current mass $m_q$ and Fermi quark-quark interactions.

The quark--hadron crossover, performed in the range 2-5 $n_0$, allows a relatively stiff EoS capable of supporting neutron stars of masses greater than 2 M$_\odot$ with a substantial quark core~\cite{Masuda:2012ed,Kojo:2015fua}. Since neither purely hadronic nor purely quark matter descriptions are reliable in the range 2-5 $n_0$, the equation of state $P(\mu)$ is constructed by a smooth interpolation between hadronic matter at $n_B \leq$ 2$n_0$ and quark matter at $n_B \geq$ 5$n_0$. 

The interpolation must satisfy the constraints related to its first derivative $\partial P/\partial \mu=n$ that must be positive and convex for all $\mu$, i.e., $\partial^2 P/\partial \mu^2 = \partial \mu /\partial n >0$; moreover, the adiabatic sound velocity, $c_s = \sqrt{\partial P/\partial \epsilon} = \sqrt{\partial \ln \mu/ \partial \ln n}$ cannot exceed the speed of light, $c$. Given the hadronic pressure at $n\leq$ 2$n_0$, these constraints allow only a limited range of quark pressures at $n\geq$ 5$n_0$. Finally, the two-solar mass constraint is imposed.

The interpolating pressure is parameterized in the 
range 2 - 5$n_0$ as a polynomial in $\mu$,
\begin{equation}\label{croosoverinterpolation}
P_{\rm inter}(\mu) = \sum_{\alpha=0}^{\alpha_{\rm max}} c_\alpha\mu^\alpha
\end{equation}
and the $c_\alpha$ coefficients are determined by matching $P_{\rm inter}$ and its first and second derivatives of the $P$ with respect to $\mu_B$ to those of hadronic and quark matter at 2$n_0$ and 5$n_0$, respectively. For the six boundary conditions, $\alpha_{max}$=5.

Generically, the NJL-EoS description of quark matter for $n\geq$5$n_0$ cannot be too stiff. But four sets of NJL model parameters are examined and yield $M_{\rm max} \approx$ 2.35M$_\odot$ at the causal boundary.
The tidal deformability is then consistent with the aLIGO--Virgo bounds.
  
This model serves as a nontrivial example of a crossover transition.

\subsubsection{Interpolating an EoS with a phase transition}.

A recent work~\cite{Tews:2019cap} has developed two extrapolation schemes in order to extend QMC results with chiral EFT interactions to higher densities from the rather well constrained crust.  In the density range from $n_0$/2 up to 2$n_0$, the NS EoS is constrained by state of the art nuclear theory models, taking as starting point the calculations of pure neutron matter (PNM) as discussed in subsection~\ref{subsec:limits}.
The choice falls on local chiral EFT interactions treated with the auxiliary-field diffusion Monte Carlo method~ \cite{Lynn:2019rdt}. Local chiral interactions are cutoff at $R_{0}$ = 1.0 fm. All models are based  on that neutron-matter EoS up to a transition $n_{tr}$ which they vary to be $n_0$ or 2$n_0$. Densities around 2$n_0$ seem to provide an upper limit to the applicability of the chiral Hamiltonians.

At high densities, two different classes of model extend EFT+QMC results: (i) the ``Minimal Model'' (MM) that is based on a density expansion about $n_0$ and  (ii) the ``Maximal Model'' based on an expansion in the speed of sound  which includes phase transitions. 
 The models are also constrained by: (i) stability ($P$ and $c_{s}$ are both positive); (ii) causality ($c_{s} < c $); (iii) the ability to support a NS maximum mass  $M_{max} \geq 1.9M_\odot$ (the centroid of the maximum observed mass minus twice the error-bar on the observation: this gives M$^{obs}_{max} \simeq 1.9M_\odot$).

The Minimal model (MM) further assumes that matter is exclusively composed of $n$, $p$, $e^-$ and $\mu$ (at the mass of the muon one should already start wondering about the presence of pions too) and corresponds to the meta-model ELFc introduced and applied to NS in Ref.~\cite{Margueron:2017eqc,Margueron:2017lup}. 
It is described in terms of the empirical parameters of nuclear matter from subsection~\ref{subsec:lab},
By varying the empirical parameters within their uncertainties, the MM can reproduce many existing purely hadronic NS EoS; 
$\beta$ equilibrium and a crust as described in~\cite{Margueron:2017eqc} are included. As for the core, 
the dense- nucleon/lepton matter EoS employed incorporates  $\beta$-equilibrium in the mean field generated by the meta-EoS. Below saturation density, the core meta-EoS is matched to the EoS for the crust based on a cubic spline.  
Once more, stability and causality are required, and also the positivity of the symmetry energy $E_{\rm sym}>0$ up to the maximal central density corresponding to $M_{max} > 1.9 M_\odot$.
The maximum density allowed to each EoS is reached either by the break-down of causality, stability, or positiveness  of the symmetry energy condition, or by the end point of the stable neutron-star branch.

The Maximal Model, or Speed-of-Sound  model (CSM), is based on an extension of the speed of sound in neutron-star matter. Starting from PNM, the NS EoS is obtained up to $ n_{\rm tr}$ by constructing a crust and extending the neutron-matter results to $\beta$ equilibrium above the crust-core transition. After giving the EoS up to $ n_{\rm tr}$, the speed of sound is computed from $c^{2}_s = \frac{\partial P(\varepsilon)}{\partial\varepsilon}$. 

Above $n_{tr}$, $c_s$ is parametrized  by randomly sampling a set of points $c_{s}(n)$,  limited by stability and causality, and interpolating between them with linear segments, just like the construction of~\cite{Oter:2019kig} but applied to $c_s$ instead of $P$.

The EoS is reconstructed step by step starting at $n_{tr}$, where $\varepsilon(n_{tr})$, $p(n_{tr})$ and $\epsilon'(n_{tr})$ are known:
$$n_{i+1} = n_i + \Delta n$$
$$\epsilon_{i+1} = \epsilon_i + \Delta \epsilon = \epsilon_i+\left(\frac{\epsilon_i+p_i}{n_i}\right)\Delta  n $$
$$p_{i+1} = p_i + c^{2}_s(n_i).\Delta \epsilon\ ,$$
where $i=0$ defines the transition density $n_{tr}$. In the second line the thermodynamic relation $p= n(\partial\delta \epsilon/\partial\delta n)-\epsilon$ has been used, valid at zero temperature. Then, the high-density EoS are iteratively obtained. 
This model represents an extension of the CSS one discussed shortly. The resulting EoS parametrization represents possible NS EoSs and may include drastic density dependences , e.g., strong phase transitions which lead to intervals with a drastic softening and or stiffening of the EoS.  

In contrast to polytropic extensions of the low--density EoS, in the CSM model the speed of sound is continuous except when first-order phase transitions are explicitly taken into account.

\subsubsection{Comparison of MM and CSM}. 

\begin{figure}
\begin{center}
\includegraphics[width=\columnwidth]{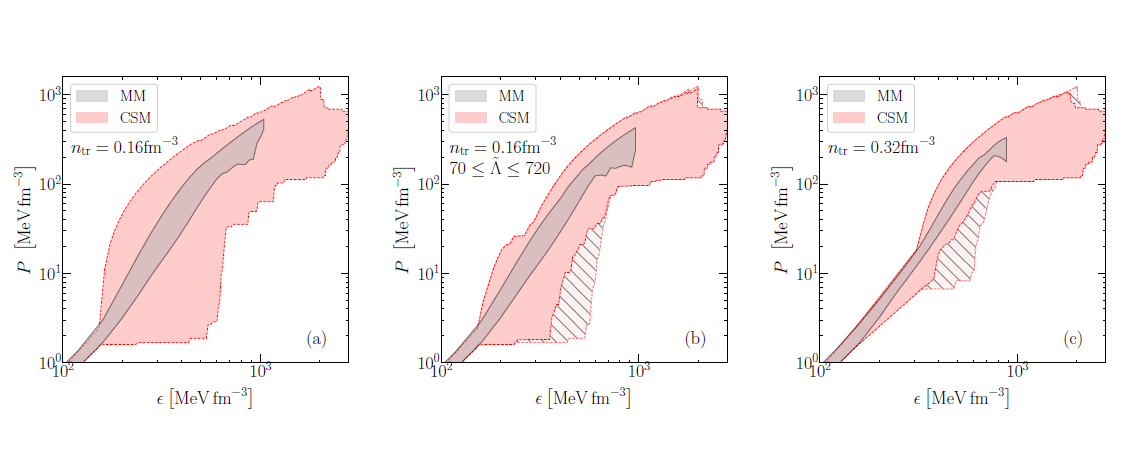}
\end{center}
\caption{
Comparison of the allowed EoS envelopes for the MM (dark grey bands) and the CSM (softer bands, red online). Three cases are visible: (a) the least restrictive, where $n_{\rm tr}$ =$n_0$ and only $M_{\rm max} \geq 1.9M_\odot$ is enforced, (b) for $n_{\rm tr} =n_0$ when enforcing 70 $\leq \widetilde{\Lambda} \leq$ 720 and (c) for $n_{\rm tr} = 2n_0$. When additionally enforcing R$_{1.6} \geq$  10.68 km, the hatched regions are excluded. 
\emph{Reproduced with author permission from preprint~\cite{Tews:2019cap}, further copyright remains with the authors}.
\label{fig:tewsgraph1}
}
\end{figure}

The comparison between both models is shown in  figure~\ref{fig:tewsgraph1}. 
One finds: (i) both models are in excellent agreement at low densities up to $n_{\rm tr} =n_0$ (unsurprisingly as both are constrained by the same low-$\epsilon$ input); (ii) the MM is a subset of the CSM above $n_{\rm tr}$; (iii) the CSM includes strong phase transitions revealed as regions of sudden stiffening or softening that are absent in the MM and (iv) although the chiral EFT  PNM EoS has sizeable uncertainties around $2 n_0$,  it can be taken to exclude a phase transition in this region.

\begin{figure}
\begin{center} 
\includegraphics[width=\columnwidth]{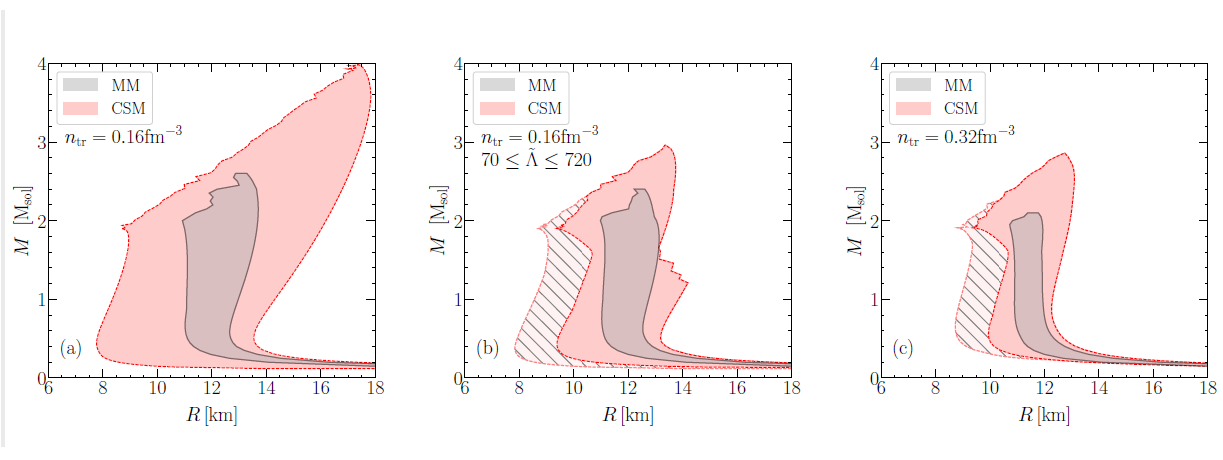}
\caption{
Comparison of the allowed $M(R)$ envelopes for the MM (dark grey bands) and the CSM (softer bands, red online). 
We show three cases (refer to figure~\ref{fig:tewsgraph1} for the corresponding EoS: 
(a) the least constraining, where $n_{\rm tr} =n_0$ and only $M_{\rm max} \geq 1.9M_\odot$ is enforced; (b) for $n_{\rm tr} =n_0$ when enforcing 70 $\leq \widetilde{\Lambda} \leq$ 720 and (c) for $n_{\rm tr}$ = 2$n_0$. When additionally enforcing $R_{1.6} \geq$  10.68 km, the hatched regions are excluded. 
\emph{Reproduced with author permission from preprint~\cite{Tews:2019cap}, copyright remains with the authors}. 
\label{fig:tewsgraph2}
} \end{center}
\end{figure}

The mass-radius (MR) relations are depicted in figure~\ref{fig:tewsgraph2}. For $n_{\rm tr}$ =$n_0$ (panel (a)), the CSM (MM) shows a radius range for the typical 1.4$M_\odot$ NS of 8.4 - 15.2 km (10.9 - 13.5 km). This range is dramatically reduced  for $n_{tr}$ = 2$n_0$ (panel (c)), where one can read off the ranges 8.7 - 12.6 km (10.9 - 12.0 km) respectively. 
CSM extends the allowed envelopes for both EoS and  $M(R)$ diagram of the MM, due to the sudden softening or stiffening of the EoS at high densities.

The small density dependence of chiral EFT constraints in the density range 1 - 2$n_0$, $n_{\rm tr}$ = 2$n_0$ along with the constraint of $M_{\rm max}$, makes  EoSs that show the emergence of disconnected compact-stars branches unlikely~\cite{Paschalidis:2017qmb}. Such EoSs need very strong first-order phase transition, 
which would soften the EoS so much that 2$M_\odot$ neutron stars could not be supported ~\cite{Alford:2015dpa}. 

In summary, the allowed variation ranges for several NS observables are larger for the CSM than the MM because 
the CSM permits regions of drastic stiffening or softening of the EoS.

To proceed, we review the confrontation of the GW observations with these two families of EoS models. 
If chiral EFT is used only up to $n_0$, the allowed range for $\Lambda$ at fixed chirp mass is found to be  $\tilde{\Lambda}_{CSM}= 60 - 2180$ {\it versus} $\tilde{\Lambda}_{MM} = 280 -1030$.  

The CSM yields as radius interval for a 1.4$M_\odot$ neutron star $9.0 {\rm km} < R_{1.4} < 13.6$ km (11.3 km $ < R_{1.4} < $13.6 km for MM).  This is consistent with~\cite{Annala:2017llu} that enforces the constraint $\Lambda_{1.4}< $ 800.

In the more restrictive case, accepting chiral EFT up to 2$n_0$
and leaving $\tilde{\Lambda}$ free, that analysis predicts the combined tidal deformability of the two neutron stars in GW170817 to be $\tilde{\Lambda}_{\rm CSM} = 80 - 580$ and $\tilde{\Lambda}_{\rm MM} = 280 - 480$
(this analysis is more constraining than the LV one). This highlights the fact that, even though the neutron-matter EoS has sizeable uncertainties at 2$n_0$, nuclear-physics calculations provide sufficient information to decrease uncertainties for  $\tilde{\Lambda}$ below current observational limits. In this case, for CSM, the radius 9.2 km $ < R_{1.4} < $12.5 km (11.3 km $ < R_{1.4} < $12.1 km for MM).

The transition density $n_{tr}$  affects the speed of sound inside the NS~\cite{Tews:2018kmu}. Employing the pure neutron-matter EoS up to 2$n_0$ requires the speed of sound to exceed the conformal limit $c_s^{2}=1/3$ to be sufficiently stiff to stabilize the observed two-solar mass NS.  For example, for chiral EFT interactions and $n_{tr}$ = 2$n_0$, the speed of sound hat to be  $c_s^{2} \geq$ 0.4.
In fact, for chiral models, the speed of sound has to increase beyond the conformal limit (1/3) for $n_{tr} >$ 0.31 fm$^{-3}$.
But in pQCD (see subsection~\ref{subsec:limits}) we know that asymptotically the value 1/3 is reached from below, not from above. This means that $c_s$, respecting pQCD and ChPT, needs to have three intervals of different curvature, two concave and one convex.

As discussed in subsection~\ref{subsec:NSmasses}, if as it appears from the later glow, the merged matter in GW170817 survived for several 100 milliseconds before collapse, that would imply~\cite{Margalit:2017dij,Rezzolla:2017aly,Shibata:2017xdx} $M_{\rm max}< 2.17\pm 0.16 M_\odot$. 
This $M_{\rm max}$ requires a stiff EoS as discussed in subsection~\ref{subsec:maxM}
(stiffness limited by the causality limit $c_{\rm s}^2 \leq 1$). Smooth EoS models with only one $M(R)$ branch (such as MM) exhibit a strong correlation between $M_{max}$ and radius, so the upper limit could be a powerful constraint. 
However, for those EoS models (such as CSM) that allow the presence of one or more phase transitions and their corresponding branches in the $M(R)$ diagram, $M_{\rm max}$ 
does not constrain NS structure and size much.
On the contrary, both models agree that an upper bound for $M_{max}$ does not strongly constrain $\widetilde{\Lambda}$. 

The EM observation of GW170817 does lead to additional constraints for radii and tidal polarizability. \cite{Bauswein:2017vtn} argued that R$_{1.6} \geq$ 10.68$^{+0.15}_{-0.04}$ km. In contrast to the $M_{max}$ constraint, this lower limit of the radius is a powerful constraint and has a sizeable impact on the CSM: in figure~\ref{fig:tewsgraph1}, the  part of the envelopes excluded by this measurement are indicated by hatched areas.

In relation to tidal polarizability, the amount of ejecta determined from the EM observations implies a lower limit of $\widetilde{\Lambda} >$  300 \cite{Radice:2018ozg}, but this limit was obtained considering only four EoS models.In general, radius and tidal polarizability are correlated and its is easy to convert this radius constraint to a limit on  $\widetilde{\Lambda}=\frac{2}{3}k_2(\frac{c^2R}{GM})^5$. ~\cite{Tews:2019cap} shows that this radius constraint implies that $\widetilde{\Lambda}\geq$ 180 (for $n_{tr}=2n_0$ and the CSM). 

However, from this analysis it seems that although there is a generic correlation between radius and tidal deformability  it is washed out by allowing phase transitions.

The behaviour of the EoS for star masses in the range relevant to 
interpret merger observations,
is of special interest to analyse the impact of phase transitions on tidal polarizability. In the case of  GW170817, M=1.4$M_\odot$, the model CSM, which includes such phase transitions,  allows small values of $\widetilde{\Lambda}$ due to the strong softening and subsequent stiffening of the EoS, but the MM does not support $\widetilde{\Lambda}<250$. 
These differences between both models permit to identify ranges of $\widetilde{\Lambda}$ for which a strong first-order phase transition is preferred, providing a poor man's means to indicate new states of matter inside NS. 

In the above example, an observation of $\widetilde{\Lambda}<$  250 would indicate a softening of the EoS that smooth (nucleonic) EoS cannot provide (an evidence for the existence of phase transitions). From these results, it is estimated that the uncertainty on $\widetilde{\Lambda}$ needs to be lowered to  $\Delta  \widetilde{\Lambda} <$ 300 to test the chiral EFT prediction in
the density range $n_0 - 2n_0$. Based on the contrast between MM and CSM, we expect that $\Delta \widetilde{\Lambda}<$ 100 is needed to reveal the possible existence of phase transitions in dense matter.     

Finally, contrasting the predictions of the models with and without phase transitions (MM and CSM) may provide useful insights on how future measurements of $\widetilde{\Lambda}$ could help to identify new forms of matter at densities beyond nuclear saturation.

\subsubsection{The Constant Speed of Sound (CSS) parametrization}.

A broadly used approach is the Constant Speed of Sound (CSS) parametrization.  Alford et al.\cite{Alford:2013aca}~assume that in hybrid stars the core of quark matter and the mantle of nuclear matter are separated by a sharp interface (first order phase transition with high enough surface tension) as opposed to a mixed phase (crossover). This is a possible scenario, given the uncertainties in the value of the surface tension \cite{Alford:2001zr,Palhares:2010be,Pinto:2012aq}. (More generic EoS that continuously interpolate between the phases to model mixing or percolation have been pursued~\cite{Macher:2004vw,Masuda:2012ed}.)

\begin{figure}
\begin{center}
\includegraphics[width=0.85\columnwidth]{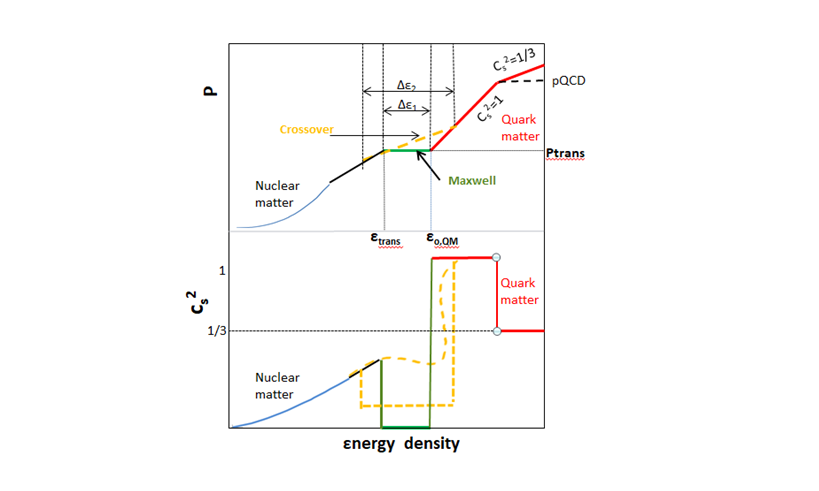}
\end{center}
\caption{
Sketch of the CSS toy model for the equation of state $P(\varepsilon)$ (top) and velocity of sound (bottom) for dense nuclear matter. The toy quark matter EoS is specified by the transition pressure $P_{\rm trans}$, the energy density discontinuity $\Delta \varepsilon$, and the speed of sound in quark matter $c_{\rm QM}$ (assumed to be density independent), explaining the variables used in the text.
\label{fig:cs}
}
\end{figure}

The CSS parametrization is applicable to high-density EoS for which (a) there is a sharp interface between nuclear and (possibly) quark matter; and (b) $c_s$ in the high-density matter is  $P$--independent for pressures ranging from the first-order transition pressure up to the maximum central pressure of neutron stars. One can then write the high-density EoS in terms of three parameters: the pressure $P_{\rm trans}$ of the transition, the discontinuity in energy density $\bigtriangleup\varepsilon$ at the transition, and the speed of sound $c_{\rm QM}$ in the quark matter phase. For a given nuclear matter EoS $\epsilon_{NM}(P)$, the full CSS EoS is then 
\begin{equation}\label{transitionCSS}
 \varepsilon(p)= \left\{
    \begin{split}
      &\varepsilon_{\rm NM}(P)  \phantom{hola}  &   P < P_{\rm trans} \\
      &\varepsilon_{\rm NM}(P) + \bigtriangleup \varepsilon+ c_{\rm QM}^{-2}(P-P_{\rm trans}) \phantom{hola}   & P > P_{\rm trans}
     \end{split}
      \right.
\end{equation}       
which is also sketched in fig.~\ref{fig:cs}.

\begin{figure}
\begin{center}
\includegraphics[width=0.75\columnwidth]{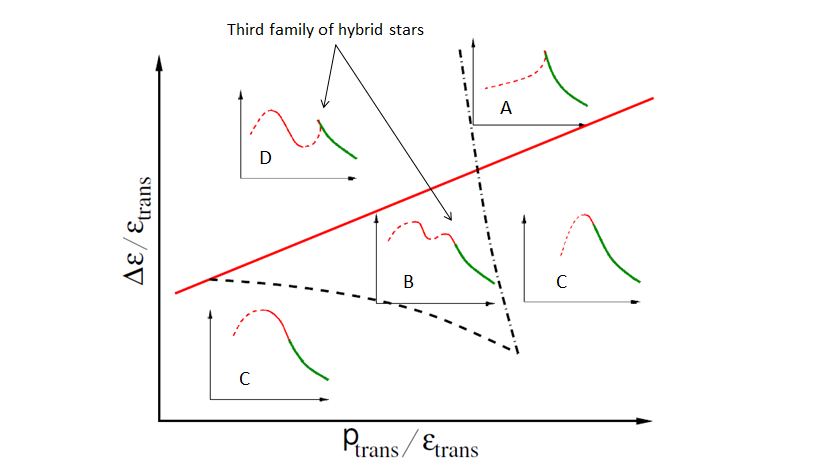}
\end{center}
\caption{
Hybrid star branches in the $M(R)$ relation of compact stars.  In each inset, the green line is the hadronic branch; solid red lines are stable hybrid stars; and the dashed red lines are unstable hybrid stars. 
The solid straight (red) line cutting the figure is the Seidov line from Eq.(\ref{pQCDPressure}): above it, stable hybrid stars are absent (A) or, if the inner core is stiff enough, a disconnected (D) ``third family'' 
sequence of stable hybrid stars emerges. Below the line, in regions B and C,  a hybrid star branch is connected to the nuclear star branch. The roughly vertical dash- dotted line marks a transition where an additional disconnected branch of hybrid stars appears/disappears. The roughly horizontal dashed curve marks a transition between $M(R)$ branches with one connected hybrid star.  The triangular-shaped region sees,  after the connected branch of hybrid stars, an instability which divides it from a third family sequence of compact hybrid stars, so that in this region the sequence is characterized by both (B): a connected and a disconnected branch of stable hybrid stars.
\emph{Reprinted from~\cite{Alford:2015dpa} with permission of the American Physical Society}.
\label{fig:Alford3}
}
\end{figure}

The “Constant-Speed-of-Sound” (CSS) parameterization  can be viewed as the lowest-order terms of a Taylor expansion of the quark matter EoS about the transition pressure  $P_{\rm trans}$. Following Ref.\cite{Alford:2013aca}, we express the three parameters in dimensionless form, as $P_{\rm trans}/\varepsilon_{\rm trans}$, $\Delta \varepsilon/\varepsilon_{\rm trans}$ (equal to $\lambda  -$ 1 in the notation of~\cite{Schaeffer:1983szh}) and $c_{\rm QM}^2$, where $\varepsilon_{\rm trans}\equiv$ ≡ $\varepsilon_{\rm NM}(P_{\rm trans}$). 
A good diagnostic for the stability of a sequence of compact stars is the study of $M(R)$, see subsubsection~\ref{subsecMofR}, with increasing central pressure. Neutron stars are stable until ${\rm max}(M(R))$ is reached, and unstable beyond. 

There are four topologies of the mass-radius curve for compact stars.  There may or may not be a second stable branch ``hybrid stars''; and if it is present, it may or may not  be connected to the hadronic branch. This is shown in figure~\ref{fig:Alford3} reproduced from \cite{Alford:2015dpa}.
These hybrid stars have central pressure above $P_{\rm trans}$, and so they contain a core of the high-density phase.

In the phase diagram the solid red line shows the threshold value $\Delta \varepsilon_{\rm crit}$  below which there is always a stable hybrid star branch connected to the neutron star branch. This critical value is given by \cite{Schaeffer:1983szh,Seidov:1971se,Lindblom:1998dp}
\begin{equation}\label{Seidovlimit}
 \frac{\Delta\varepsilon_{\rm crit}}{\varepsilon_{\rm trans}}= \frac{1}{2} + \frac{3}{2}\frac{ P_{\rm trans}}{\varepsilon_{\rm trans}}
\end{equation} 
and was obtained by performing an expansion in powers of the size of the core of high-density phase. This equation  is an analytic result, independent of $c_{\rm QM}^2$ and the nuclear matter EoS.

This condition is sometimes referred to as Seidov-limit and, when it is satisfied, the corresponding compact star will suffer an instability
of the same type as the maximum mass star of a stellar sequence. However,  in the corresponding EoS it is possible to  again reach stability resulting in a second stable sequence of stars if the parameters $P_{\rm trans}$ and $\Delta\varepsilon_{\rm crit}$ are chosen appropiately. This second branch is usually referred to as the ``third family'' for its
property of being the third stable solution of the TOV equations.
Such limit appears for first order phase transitions but not in the case of crossovers.

A strong first order phase transition in a hybrid star causes a compaction of the star that leads to a reduction of the gravitational mass.The hybrid star escapes gravitational collapse only if  the quark matter is  sufficiently stiff  after the transition.

The dashed and dot-dashed black lines mark the border of regions where the disconnected hybrid star branch exists. In the event that the phase transition takes place at low densities such as 1 $\leq n_{trans}/n_0 \leq$ 2, most observed neutron stars are likely to be above the onset mass $M_{\rm trans}$ for phase transitions, and accordingly the  hadronic branch is less relevant.

\subsubsection{Tidal distortion and the CSS parametrization}

In this paragraph, we jointly comment on (a) the analysis of the tidal deformability in presence of one or two first order transitions within this CSS parameterization~\cite{Han:2018mtj,Han:2019dpq}; and
(b) the impact that one and two first phase transition have on the neutron star structure. 10 SFHo \cite{Steiner:2012rk}  parameterizations of the EoS and 13 DBHF \cite{GrossBoelting:1998jg} ones, fullfilling the 2 $M_\odot$ maximum mass constraint and properties of symmetric nuclear matter (the DBHF EoS become superluminic for $n_M \sim 6n_0$).

Although at asymptotically high densities all EoS should approach the QCD limit $c_{QM}^2\approx 1/3$   \cite{Kurkela:2014vha,Bedaque:2014sqa}, uncertainties in the speed of sound at intermediate densities are still large. The constraint $M_{max}\geq$ 2$M_\odot$ implies that a first-order transition at the quark-hadron interface occurs more readily (i.e. allowed phase space is larger) if the speed of sound in quark matter is high, and taking the asymptotic  $c_{QM}^2\approx$  1/3  to be valid through the quark matter range, almost no detectable hybrid configurations are present \cite{Alford:2015gna}. Therefore, calculations are performed assuming that quark matter is maximally stiff $c_{QM}^2$ = 1 (from which a conservatively high limit on $M_{\rm max}$ follows).
Different values of  $n_{\rm trans}/n_0 \in [1 - 3.5]$ and 
 $\Delta \varepsilon/\varepsilon_{\rm trans} \in [0.2 - 1.8]$.~\cite{Han:2018mtj} have been explored.

An EoS with two first--order transitions might be realized in nature if hadronic matter is followed by two-flavor color superconducting (2SC) quark matter at moderate densities and color-flavor-locked (CFL) quark matter at the highest densities in the core \cite{Alford:2017qgh}. The presence of an additional phase transition introduces three new parameters $n_{\rm trans_2}/n_0, \Delta \varepsilon_{\rm trans_2} / \varepsilon_{\rm trans_2}$  and  $c_{\rm QM_2}^2$.

To avoid the discontinuity in $c_s^2$ from causing numerical difficulties with the system of equations equivalent to Eq.~(\ref{basicTOV},\ref{computeTidal}) for tidal deformability calculations, the sharp phase transition can be  slightly smoothened~ \cite{Alford:2017vca}:

\begin{equation}\label{crossover}
 \varepsilon (P) = \frac{1}{2}\left[1-\tanh\left(\frac{P-P_{\rm trans}}{\delta P}\right)\right]\varepsilon_{\rm NM}(P)
+  \frac{1}{2}\left[1+\tanh\left(\frac{P-P_{\rm trans}}{\delta P}\right)\right]\varepsilon_{\rm QM}(P))
\end{equation}  
where the nuclear matter EoS $\varepsilon_{\rm NM}(P)$ and quark matter EoS $\varepsilon_{\rm QM}(P)$ take the same form as in the standard CSS  parameterization Eq.~(\ref{transitionCSS}). 
In the vicinity of the transition
region where $P\in [P_{\rm trans}-\delta P, P_{trans}+\delta P]$, the sound speed is changing rapidly (as opposed to jumping).  

The limiting case $\delta P \rightarrow 0$ restores a true discontinuity (sharp transition) in the EoS, for which matching conditions at the phase boundary are necessary. Three regulated smooth EoS (technically, a crossover) are compared, putting $\delta P =$ 0.2,0.1,0.05. The extrapolation to a sharp boundary is then assessed.

It appears that when the central pressure of the star exceeds $p_{\rm trans}$, $k_2(M)$ and $\Lambda (M)$ decrease below the values obtained in a purely hadronic star.In addition, large $\Delta \varepsilon$ leads to changes of $k_2(M)$. If this decrease of $\Delta \varepsilon$ would be large enough, hadronic stars and hybrid stars might be distinguishable with a good sample of future BNS mergers event. However, such high densities might only be reached in very massive stars that have small tidal deformabilities, and it will be a challenge to measure $\Lambda$ precisely enough to recognize the effect of a phase transition.  

Unfortunately, once the central pressure greatly  exceeds $p_{\rm trans}$ (stars with large quark cores) global quantities such as the tidal deformability become insensitive to the exact nature of phase transition being sharp first-order or rapid crossover. 

A relevant result is that the minimal $\Lambda_{1.4} \approx$ 60 is associated with $n_{trans}=n_0$ and extremely large $\Delta \varepsilon$ ($\Delta \varepsilon/\varepsilon=$ 3.1). The minimal $\Lambda_{1.4} \approx$ 40 ($n_{trans_1}$ = 1 $\sim$ 2$n_0$) for EoSs with two sequential phase transitions. The results also show a drastic change in the tidal deformability from below to above the threshold for phase transition. If the phase transition has large enough $\Delta \varepsilon$ and takes place at low enough density $n_{trans}$, there can exist a hybrid star branch on the M-R diagram that is disconnected from the normal branch, that is, the ``third-family'' of stars. If a second phase transitions occurs, then there might be a  ``fourth-family'' of hybrid stars with smaller radius and tidal deformability.    

Assuming  both neutron stars in GW170817 obey the same nuclear matter EoS, all possible combinations  for the primary mass $m_1$  $\in$ [1.36, 1.60]$M_\odot$ and secondary mass $m_2  \in$ [1.17, 1.36]$M_\odot$, summing $m_{\rm tot}$ = 2.74$^{+0.04}_{0.01}M_\odot$ and quotienting to $q \in$ [0.7,1.0] for low spin priors, were scanned.  $\Lambda_1$, $\Lambda_2$ and eventually $\tilde{\Lambda}$ were then computed; for all hybrid star configurations $c^2_{QM}$ = 1. The results in ~\cite{Han:2018mtj} show a stronger dependence of $\tilde{\Lambda}$ on $q$ in the presence of a strong first order phase transition. Therefore, it may possible to observe different values for $\tilde{\Lambda}$ mergers with identical chirp masses but different mass ratios, if the EoS had such a phase transition.

Particularizing to two quark phases, $\rightarrow$ 2SC $\rightarrow$ CFL with  sequential phase transitions~\cite{Alford:2017qgh}, it seems that the $M_{max}\geq$ 2$M_\odot$ bound necessitates a rather stiff hadronic EoS and the first phase transition onset density $n_{trans_1}$ very close to the central density of a 2$M_\odot$ star. The high transition density from hadrons to quarks implies that for typical component masses (1.1 - 1.62$M_\odot$) observed in a binary, quark matter is nonexistent even in the densest cores. 

As a result, in the pre-merger stage, tidal deformabilities $\Lambda_1$ and $\Lambda_2$ (and other observables) are entirely determined by the nuclear matter EoS. By contrast, the postmerger remnant (if it survives as a supramassive or hypermassive neutron star) might attain densities above the phase transition threshold, so that gravitational-wave signatures in the postmerger stage potentially probe the densest quark matter~\cite{Most:2018eaw,Bauswein:2018bma}.

Another promising global observable is the moment of inertia. In the slow rotation approximation, the dimensionless moment of inertia $\tilde{I}=I/M^3$ and the dimensionless deformability $\tilde{\Lambda}$ of NS without phase transitions are related by EoS-independent universal relations to within 1$\%$ \cite{Yagi:2013awa,Yagi:2016bkt}, that is, the strong correlation between $\tilde{I}$ and $\tilde{\Lambda}$ is found to be accurate within 1$\%$. However, with sequential first order phase transitions into quark matter, the deviation can be as large as 9$\%$ \cite{Han:2018mtj}.     

\subsection{Twin stars}

Twin stars are two compact stars with approximately the same mass but quite different radii~\cite{Glendenning:1998ag}. Such configurations are obtained if there is a strong first order phase transition in the interior of a compact star. In this case, the $M(R)$ diagram exhibits a third branch of gravitationally stable stars and this branch is separated from the second family of NS by a sequence of unstable configurations topologies classified as ``D'' and ``B'' in  \cite{Alford:2013aca} (see figure~\ref{fig:Alford3}). The "D" topology consists of a hadronic and a hybrid star branch, both of which are gravitationally disconnected from each other. The ``B'' topology consists of a branch of stable hadronic stars followed by stable hybrid stars, which are gravitationally disconnected from a second branch of stable hybrid stars.

The discovery of only about two stars on each branch with a radius resolution better than a kilometre by current estimates would suffice to establish the twin branches. It has been speculated that GW170817 could be explained in terms of the coalescence of one neutron star and one hybrid star \cite{Paschalidis:2017qmb,Most:2018hdf,Burgio:2018yix,Nandi:2017rhy,Alvarez-Castillo:2018pve,Sieniawska:2018zzj,Li:2018ayl,Christian:2018jyd,Han:2018mtj}. Depending on the features of the phase transition of the hybrid star, twin-star solutions and the two branches in $M(R)$ might appear.

There is a large number of EoS parameterizations for analysing third familly sequences fulfilling the constraint $M_{\rm max}\geq$  2$M_\odot$, which are based mainly on CSS models and multipolytropes approaches \cite{Alford:2017qgh,Christian:2017jni,Paschalidis:2017qmb,Christian:2018jyd,Han:2018mtj,Montana:2018bkb}.

Depending on the features of the phase transition of the hybrid star, twin-star solutions and the two branches in $M(R)$ might appear. Indeed, information from gravitational waves is being exploited to better understand the twin-star scenario \cite{Burgio:2018yix,Paschalidis:2017qmb,Alvarez-Castillo:2018pve,Sieniawska:2018zzj,Christian:2018jyd}.

An extensive analysis of the features of the hadron-quark phase transition that are needed to obtain the twin branch, enforcing, $M_{\rm max}\geq$  2$M_\odot$ and incorporating the information from GW170817 has recently appeared~\cite{Montana:2018bkb}. 
Two models are employed, with equal hadronic EoS (crust from~\cite{Sharma:2015bna} and hadron EoS from a relativistic mean field model, FSU2H as in \cite{Tolos:2016hhl,Tolos:2017lgv}) 
but either a parametrized first-order or a crossover transition, followed by a (CSS) parametrization of the quark phase.

The crossover between the hadronic and the quark phases is mocked by a polytrope $p(\rho)= K_m\rho^{\Gamma_m}$. The first order transition, with a jump in $\varepsilon$, can be cast as a $\Gamma_m$ = 0 polytrope (this can be slightly softened as needed).
Therefore, the two model EoS are
\begin{itemize}
\item Model-1: FSU2H + Maxwell + CSS
\begin{equation}\label{pQCDMaxwell}
 \varepsilon(P)= \left\{
    \begin{split}
      &\varepsilon_{\rm FSU2H}(P)
        & P < P_{\rm tr} \\
      &\varepsilon_{\rm FSU2H}(P) + \Delta\varepsilon + c_s^{-2}(P-P_{\rm tr})
        & P > P_{\rm tr}
     \end{split}
      \right.
\end{equation}
with c$_s^2$ = 1.
\item Model-2: FSU2H + Gibbs + CSS
\begin{equation}\label{pQCDGibbs}
 \varepsilon(P)= \left\{
    \begin{split}
      & \varepsilon_{\rm FSU2H}(P)
        & P  \leq P_{\rm tr} \\     
      &(1+a_m)(P/K_m)^{1/\Gamma_m - 1)}+ P/(\Gamma_m -1)
        &P_{\rm tr} \leq  P \leq P_{\rm CSS} \\
      &\varepsilon_(P_{\rm CSS}) + c_s^{-2}(P-P_{\rm CSS})
        & P \geq P_{\rm CSS}
     \end{split}
      \right.
\end{equation}
\end{itemize}
with $c_s^2$= 1 and $\Gamma$ = 1.03.
The values of the polytropic constant $K_m$ and its coefficient were obtained by ensuring that $P$ and $\varepsilon$ are continuous at the transition points. Since $\Delta \varepsilon$ does not jump during a crossover, increasing values of $\varepsilon(P)$ are just sampled along the mixed phase.

To allow for a wide range of EoSs, the sampled  parameter space includes $n_{tr} \in [1.4 - 6.5]n_0$ and $\Delta n \in [0.2 - 3.0]n_0$.

The maximum masses in the two branches allow a simple classification to order the discussion
\begin{itemize} 
\item  Case I: $M_{TOV} \geq 2.0M_\odot$ and $M_{TOV; T} \geq 2.M_\odot$
\item  Case II: $M_{TOV} \geq 2.0M_\odot$ and $M_{TOV; T} < 2.0M_\odot$
\item  Case III: $1.0 M_\odot\leq M_{TOV} < 2.0 M_\odot$ and $M_{TOV; T} \geq 2.0M_\odot$.
\item Case IV: $M_{TOV} \leq 1.0M_\odot$ and $M_{TOV; T} \geq 2.0M_\odot$,
\end{itemize}

where $M_{\rm TOV}$ and $M_{\rm TOV; T}$ are the maximum masses of the normal and twin branches, respectively.

The results from Model-1 [Model-2] EoSs show that, in order to reach 2$M_\odot$ in the normal neutron star branch for Cases I-II, we need $P_{\rm tr} >$ 180 MeV fm$^{-3}$ [160 MeV fm$^{-3}$]. These two case are difficult to differentiate since the values of $M_{\rm max}$ for the two branches lie within a rather small range, {\it i.e.}, 1.95$M_\odot\leq$ $M_{\rm TOV; T}\leq$  2.05 $M_\odot$.

Twin-star solutions of Case IV appear only for very low values of $P_{\rm tr}$ (i.e., $P_{tr}\leq$ 25 MeVfm$^{-3}$ [15 MeVfm$^{-3}$]). They are unlikely to exist because the mass in the normal-neutron-star branch is much lower than detected (see section~\ref{subsec:NSmasses}).

Most twin-stars are in  Case III, with 25 MeV fm$^{-3}\leq$  $P_{\rm tr}\leq$  180 MeV fm$^{-3}$ [15 MeV fm$^{-3}\leq$ $P_{\rm tr}\leq$ 160 MeV fm$^{-3}$]; this is also the  case easiest to test/falsify from an astrophysical point of view, as it accommodates twin stars of masses around the canonical 1.4M$_\odot$ value, where the NS population in figure~\ref{fig:massdistribution} peaks.

The tidal deformability also displays two distinct branches having the same mass~\cite{Burgio:2018yix,Christian:2018jyd}, an additional smoking gun signature for the existence of twin stars.

The radius difference $\Delta  R$ among the two equal-mass twin stars has been found to be $\Delta  R \sim$ 1.6 km for Case II, Model-1,  which is softer than that in~\cite{Christian:2017jni}  where  $\Delta $R as large as  4.0km  was claimed possible for $M\sim 1.6 M_\odot$, for the parameters  $P_{\rm trans}$ = 118 eV/fm$^3$ and $\Delta \varepsilon$ = 690MeV/fm$^3$). 

For the most interesting case (twin stars of Case III with masses around 1.4 M$_\odot$), $\Delta  R \sim$ 0.8 km, thus making it more difficult to distinguish the two types of stars without very good data: 5\% precision on the determination of the radius will be necessary. 

As for the tidal deformability $\Lambda$,  it spans several orders of magnitude for different EoSs.
There is considerable difference between EoSs that exhibit a phase transition at low densities (such as Case IV) and at high densities (Cases I-II). More specifically, a reference 1.4  M$_\odot$ star with a dense core of quark matter in Case IV has $\Lambda_{1.4}$ ranging from a few tens to a few hundreds, while a 1.4  M$_\odot$ pure hadronic neutron star in Cases I-II has $\Lambda_{1.4}$ = 760.
This can be explained by the different compactness of the stars $\Lambda \propto \mathcal{(R/M)}^5$~ \cite{Hinderer:2009ca}. Therefore, for the same  1.4 M$_\odot$ mass, stars with a quark-matter core have smaller radii and, hence, smaller values of $\Lambda_{1.4}$.

In Case III, with masses around 1.4 M$_\odot$, the configurations have a core of mixed or pure quark matter with a radius for the star intermediate between the radius for Cases I-II and Case IV. Thus,  $\Lambda_{1.4}$ lies between that found in Cases I-II and Case IV.

The  upper bound $\Lambda_{1.4} <$ 800 in GW170817 was obtained by expanding $\Lambda$(M) linearly about M = 1.4 M$_{\odot}$, and  if the twin branch appears at M = 1.4 M$_{\odot}$ such approach is no longer valid and the upper bound on $\Lambda_{1.4}$ could be further decreased.

To constrain twin stars with GW170817, the values of $\Delta \varepsilon$ and $P_{\rm tr}$ have been studied as a function of $\tilde{\Lambda}_{1.4}$ showing that a hadron-quark phase transition softens the EoS so as to make it compatible with the GW170817 event.

As for the limits on the radius of a canonical 1.4 M$_{\odot}$ star, several works have reported values for stars of a given mass and we have collected them in Table~\ref{tab:Constrainsontheradius}. 
\begin{table}
\caption{\label{tab:Constrainsontheradius} 
Constraints on the radius of neutron stars from
GW170817 for models without a phase transition, works considering the possibility of a transition to quark matter (middle) and for EoSs of Case III in \cite{Montana:2018bkb}.
}
\begin{center}
\begin{tabular}{cc}\hline
 Reference                  &          $R_i$ [km]              \\
\hline
$Without \ a \ phase \ transition$                                  \\
Bauswein et al.\cite{Bauswein:2017vtn}&$10.68^{+0.015}_{-0.03} \leq R_{1.6}$\\
Most et al.\cite{Most:2018hdf}&$12.00\leq R_{1.4} \leq 13.45$\\
Burgio et al. \cite{Burgio:2018yix}&$11.8 \leq R_{1.4} \leq 13.1$\\
Fattoyev et al. \cite{Fattoyev:2017jql} &$R_{1.4}\leq$ 13.76\\
Tews et al.\cite{Tews:2019cap} &                            \\
$n_{tr} = n_0$          &$10.9 \leq R_{1.4} \leq 13.5$  \\   
$n_{tr} = 2n_0$         &$10.9 \leq R_{1.4} \leq 12.0$  \\ 
De et al. \cite{De:2018uhw} & $8.9 \leq R_{1.4} \leq 13.2$  \\
Ligo/Virgo \cite{Abbott:2018exr}    & $10.5 \leq R_{1.4} \leq 13.3$ \\
Lim et al. \cite{Lim:2019som} $10.0 \leq R_{1.4} \leq 12.7$ \\ 
\hline
$With \ a \ phase \ transition$ &                           \\ 
Annala et al. \cite{Annala:2017llu}& $R_{1.4} \leq 13.6$     \\
Most et al.\cite{Most:2018hdf}&$8.53 \leq R_{1.4}\leq 13.74$ \\
Burgio  et al. \cite{Burgio:2018yix}& $R_{1.5} = 10.7$       \\ 
Tews et al.\cite{Tews:2019cap}&                               \\
$n_{tr} = n_0$          &$8.5 \leq R_{1.4} \leq 15.2$    \\   
$n_{tr} = 2n_0$         &$8.7\leq R_{1.4} \leq 12.6$     \\
\hline
Montana et al. \cite{Montana:2018bkb}&                       \\
NS                         &           $R_{1.5} = 13.11$     \\ 
HS Model-2                 & $12.9 \leq R_{1.4} \leq 13.11$ \\
HS$_T$ Model-1             & $10.1 \leq R_{1.4} \leq 12.9$  \\
HS$_T$ Model-2             & $10.4 \leq R_{1.4} \leq 11.9 $\\
\hline
\end{tabular}
\end{center}
\end{table}
Here, the  lowest values of R$^{\downarrow}_{1.4} \simeq$ 10 km and $\Lambda^{\downarrow}_{1.4} \simeq$ 100 are produced for transition pressures well below P$_{\rm tr}$ = 50MeV fm$^{-3}$, i.e., $n_{\rm tr}$ = 2.4$n_0$. 

If such small radii and tidal deformabilities are confirmed by future measurements, perhaps hyperons could exist in a very narrow  region of the interiors of neutron stars. In the FSU2H model, only the $\Lambda$ particle would be present, since it appears at $n$ = 2.2 $n_0$.

Therefore, these results indicate that if future detections of gravitational waves from LIGO/Virgo determine values of $\Lambda_{1.4} \leq$ 400 and, at the same time, chirp masses $\mathcal{M} \leq$ 1.2 M$_{\odot}$, the inspiral phase will have the imprint of a hadron-quark phase transition with a low transition pressure. Otherwise, it will be difficult to distinguish during the inspiral whether one of the components of the binary is a Hybrid Star, as a HQPT with transition pressures above 50MeV fm$^{-3}$ might be indistinguishable from one without phase transition.

Other recent work on twin stars~\cite{Alvarez-Castillo:2018pve}
 has investigated  a class of EoS that jumps from a low-density phase of hadronic matter, modelled by a relativistic mean field approach with excluded nucleon volume, via a first order phase transition, to a  high-density phase of color superconducting two-flavor quark matter, described within a nonlocal covariant chiral quark model. The results show that these hybrid stars do not form a third family disconnected from the second family of ordinary neutron stars unless additional (de)confining effects are introduced with a density-dependent bag pressure. If this is artificially done, then a third family arises~\cite{Kaltenborn:2017hus} which seems relatively robust against varying the introduced parameters.

\begin{figure}
\begin{center}
\includegraphics[width=0.75\columnwidth]{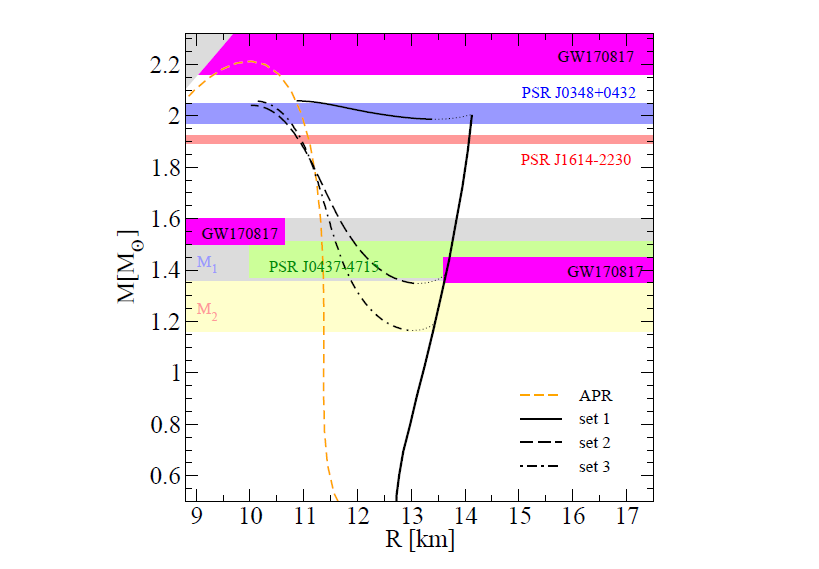}
\end{center}
\caption{
$M(R)$ diagram for sequences of compact stars with the hybrid EoS of~\cite{Alvarez-Castillo:2018pve}  corresponding to different onset masses for the
deconfinement transition. The dotted lines denote the unstable
configurations that should not be realized in nature but
guide the eye to the corresponding stable hybrid star sequence (third family) disconnected from the neutron star one (second family). \emph{Reproduced from~\cite{Alvarez-Castillo:2018pve} under the} 
\href{https://creativecommons.org/licenses/by/4.0/}{\tt Creative Commons 4.0 License}
\label{fig:AlvarezCastilloMR} 
}
\end{figure}

Figure \ref{fig:AlvarezCastilloMR} shows the mass-radius relationships for  three different hybrid EoS parametrizations, given in table ~\ref{tab:hybridEoSParametersACastillo}. The dotted lines denote the unstable configurations that should not be realized in nature but guide the eye to the corresponding stable hybrid star sequence (third family) disconnected from the neutron star one (second family). The blue and red horizontal bands denote the mass measurement for PSR J0348+432 and PSR J1614-2230, respectively. The grey and orange bands labelled ``M1'' and ``M2''
are the mass ranges for the compact stars in the binary merger GW170817 for which~\cite{Bauswein:2017vtn} has excluded radii smaller than 10.68 km of 1.6 M$_\odot$ stars and~\cite{Annala:2017llu} excludes radii exceeding 13.4 km at 1.4 M$_\odot$. The green band denotes the mass range 1.44 $\pm$ 0.07 M$_\odot$ of PSR J0437-4715. The upper limit to $M_{\rm max}$ has been taken from the conjecture that GW170817 did not lead to a prompt black hole formation after the merger~\cite{Margalit:2017dij,Rezzolla:2017aly,Shibata:2017xdx}.

\begin{table}
\begin{center}
\caption{\label{tab:hybridEoSParametersACastillo} 
NS parameters calculated for a hybrid EoS with a first-order phase transition from hadronic matter (described by the DD2 p40 EoS \cite{Typel:2009sy,Typel:2016srf}) to quark matter (described by the interpolated nonlocal NJL model) obtained by a Maxwell construction for  three parametrizations of sets 1-3: the critical chemical potential $\mu_c$, the critical pressure $p_c$, and the values of energy density and baryon number density corresponding to the onset of the first-order phase transition, $\varepsilon_c$ and $n_c$, respectively. Solving the TOV equations with the hybrid EoS for $M(R)$ allows to obtain the maximum $M_{\rm max}$ and minimum masses $M_{\rm min}$ of the hybrid star branch as well as a mass $M_c$ at the onset of the phase transition in the center of the compact star.
}
\begin{tabular}{|c|c|c|c|}\hline
                          &  Set 1   &  Set 2   &   Set3   \\
\hline
$\mu_c$ (MeV)             & 1214  & 1100  &  1080  \\
$p_c$ (Mev/fm$^3$)        & 67    & 33.1  &  26.1  \\
$\varepsilon_c$(MeV/fm$^3$)& 339   & 277   &  260  \\
$n_c$ (fm$^{-3}$)          & 0.334 & 0.281 & 0.266  \\
\hline
$M_c$ (M$_\odot$)         &   2.00  &   1.39  &   1.20    \\
$M_{\rm min}$ (M$_\odot$)       &  1.99   &  1.35   &  1.17    \\
$M_{\rm max}$ (M$_\odot$)       &  2.06   &  2.04   &  2.06    \\
\hline
\end{tabular}
\end{center}
\end{table}

\subsection{Identifying a phase transition in a gravitational wave signal}
From these long exposition of extant work, let us extract as key questions whether there is a feature in the possible $M(R)$ sequences that would signal a first order phase transition in the corresponding EoS, and 
whether the tidal deformability and postmerger GW frequencies could act as telltale signals of strong first order phase transitions. This seems to be the case, but theoretical models or simulations should provide sufficient precision to allow for proper interpretation of the measurements, and good data is a must.

In~\cite{Bauswein:2018bma,Bauswein:2019skm} an observable imprint of a first-order hadron-quark phase transition at supranuclear
densities on GW emission of NSs is identified, based on the fact that the tidal deformability during inspiral and the  oscillation frequencies of the postmerger remnant can be determined with relatively high reliability.

On one hand a specific nucleonic reference model DD2F \cite{Typel:2005ba,Typel:2009sy} describing the EoS at all densities is considered.  On the other hand, hybrid EoSs are constructed by employing the same EoS but only at densities below a  first order phase transition to deconfined quark matter.

NS merger simulations are performed  with the novel temperature-dependent, microscopic hadron-quark hybrid EoS DD2F-SF of Ref. \cite{Fischer:2017lag}. Among the purely hadronic EoS models, DD2F EoS were used \cite{Typel:2005ba,Fischer:2017lag,Alvarez-Castillo:2016oln} and  the corresponding hybrid EoSs with a first order phase transition to deconfined quark matter (DD2F-SF) of \cite{Typel:2005ba}. The acronym DD2F-SF refers to all seven hybrid models. Three of the purely hadronic EoSs include a 2nd order phase transition to hyperonic matter. Additionally, the two EoSs ALF2 and ALF4 are employed  which resemble models with a more continuous transition to quark matter (with vanishing latent heat).

Merger simulations focused on symmetric nonspinning systems with a total mass of $M_{\rm tot}$ =2.7 $M_\odot$, comparable to GW170917; the starting point is 
a circular quasi--equilibrium orbit with non-spinning stars in $\beta$ equilibrium and with $T=0$ a few revolutions before merging.

 During the evolution, temperature effects are taken into account selfconsistently if provided by the
EoS. For some EoSs where the temperature dependence is not available, an approximate treatment of
thermal effects is employed, which requires to choose a  “ideal-gas index” $\Gamma^{th}$ (subsection~\ref{subsec:finiteT}). This regulates the strength of thermal pressure supporting the matter. The value of the coefficient $\Gamma^{th}$ = 1.75 reproduces results with fully temperature--dependent EoSs relatively well.

\begin{figure}
   \begin{minipage}{0.46\textwidth}
    \includegraphics*[width=5in]{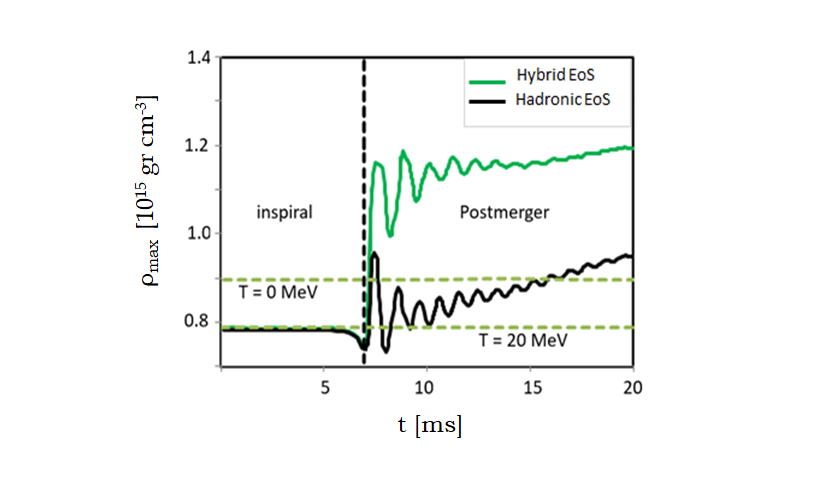} 
    \end{minipage}\hspace{0.2cm}
    \begin{minipage}{0.46\textwidth}
     \includegraphics*[width=5in]{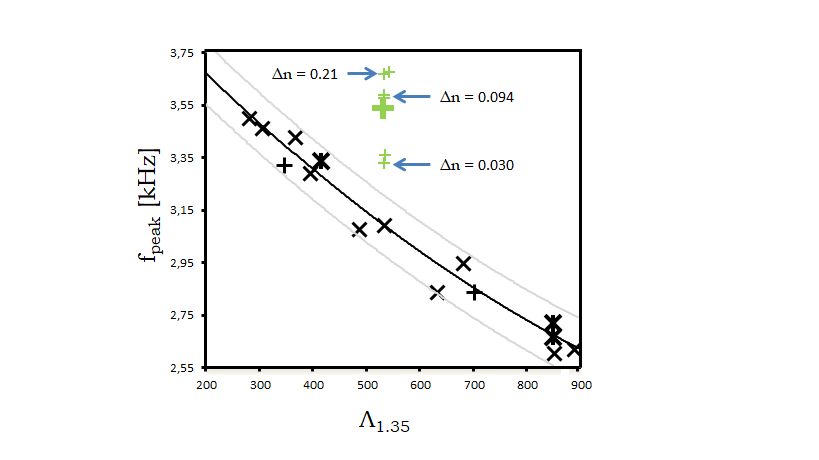}
   \end{minipage}
  \caption{Left panel: Evolution of the maximum rest-mass density comparing hadron-quark hybrid (green) and hadronic EoS (black) for symmetric--binary 1.35-1.35 $M_\odot$ mergers (solid curves); horizontal dotted green lines mark the
onset density $\rho_{onset}$ of the phase transition for hybrid EoS at T = 0 and at 20 MeV. Right panel: Dominant post-merger GW frequency $f_{\rm peak}$ as a function of tidal deformability. Black symbols display results for purely hadronic EoSs, black plus signs which correspond to hybrid models of \cite{Bauswein:2018bma}  text). Green markers show the peak frequency for EoS models with a first-order phase transition to quark matter. The solid black curve is a fit to the data excluding hybrid models and the two solid grey curves indicates the upper and lower deviation from the fit among the purely hadronic model (data from  \cite{Bauswein:2018bma}). 
 \label{fig:Evolutionpostmerger}}
\end{figure}

Some results from these simulations are shown in Figure \ref{fig:Evolutionpostmerger}. Its left panel displays the evolution of the maximum rest-mass density as function of time for symmetric binary 1.35--1.35 M$_\odot$ simulations with the hybrid DD2F-SF-1 (green) and the purely hadronic counterpart DD2F (black). The dotted horizontal green lines indicate the onset density $\rho_{onset}$ of the phase transition at T = 0 and 20 MeV ($\beta$--equilibrium is imposed).

During the inspiral phase, the stars central density is below the transition one and the evolution with/without phase transition is identical. The two stars merge at about 7 ms and form a single central object with a steep increase of the maximum rest--mass density. For  EoS that includes quark matter, $\rho$ rises above the threshold for the phase transition into the pure quark matter phase. A quark core forms in the center of the merger remnant, and its mass amounts to about 20 $-$ 30$\%$ of the total one; but the maximum density in the calculation with the purely hadronic EoS always remains below that of DD2F-SF-1. The stronger density increase in the model with quark matter is a direct consequence of the density jump across the phase transition. 

The right panel displays a significant deviation from the empirical relation between the dominant postmerger gravitational wave frequency $f_{\rm peak}$ and the radius/tidal deformability of a star of fixed mass. This deviation is observed if a strong first-order phase transition occurs.

Employing this diagnostic to decide on a phase transition in hadron matter requires a precise measurement of the tidal deformability and of $f_{\rm peak}$ which may be achieved in the near future using instruments with higher sensitivity (in the range $\leq$ kHz). 

In conclusion, a good population of detected mergers would help, via knowledge of the tidal deformability  $\Lambda(M)$, to constrain the equation of state, or given this, to constrain General Relativity; but deeper information would be obtained if the gravitational postmerger signal would be reported for one strong, nearby event, since the deformability is just one integrated quantity over the entire star.

\newpage
\section{Further observables} \label{sec:otros}

\subsection{Cooling, damping and transport} \label{subsec:transport}

Though much of the community's interest is focused on the Equation of State, which is an integrated quantity averaging many particle interactions, and gives information mostly on the bulk of the star in its ground state, there are further physical quantities that can be studied in neutron stars.
Transport coefficients in particular give access to the effective excitations above the ground state.
Low-energy excitations control the emission and propagation of neutrinos, the star cooling, and transport.
Theory studies proceed, for example, by computing spectral functions.

In the neutron (and proton) superfluid that is often assumed to constitute part of the star, the typical low-energy excitations would be a Goldstone boson near zero excitation energy over the ground state.
The fermion excitations corresponding to the breaking of an nn or a pp pair are excitations at twice their respective mass gap~\cite{Baldo:2019rzt} $\Delta$, and for temperatures smaller than that gap, their contribution to transport is suppressed by $e^{-\Delta/T}$. 
Likewise, in the asymptotic high-density regime, the CFL phase transport coefficients are dominated by the Goldstone boson of $U(1)$ symmetry breaking~\cite{Manuel:2004iv,Manuel:2007pz}. 

In intermediate quark phases, the unpaired quarks are calculated to dominate transport, although totally unpaired quark matter is thought to appear only for larger temperatures $T>10$MeV. The extent of Cooper pairing at non-asymptotic density and finite temperature depends on dynamical details and there is no model independent assessment. Finally, in the crust, free electrons and free neutrons are probably the leading degrees of freedom.

An exception is the thermal conductivity, since heat is efficiently transferred over large distances by neutrinos, so we will dedicate a paragraph to discussing them.
A lucid review has been recently put together by Schmitt and Shternin~\cite{Schmitt:2017efp}.

In all such regimes where quasiparticles are the carriers of conserved charges and transport proceeds by their diffusion, the transport coefficients are inversely proportional to the scattering cross section, $\eta\propto 1/\sigma$. This is typical of a gas, in which the momentum or charge from an excess concentration  is carried away in the ballistic motion of the particles leaving that concentration, hampered by collisions.

For a Fermi liquid, in which temperature fuzzes the Fermi surface and augments the phase space for scattering, the quasiparticles travel more freely at low temperatures, so that
\begin{equation}
\eta \propto \frac{1}{T^2}\ ;\ \ \ \  \kappa\propto \frac{1}{T}\ ;\ \ \ \  \sigma_{\rm electric}\propto \frac{1}{T^2}\ .
\end{equation}

In the case of the crust, where the pasta phases are anisotropic, or in the presence of a magnetic field, transport is considerably more complicated as the transport quantities become tensors under rotations, and we will not review the efforts in this direction.

The size of transport coefficients is one more piece of information helping to classify a given GW event as having its origin (or not) in a neutron star merger. For example, if we ignored gravitational wave damping $\tau^{-1}_{GW}$, $\eta$ for the event GW150914 would be estimated at  about 4$\times 10^{28}$ Poise, from the damping time in subsection~\ref{subsec:anatomy}; this is extracted for a Newtonian sphere of fluid~\cite{Yunes:2016jcc}, $\eta \sim 5\rho R^2 /\tau_\eta$. This figure is way larger than the $\sim 10^{15}$ Poise relevant for neutron matter at T=10 MeV (though perhaps commensurate with the Alfven viscosity in a huge magnetic field $B\sim 10^{16}$ Gauss).  Note that beyond General Relativity, the masses of NSs can be made larger as in figure~\ref{fig:Resco} above, so that knowledge of the merging $m_1$ and $m_2$ 
is not sufficient to classify an event. Other observables, such as damping times, are also affected by the modifications of GR, but with different systematics, so they can assist in the classification.

\subsubsection{Neutrino emissivity and cooling}

Once neutron stars are formed, their temperature steadily drops (although some examples of seemingly old but still warm neutron stars have been reported; they must have some source of heat, perhaps from $\beta$-decay, perhaps from accretion). A well known example~\cite{Page:2010aw} is Cassiopeia A, believed to be only 330 years old and whose surface temperature dropped between 2000 and 2009 from 2.12 to 2.04 million Kelvin.
The loss of internal energy brought about by this cooling is balanced by the star's luminosity (power)
\begin{equation}
C_V \dot{T} = -L \ .
\end{equation}
(The heat capacity $C_V$ can be evaluated from the Equation of State once its temperature dependence is known as the derivative of the internal energy respect to the temperature. A back of the envelope estimate~\cite{Page:2010aw} is $C_V\sim 10^{39}\left(\frac{T}{10^9\rm K}\right)$ erg/K.)

Neutrinos have the longest mean free path in matter among the Standard Model fundamental fields.
For example, the charged current interaction $\sigma(\nu N\to e N)\simeq \frac{G_F^2}{\pi}\arrowvert{\mathcal M}\arrowvert^2 \frac{p^2}{v_iv_f}\sim \frac{G_F^2}{\pi}E^2$ implies a mean free path $\lambda = \frac{1}{n\sigma}$ that, for a 1 MeV neutrino and at $n\simeq 2n_0\sim 0.3 {\rm fm}^{-3}$, becomes of the order of the star radius, $\lambda\simeq 2$km.

This means that neutrinos emitted all through the crust certainly escape the star and cool it, and a fraction of those produced at the star's interior also manage to find their way out.
Neutrinos can be produced by the direct Urca processes (simple $\beta$ or induced $\beta$ decay) in which only one nucleon in the initial state participates, 
\begin{equation}
N_1\to N_2 + l +\bar{\nu}\ ; \ \ \ N_2+l \to N_1 +\nu\ . 
\end{equation}
But in cold, dense nuclear matter, because of Pauli blocking, the two nucleons have to be at the Fermi surface
with $|{\bf p}|_i\sim k_F$ which restricts the possible reaction kinematics, forcing even a threshold to the proton fraction in the star matter. 
In the case in which neutrons and protons form Cooper pairs (superfluid phase), the neutral current process
\begin{equation}
N_1 \to N_1 + \nu + \bar{\nu}
\end{equation}
becomes possible (opening or closing the pairing gap helps satisfy energy/momentum conservation).

If that threshold is not met, an additional nucleon coupled by the strong nuclear force needs to be present to absorb the momentum imbalance, at a cost of a $(T/\mu)^2$ factor, according to
\begin{equation}
N_1 + N_3 \to N_2 + N_3 + l +\bar{\nu}\ ; \ \ \ N_2+N_3+l \to N_1 +N_ 3 + \nu \ ; \ \ \ N_1+N_3\to N_1+N_3+\nu+\bar{\nu}\ .
\end{equation}
$N_1$ and $N_3$ should be coupled with the full chiral interaction but, once more, the most used approximation is a simple potential interaction based on one-pion exchange~\cite{Friman:1978zq}.
The additional particle can also be a lepton (``electromagnetic neutrino-brehmsstrahlung'') as in
\begin{equation}
l+N\to l+N +\nu +\bar{\nu}\ .
\end{equation}
Table~\ref{tab:emisivities} collects  these calculations as quoted by~\cite{Schmitt:2017efp}.

\begin{figure}
\begin{center}
\includegraphics[width=10cm]{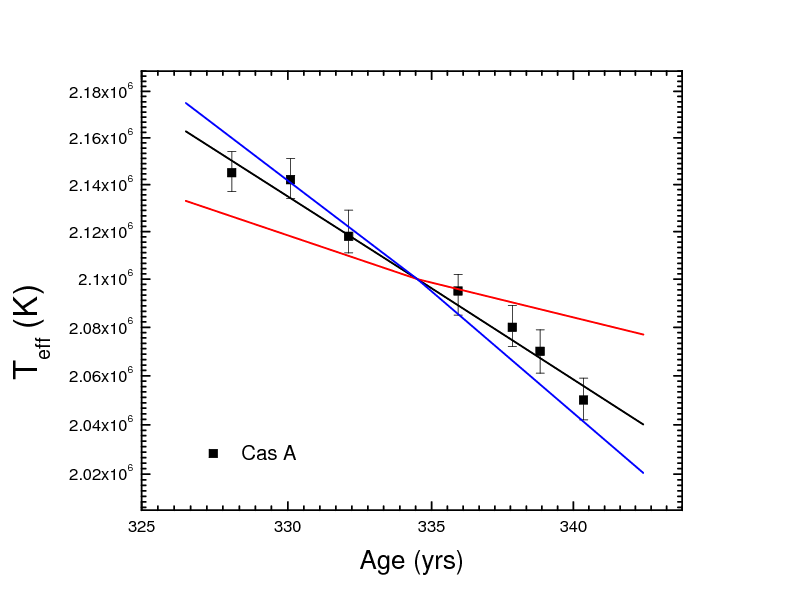}
\caption{\label{fig:Casiopea} Surface temperature measurements of the Cas A neutron star.
Lines indicate the best fit and the upper and lower limits on the cooling rates of Cas A when all \emph{Chandra} detectors and modes are included: the decline varies between $\sim 2\%$ and $\sim 5.5\%$ in the decade 2000-2009. \emph{(Reprinted from~\cite{Newton:2013zaa} with permission.)}}
\end{center}
\end{figure}

\begin{table}
\begin{center}
\caption{Estimates of the neutrino emissivities in neutron star matter~\cite{Schmitt:2017efp} in erg/(cm$^3$s) (except for factors close to 1 related to the matrix elements).$\Delta$: pairing gap. $T$: temperature. $n_b$: baryon density. $n_0$: 0.16fm$^3$, saturation density. $x_l$: lepton fraction. $m^*$: {\it in medio} mass. $P_F$: Fermi momentum. $\mu_i$: chemical potential. $G_F$: Fermi constant.
\label{tab:emisivities}}
\begin{tabular}{|cc|} \hline
Direct Urca ($\beta$) if above threshold & $4\times 10^{21} \left( \frac{n_B}{n_0}\right)^{1/3} x_l^{1/3}\frac{m_p^* m_n^*}{m_N^2} \left(\frac{T}{10^8{\rm K}}\right)^6$ \\
Nucleon Cooper pair breaking& $3.6\times 10^{14} \frac{m_N^*}{m_N} \frac{p_{FN}}{m_Nc} \left(\frac{T}{10^8{\rm K}}\right)^7 F(\Delta/T) $   \\
Electromagnetic brehmsstrahlung & $5.1 \times 10^{12}  \left( \frac{n_l}{n_0}\right)^{2/3} 
\left( \frac{m_{\rm Debye}c^2}{2T}  \right)\left(\frac{T}{10^8{\rm K}}\right)^7$ \\
Urca + additional $N$ & $8\times 10^{13}  \left( \frac{n_B}{n_0}\right)^{1/3}
\left(\frac{m_p^*}{m_p}\right)^4\left(\frac{m_n^*}{m_n}\right)^3 
\left(\frac{p_{Fl}c}{\mu_l}\right)  \left(\frac{T}{10^8{\rm K}}\right)^8 $ \\ \hline
Unpaired quark matter & $\frac{457\alpha_s}{630} G_F^2 \mu_e\mu_d\mu_u T^6$ \\
\hline
\end{tabular}
\end{center}
\end{table}
The luminosity is then proportional to the volume $L\propto V$; the typical volume is of order $4\times 10^{18}$cm$^3$.
Considering that the direct $\beta$ decay channel is kinematically forbidden,  the estimate~\cite{Page:2010aw} of the modified Urca reaction with an additional nucleon (fourth entry in table~\ref{tab:emisivities}) is around $10^{40} \left(\frac{T}{10^9\rm K}\right)^8$ erg/s. This seems insufficient to explain the cooling of the Cassiopeia A neutron star (see figure~\ref{fig:Casiopea}).
The modified Urca process suggests $T=10^9{\ \rm K} (1{ \rm yr}/t)^{1/6}$ and the faster cooling rate in the data is taken to come from another phase.

Then we mention first the effect of strangeness~\cite{Tolos:2019wea}: if some hyperons are formed by elevating neutrons to have a strange quark, \emph{(a)} the neutron density decreases for equal total baryon density and \emph{(b)} the central density is higher (as the EoS is softened). Both effects enhance direct URCA and thus accelerate cooling.

In turn, phases with ungapped quarks have a high heat capacity$\propto \mu^2 T$ and the relativistic quarks have phase space to participate in direct Urca processes. On the contrary, CFL, where all the quarks are paired, has low heat capacity $\propto T^3$ from the superfluid phonon only, and has low emisivity~\cite{Alford:2019oge}.

Ignoring the intermediate--density phases, and concentrating on the low--density part alone, 
the slope parameter $L$ of the symmetry energy, defined in Eq.~(\ref{symmetry}), is correlated with the cooling efficiency~\cite{Tolos:2019wea}.  If it increases, it forces a higher number of protons; this increases the probability of direct Urca $\beta$ reactions taking place; and thus $\nu$ cooling is facilitated.

\begin{figure}
\begin{center}
\includegraphics[width=8cm]{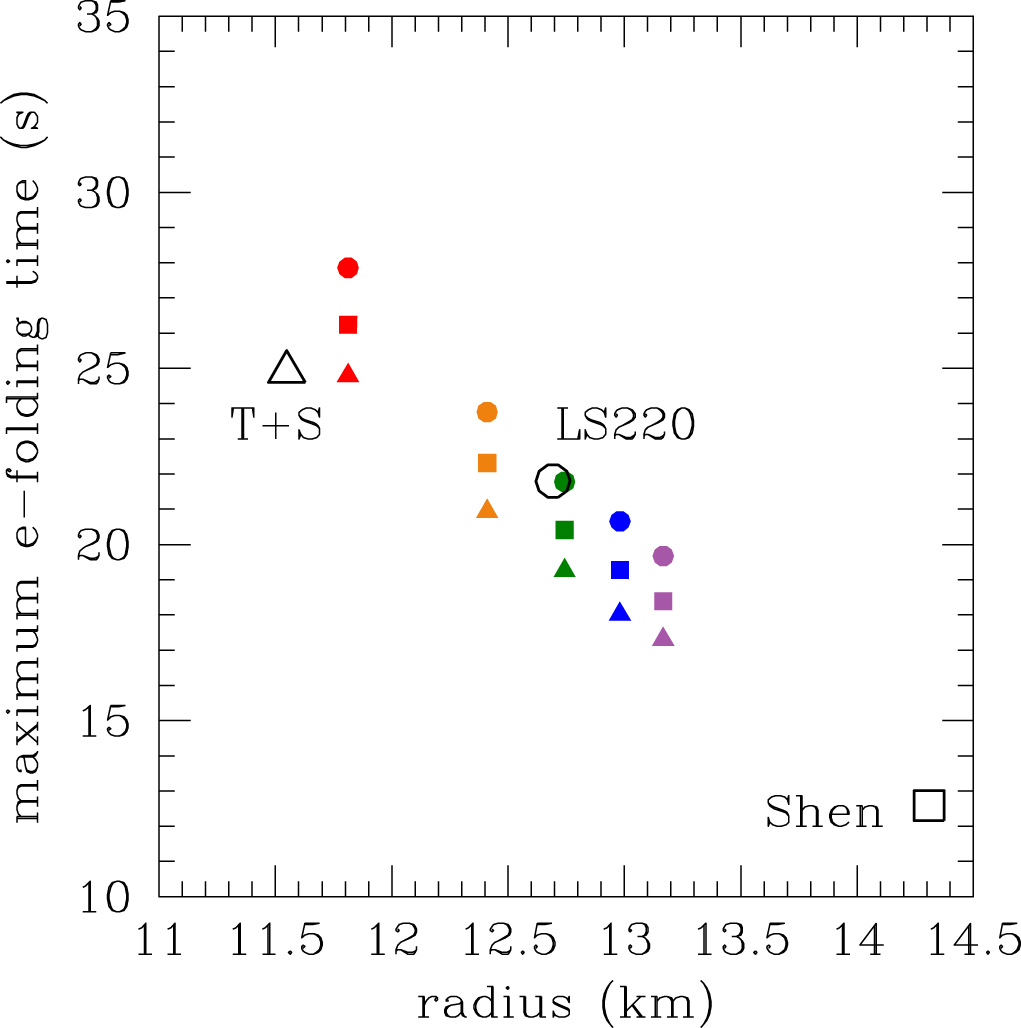}
\caption{\label{fig:Nakazatocooling} The maximum cooling time (exponential scale, y--axis, defined from the luminosity via $L(t+\tau)=e^{-1}L(t)$: the $e$--folding $\tau$ is plotted) is correlated with the neutron star radius (x--axis). The actual correlation depends on the EoS chosen so that progress in its determination can be used to compare two totally different measurements, the size of a protoneutron star and its cooling speed. \emph{Reprinted from~\cite{Nakazato:2019ojk}, with permission.}}
\end{center}
\end{figure}

To conclude this paragraph, let us show a correlation between the cooling by neutrino emission and the radius of a proto--NS is shown in figure~\ref{fig:Nakazatocooling}, taken from~\cite{Nakazato:2019ojk}. The computational data, that finds temperatures up to 35--40 MeV,  spans a band whose width is related to the spread in the EoS models employed. 

\subsubsection{Diffusive transport coefficients}

If equilibrium in the star is only local, microscopic transport tends to decrease the gradients of conserved quantities and produce entropy. 
Most of the work has been carried out in the quasiparticle regime in which one can identify simple degrees of freedom with a particle-like distribution $f$. Then kinetic theory tools can be deployed. (The derivation from more general setups has recently been explored with the on-shell effective theory~\cite{Manuel:2014dza}.)
Firstly, the distribution function relaxes with a Boltzmann-like equation in kinetic theory
\begin{equation}
p^\mu \frac{\partial f}{\partial x^\mu} - \Gamma^\mu_{\nu\rho} p^\nu p^\rho \frac{\partial f}{\partial p^\mu}= {\mathcal C}[f]\ ;
\end{equation}
there, the Christoffel symbols $\Gamma$ are needed because the fluid is not in free fall (but held in place by the pressure instead). However, because they are derivatives of the metric, and the Schwarzschild one is smoothly varying with a km scale, it is often simplified for shorter distances to a simple advective derivative 
\begin{equation} \label{NRBoltzmann}
\frac{\partial f}{\partial t} + {\bf v}\frac{\partial f}{\partial {\bf x}} = {\mathcal C}[f]
\end{equation}
as in conventional fluids~\footnote{A salient application of the GR Boltzmann equation is for transport in rotating systems, which could prove necessary for binary mergers~\cite{Miralles:1993bwb}.}. The integral collision term is such that ${\mathcal C}[f_{\rm equilibrium}]=0$. As Eq.~(\ref{NRBoltzmann}) is an integrodifferential equation, usually its solution is avoided; instead we project the equation appropriately over a basis of functions of ${\bf p}$ and other conserved quantities to obtain variational approximations useful to obtain transport coefficients. See~\cite{Torres-Rincon:2012sda} for complete computations in a meson gas. 

The ideal fluid of Eq.~(\ref{stressT}) then receives dissipative contributions $T=T^{\rm ideal}+\tau$, that in terms of $\Delta^{\mu\nu}=g^{\mu\nu}-v^\mu v^\nu$, $T=T^\mu_\mu$ the thermal conductivity $\kappa$, the shear viscosity $\eta$ and the bulk viscosity $\zeta$ read
\begin{equation}\label{dissipativeT}
\tau^{\mu\nu} = \kappa (\Delta^{\mu\gamma}v^\nu)(T_{,\gamma}+T v^\sigma  v_{\gamma,\sigma})
+\eta \Delta^{\mu\gamma}\Delta^{\nu\delta}(v_{(\gamma,\delta)}-\frac{2}{3}g_{\gamma\delta}v_{\gamma,\gamma})
+\zeta \Delta^{\mu\nu} v_{\gamma,\gamma}
\end{equation}
that generalizes the well known equation from nonrelativistic fluid mechanics.

The poster calculation is that of the shear viscosity because of its presumed role in damping the instability of the $r$-modes that would strongly feed gravitational radiation~\cite{Andersson:1997xt}, see subsec.~\ref{subsec:r}. First, let us quote the result for the asymptotic CFL phase (that may not necessarily be realized in neutron stars, but nonetheless is interesting as a limit of QCD). We accompany it by the viscosity computed in unpaired quark matter from~\cite{Schmitt:2017efp}; we do not know, if this phase is reached at all, at what densities and temperatures it might be present. But it is natural that its properties come intermediate between the two known asymptotic limits of the CFL and the nucleon-lepton matter.

\begin{equation}\label{visco:CFL}
\eta_{\rm CFL} = 7 \times 10^{22} \left( \frac{\mu}{500\rm MeV}\right)^{8} 
\left(\frac{1\rm MeV}{T}\right)^{-5}\ {\rm Poise}
\end{equation}

\begin{equation}\label{visco:unpairedQM}
\eta_{\rm unpaired\ QM} = 3 \times 10^{15} \left( \frac{\mu}{500\rm MeV}\right)^{14/3} 
\left(\frac{1\rm MeV}{T}\right)^{5/3}\ {\rm Poise}\end{equation}

\begin{figure}
\begin{center}
\includegraphics[width=10cm]{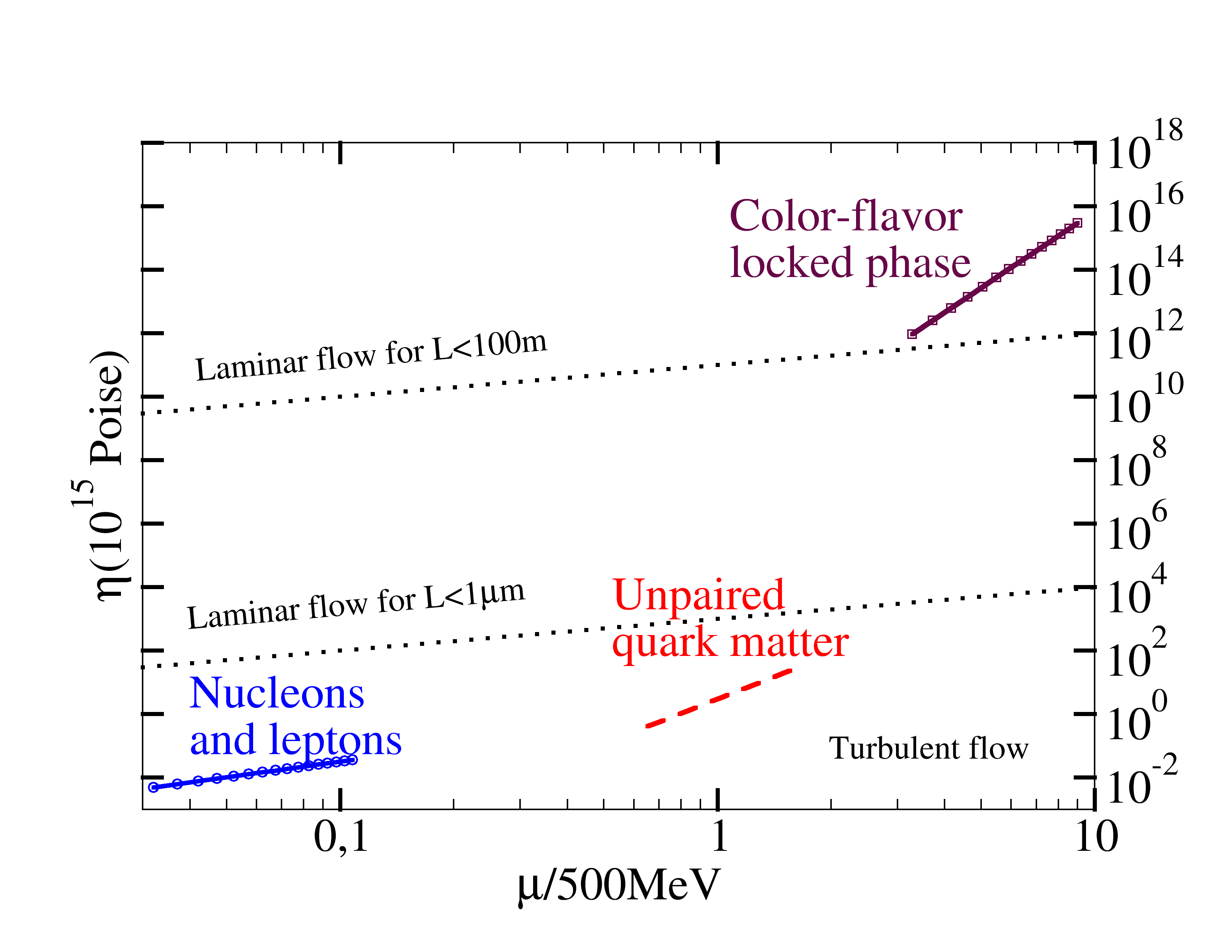}
\caption{\label{fig:viscosity}
Viscosity in the two extreme phases of neutron matter (the asymptotic CFL phase dominated by the Goldstone boson of $U(1)$ breaking; and the hadron-lepton gas at low density) as well as one of the possible intermediate density phases, unpaired quark matter. We have taken $T=1$MeV (though this might be a bit high for the CFL phase, it is an easy reference point; the mean free path according to the scaling in~\cite{Manuel:2004iv} is about 80 meters, much shorter than the typical NS scale of $\sim$km, so that the fluid is well coupled). We have also marked the lines of Reynolds number $Re=10^3$ below which the flow is expected to be turbulent at the scale indicated for a viscosity of $10^{15}$ Poise.}
\end{center}
\end{figure}

And thirdly, we turn to the lowest density within hadron physics ({\it i.e.} ignoring the nuclear phases in the crust). We have (perhaps over-) simplified Eq.~(75b) of~\cite{Schmitt:2017efp} and turned it into an equation in terms of $\mu=\partial \epsilon/\partial n$ instead of $n$ with a quick estimate based on the EoS of~\cite{Holt:2016pjb} (basically, we have taken some ballpark values 
$\epsilon=\epsilon_0 n^2/n_0^2$, $n_0\simeq 16\ {\rm fm}^{-3}$, yielding $E(n_0)=16$ MeV per nucleon, $\mu_0=32$ MeV). Thus, the shear viscosity in the nucleon and lepton phase becomes:
\begin{equation}\label{visco:n}
\eta_n = 1.5\times 10^{15} \left( \frac{\mu}{500\rm MeV}\right)^{5/3} \left(\frac{1\rm MeV}{T}\right)^2
\ {\rm Poise}
\end{equation}
which is directly comparable with Eq.~(\ref{visco:CFL}) and~(\ref{visco:unpairedQM}).
All three are plotted in a cartoon in figure~\ref{fig:viscosity}.

In the figure we have also shown lines of equal Reynolds number $Re=\frac{\rho v L}{\eta}=10^3$. For the estimate, we set $\eta=10^{15}$ Poise, $v\sim 1$, $\rho\sim 2.3\times 10^3 \frac{\rm MeV}{\rm fm^3} \left(\frac{\mu}{500\ \rm MeV}\right)$, and taken $L=100$ m (transport at macroscopic distances in a star of radius of order 11 km) and $L=1\mu$m (microscopic transport). Below the marked lines the flow should be largely turbulent at the scale given, and above them, laminar.

It appears that viscosity controls the flow only in the CFL phase (if ever realized; one should think that it is the asymptotically high--density phase of QCD). This means that turbulence is likely to provide the largest damping of the $r$-mode and other instabilities; and it is clear that its study is one of the promising future directions in the field. A decade--old study~\cite{Melatos:2009mz} showed that, for relatively low frequencies, the hydrodynamic turbulence in a core collapse supernova setting off in our galaxy (which happens a couple of times per century) could be detectable by the emission of Gravitational Waves with a strain of order $h\sim 10^{-20}$ on Earth. Also, it has been shown~\cite{Andersson:2007uv} that the superfluid part of the neutron star is also very likely to be turbulent (in a superfluid, the erratic motion is best followed by studying vortex mixing and evolution).

Moving on to other transport coefficients, we have performed a quick scan of the literature and have not been able to find convincing calculations of flavor diffusion (isospin and strangeness being conserved charges). Thermal conductivity $\kappa$ of the various phases has been calculated in several phases (the various results are collected in~\cite{Schmitt:2017efp}) and helps diffuse heat within the star.  But since much of the cooling is effected by neutrino emission, already discussed, let us instead focuse a last comment on the bulk viscosity~\cite{Manuel:2007pz}.

Bulk viscosity acts to diminish the pressure in an inhomogeneous dilatation: spherical shells moving faster drag along the ones moving slower.
In a scale--invariant theory a dilatation of the system does not relax the interaction, so that there is no bulk viscosity. Therefore, this coefficient signals the breaking of scale invariance (or the splitting of particles changing their number). It has been calculated in the CFL and various other phases. Some of these are superfluid and significantly more complicated in their hydrodynamic description, for example the proton/neutron superfluid ones. 

Following Khalatnikov's theory, that in addition to the conventional bulk viscosity scaling as $\zeta=\zeta_2\sim m_s^4/T$) presents additional terms relating to the friction between the superfluid and the normal fluid components of the medium, the Barcelona group \cite{Manuel:2013bwa,Mannarelli:2009ia} has estimated the subleading bulk viscosities to scale as $\zeta_1\sim m_s^2/(T\mu)$ and $\zeta_3\sim m_s^2/(T\mu^2)$.

       \subsection{Vibration modes of neutron stars \emph{(or Titan resonances)}.}
       \label{subsec:vibration}
As composite objects, neutron stars can be excited above their classical rest (ground) state.
The perturbation affects the energy density in the star $\epsilon+\delta \epsilon$ (and consequently the pressure), the velocity field $u+ \partial \xi / \partial t$ (with $\xi$ the displacement of a fluid element) and, saliently in General Relativity, the metric $g=g^{(0)}+h$, where the unperturbed quantities correspond to the static, spherically symmetric Tolman-Oppenheimer-Volkoff solutions.

\begin{figure} \begin{center}
\includegraphics[width=10cm]{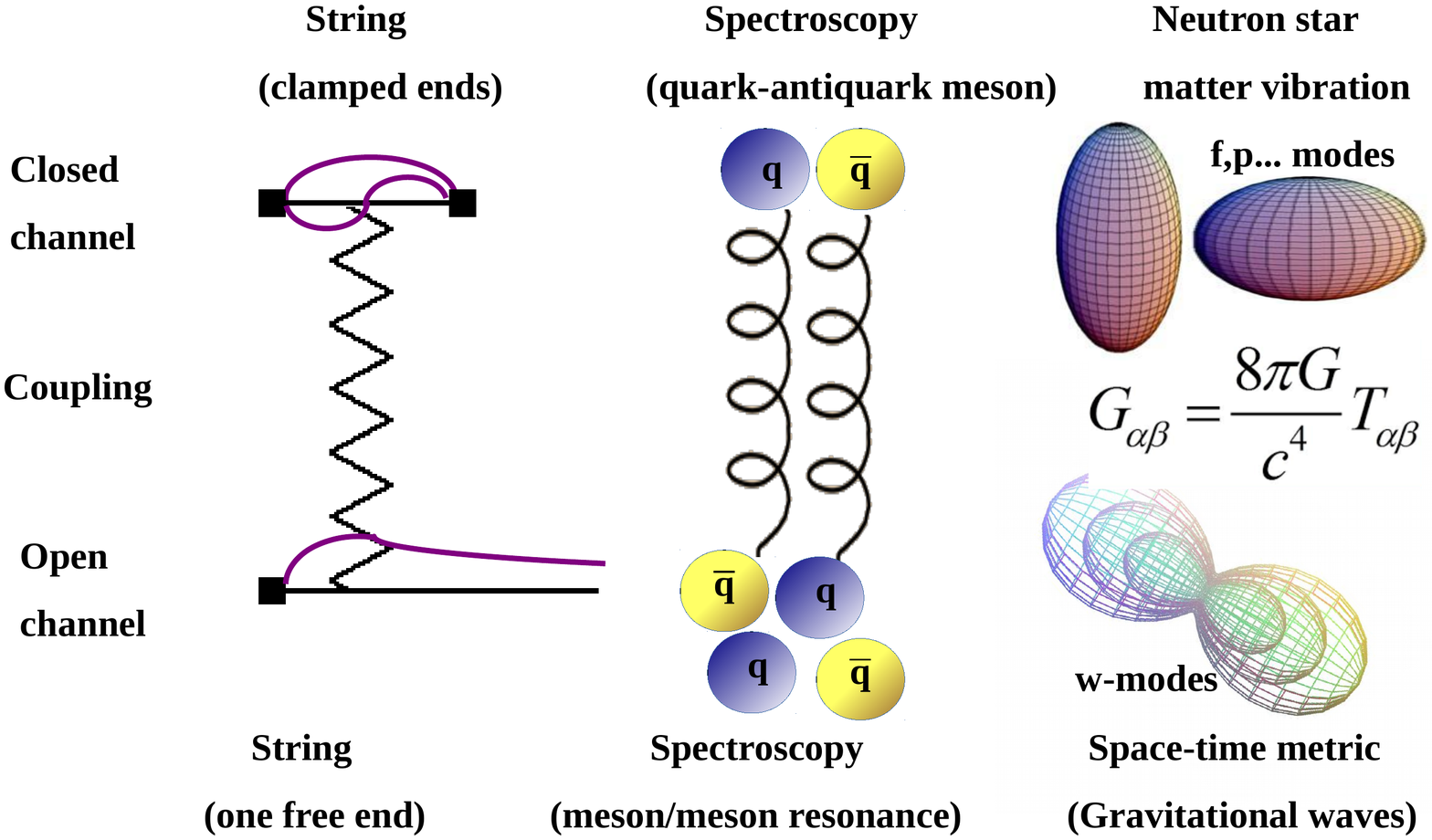}
\caption{\label{fig:modecartoon}
Neutron star quasinormal modes (right) can be explained by analogy with a mechanical system of two coupled vibrating strings~\cite{Kokkotas:2003mh}, one clamped and one semiopen (left) or with any spectroscopic system with a closed channel and an open channel, such as quark-antiquark and meson-meson resonances (center). In all cases, the closed channel has a vibration spectrum (in the case of the neutron star, its modes are analogous to Newtonian theory such as $f$-, $p$-, $\dots$ modes). But the open channel can also develop resonances due to the coupling of the closed one. In the case of the neutron-star, these are the $w$-modes of the space-time metric. They are strongly damped due to gravitational wave radiation.}
\end{center}\end{figure}

Several ``polar'' (or ``fluid'') modes, in which the perturbation to a physical quantity is of the form
$\sum_{lm}W_l(r) \hat{r} + V_l(r) \nabla Y_l^m $ (yielding natural parity $(-1)^l$)
directly couple the metric to first order perturbations of the stress-energy tensor; they closely relate to the behavior of the neutron matter in the star and have Newtonian analogues in seismology, oceanography, etc.  
Axial modes are related to perturbations of the form $\sum_{lm} U_l(r) \hat{r}\times \nabla Y^l_m$, and have unnatural parity $(-1)^{l+1}$.  They are best seen as vibrations of space-time, though they are also (more weakly) coupled to the star's matter.
The model of the two vibrating strings~\cite{Kokkotas:2003mh} illustrates their nature beautifully, and is familiar to nuclear and hadron physicists from spectroscopy. 
This we depict in figure~\ref{fig:modecartoon}.

All these perturbations can leave an imprint in X-ray spectra (and in fact are often discussed in that context).
But if neutron stars have a varying quadrupole (or higher) moment, their vibration modes can additionally radiate gravitational waves.
These modes are reminiscent of the Giant Resonances in ordinary nuclear physics: whereas in nuclei the restoring force is provided by nuclear potentials (that respond to charge separation, for example),
the richness here comes from the presence of the additional gravitational interaction. The size of neutron stars when compared to ordinary nuclei suggests the nickname ``Titan resonances'' (also because Hesiod attributes the naming of the ``Titans'' to ``strain'').

It is not reasonable to expect that aLIGO-Virgo will be able to detect these vibration modes in gravitational wave signals, but third-generation (proposed) detectors such as the Einstein Telescope and the Cosmic Explorer, reaching far objects out to redshift $z\simeq 2-3$ and $6$ respectively will increase the statistics to the point that the low probability (0.1\%) in any one event~\cite{Bauswein:2011tp} turns into a real chance.  This could be up to 0.1-10 yr$^{-1}$\cite{Chirenti:2016xys}.

Of course, the lowest multipole titan-dipole resonances cannot radiate GWs and are thus not accessible to this new astronomy. The logic ones to search for are the next-to-lowest  titan-quadrupole resonances. These we briefly review here, though we will also quickly mention some new questions concerning radial modes: the two lower-$l$ modes, such as the monopole, are accessible to more traditional observables such as X-ray intensities accessible from satellites.

Concerning the characteristic frequency of these collective modes, the order of magnitude in a neutron star is at most $c/R_{NS}\sim 30$ kHz. 
Calculations of the normal modes in the literature quoted through this section find, quite naturally, significantly smaller frequencies of order 0.5-5 kHz and only very exceptionally a mode above 10kHz. This probably just reflects the relative smallness of the sound speed $c_s<c$. 

As for the technical tasks to theoretically solve, take as an example the axial modes that satisfy an eigenvalue equation akin to that of a Schr\"odinger's wavefunction,
\be \label{normalmodes}
\frac{d^2}{dr^{*2}} \Psi(r^*) + (\omega_n^2-V(r))\Psi(r^*) =0
\ee
with effective potential
\begin{equation} \label{normalmodepotential}
V(r) = g_{00}(r) \left( \frac{l(l+1)}{r^2}+ 4\pi (\epsilon(r)-P(r))\right) - \frac{6m(r)}{r^3}
\end{equation}
where $r^*=\int^r \sqrt{g_{11}(r')/g_{00}(r')}dr'$ is the tortoise coordinate. 
Its numerical solution (or semianalytical solution, by the WKB method~\cite{Volkel:2019gpq}) provide the $\omega_n$ eigenmode frequencies.
an interesting feature of Eq.~(\ref{normalmodepotential}) is its dependence on the combination 
$(\epsilon(r)-P(r))$ instead of the $(\epsilon(r)+P(r))$ one that enters the TOV equation.

The observables are, in the first place, the intensity of the signal (whether X-ray or gravitational waves) but more characteristically, the angular frequency of the periodic phenomenon, $\omega=\omega_R+\omega_I$ (the imaginary part providing the inverse of the mode's damping time; both real and imaginary part are positive numbers).

A few modes of interest have been collected in table~\ref{tab:modes}. In the following, we discuss the $f-$ and $w-$ modes that are more promising for the physics of gravitational waves.

\begin{table}
\begin{center}
\caption{\label{tab:modes}A few simple quasinormal modes of neutron stars. Additional modes have been described: r-modes in rotating stars, see subsection~\ref{subsec:rot}, torsion and shear $t$- and $s$- modes of the crust, interfacial modes at any hypothetical surface connecting two different phases of nuclear matter, etc.}
\begin{tabular}{|ccccc|} \hline
Lettering               & $f$        &    $p$      &  $g$       & $w$               \\
Name                    &fundamental & pressure    &``gravity'' & grav. waves       \\
Physical characteristic & shallow    & bulk-sound & buoyancy   & strong GW damping \\
Type                    & polar      & polar       & polar      & axial \\
\hline
\end{tabular}
\end{center}
\end{table}

\subsubsection{f-modes}

The presence of a binary companion (or other phenomena) can excite, via tidal forces, shallow-depth excitations of the neutron star, or (``fundamental'') f-modes (and their excitations that reach deeper into the star, and are therefore sensitive to the nuclear matter at higher densities, the p-modes).

The numerical extraction of the gravitational wave signal~\cite{Rosofsky:2018vyg} uses an auxiliary complex (gravitational-gauge invariant field) in terms of the two gravitational wave polarizations, and it admits a spin-weighted multipole expansion as
$$
\psi_4:= \ddot{h}_+ +i\ddot{h}_\times= \sum_{l=2}^\infty\sum_{m=-l}^l \psi_4^{lm} \phantom{a}   _{-2}Y_{lm}(\theta,\varphi)\ .
$$
(In the computation of~\cite{Rosofsky:2018vyg}, the extracted wave is linearly polarized and thus $\psi_4$ is real.)
This GW is sourced, in an ansatz for the lowest mode, by a quadrupole pressure distortion proportional to the density $\rho$ and of maximum intensity at the surface, $\delta P=\alpha\ \rho\ \left(\frac{r}{R}\right)^2 Y_{22}$ in the stress-energy tensor.

\begin{figure}
\centerline{\includegraphics[width=7cm]{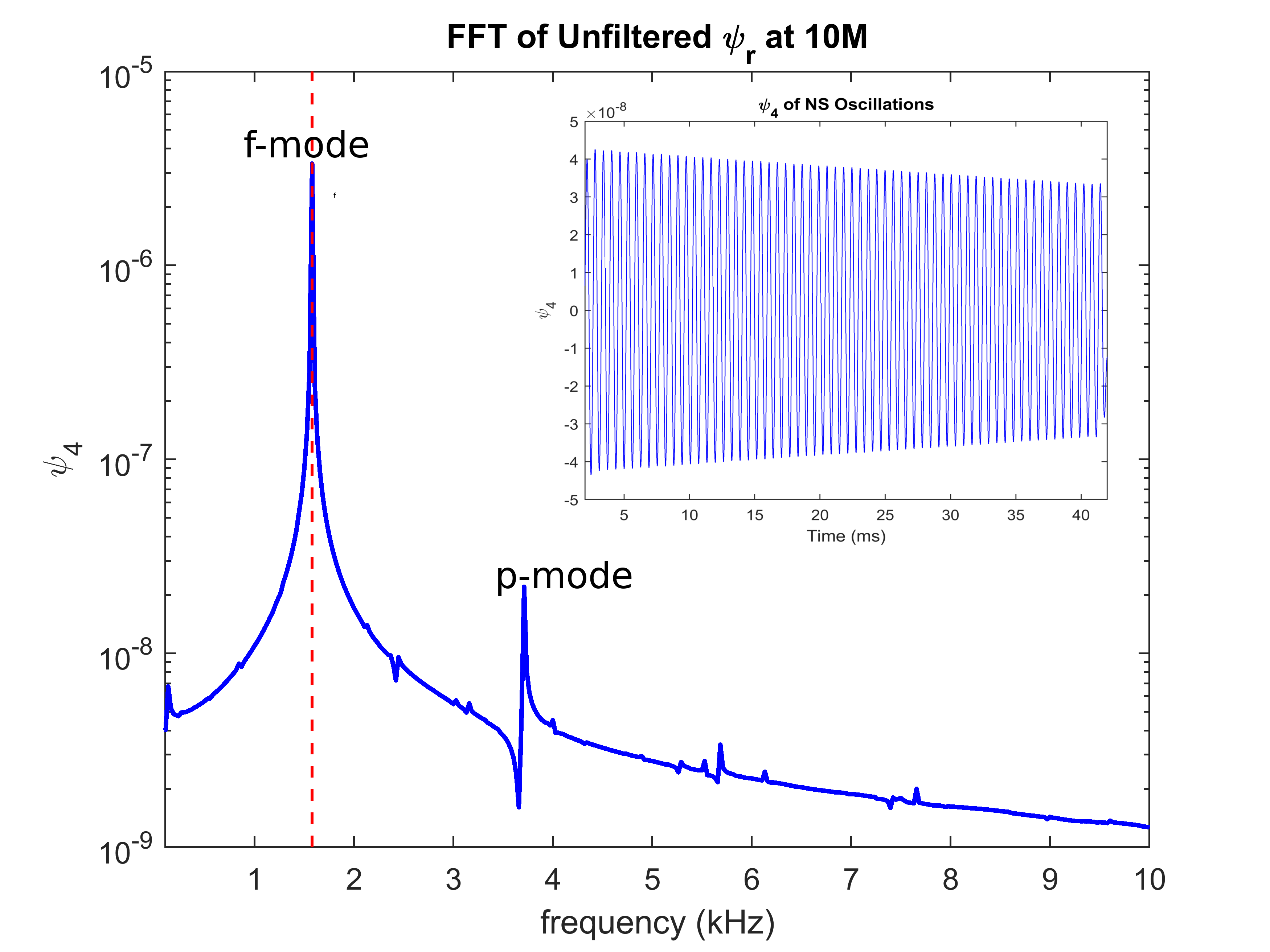}}
\caption{\label{fig:Rosofsky} Spectrum of quadrupole excitation modes and their amplitude damping (in the inset). The gravitational wave strain is measured at a distance $r=10M$ from the star's center.
The fundamental mode (shallow mode concentrated near the surface) provides by far the largest signal. The deeper $p$-modes appear as overtones.
\emph{Modified from~\cite{Rosofsky:2018vyg} with author permission.}}
\end{figure}

The perturbation oscillates with angular frequency $\omega_f$ and is damped with a characteristic exponential decay $\tau_f$ as depicted in figure~\ref{fig:Rosofsky} calculated in~\cite{Rosofsky:2018vyg}.

The calculated $f$-mode frequency comes to be 1.58 kHz, with $p$-excitations at 3.71, 5.68 and 7.66 kHz approximately. As the aLIGO reported GW signals do not seem to extend beyond the 500-600 Hz, their extraction lies in the future. According to~\cite{Wen:2019ouw}, an uncertainty of 9\% in the extraction of the mode frequency would provide a measurement of the nuclear symmetry energy $E_{\rm sym}(n=2n_0)$ with uncertainty 22\%.

The damping times calculated with various numeric methods and physical details such as the atmosphere around the neutron star are given in table~\ref{tab:fdamping}.
\begin{table}
\caption{\label{tab:fdamping} Approximate damping times, in ms, of the $f$-mode from~\cite{Rosofsky:2018vyg} for various simulations described there (the largest one has a disproportionately massive atmosphere; it is fair to state that these calculations point out to a damping time of order 300 ms). The last entry is the estimate from~\cite{Wen:2019ouw} that is slightly smaller.}
\begin{center}
\begin{tabular}{|cccccc|c|}\hline
PPM big $\alpha$ & PPM big $\rho_{\rm atm}$ & PPM CCZ4 & PPM & WENO & WENO HLLE & Wen et al.\\ 
65               &  430                     & 375      & 310 & 280  & 290       & 155-255\\
\hline
\end{tabular}\end{center}
\end{table}
Such damping times correspond to the emission of gravitational wave radiation. The presence of an artificially massive atmosphere entails a larger damping time. The bulk of the calculations point out to a damping time 
$\tau\simeq  300$ms.

The mode can be excited, for example, by an orbital resonance. An eccentric orbit can resonate with the mode and transfer energy to it~\cite{Yang:2019kmf} when $3 \omega_{\rm orbital} = \omega_{f}$. This is because, for eccentric binaries, the star's quadrupole ceases to point towards the companion, so that it is misaligned with the tidal field.

If these modes are observed, the frequency and damping time carry information on the mass and radius (or moment of inertia); the orbital energy that the binary companion deposits in the mode can in turn help extract the deformability.
An interesting relation with reduced nuclear--model dependence has been presented in~\cite{Chakravarti:2019sdc}. The observation is that the star mass times the frequency of the mode, $\nu_f$, is proportional to the square root of the mean star density, so that $M_{NS} \nu_f \propto M_{NS} \sqrt{\bar{\rho}}$. But the mean density is in turn  $\sqrt{\bar{\rho}}\propto M_{NS}/R_{NS}^3$, so that 
$M_{NS} \nu_f \propto (M_{NS}/R_{NS})^{3/2}$.
Because the Love number associated to the tidal deformability scales with the compactness as $k_2\propto (R_{NS}/M_{NS})^5$, one obtains
\begin{equation}\label{fmodeandk2}
M\nu_f \propto \frac{1}{k_2^{3/10}}\ .
\end{equation}
With more precision than this scaling estimate~\cite{Chakravarti:2019sdc}, the exponent for cold neutron stars is 0.22 instead of 3/10; and for the hot remnant of a merger, it is reported as 0.28.

\begin{figure}
\begin{center}
\includegraphics[width=8cm]{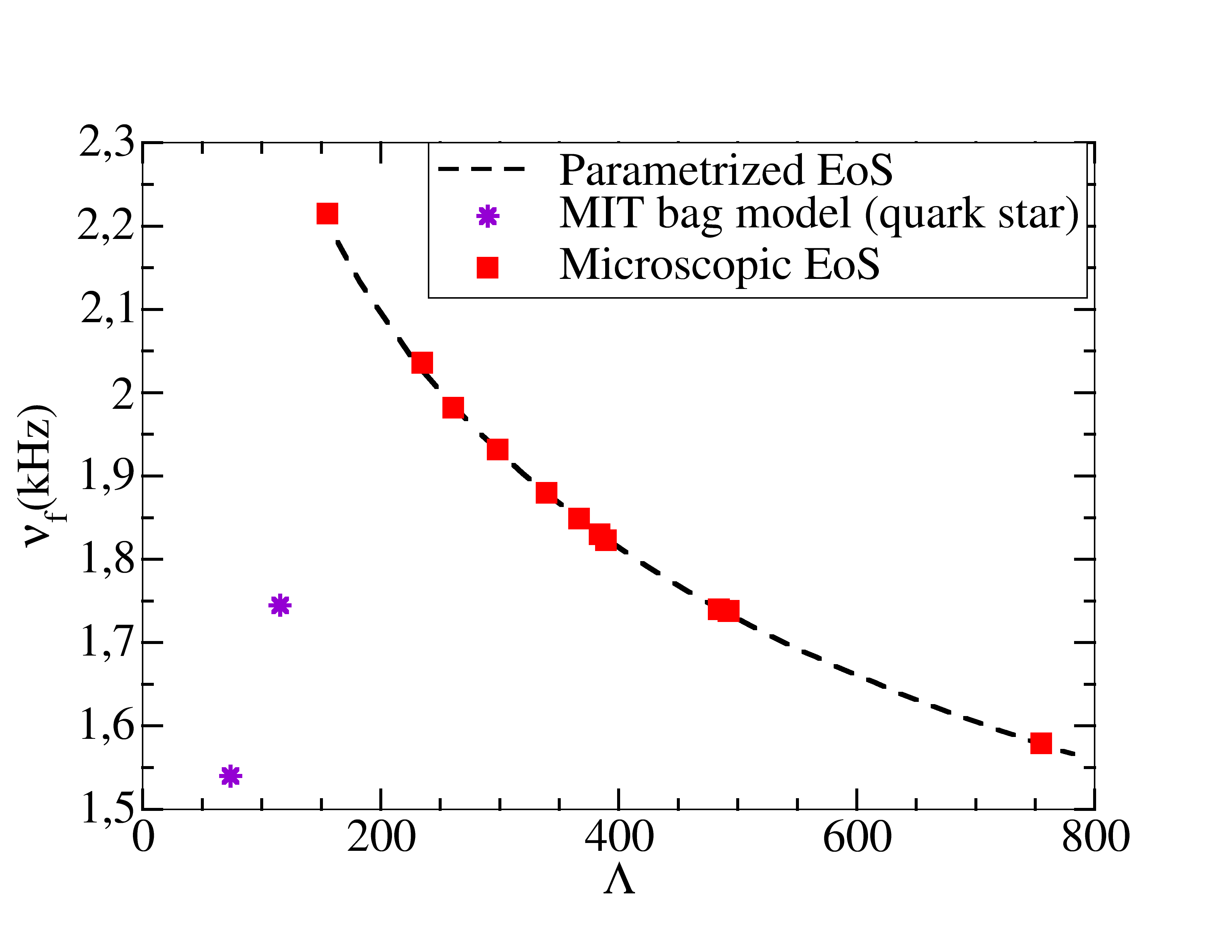}\ \ \ \ \ 
\includegraphics[width=8cm]{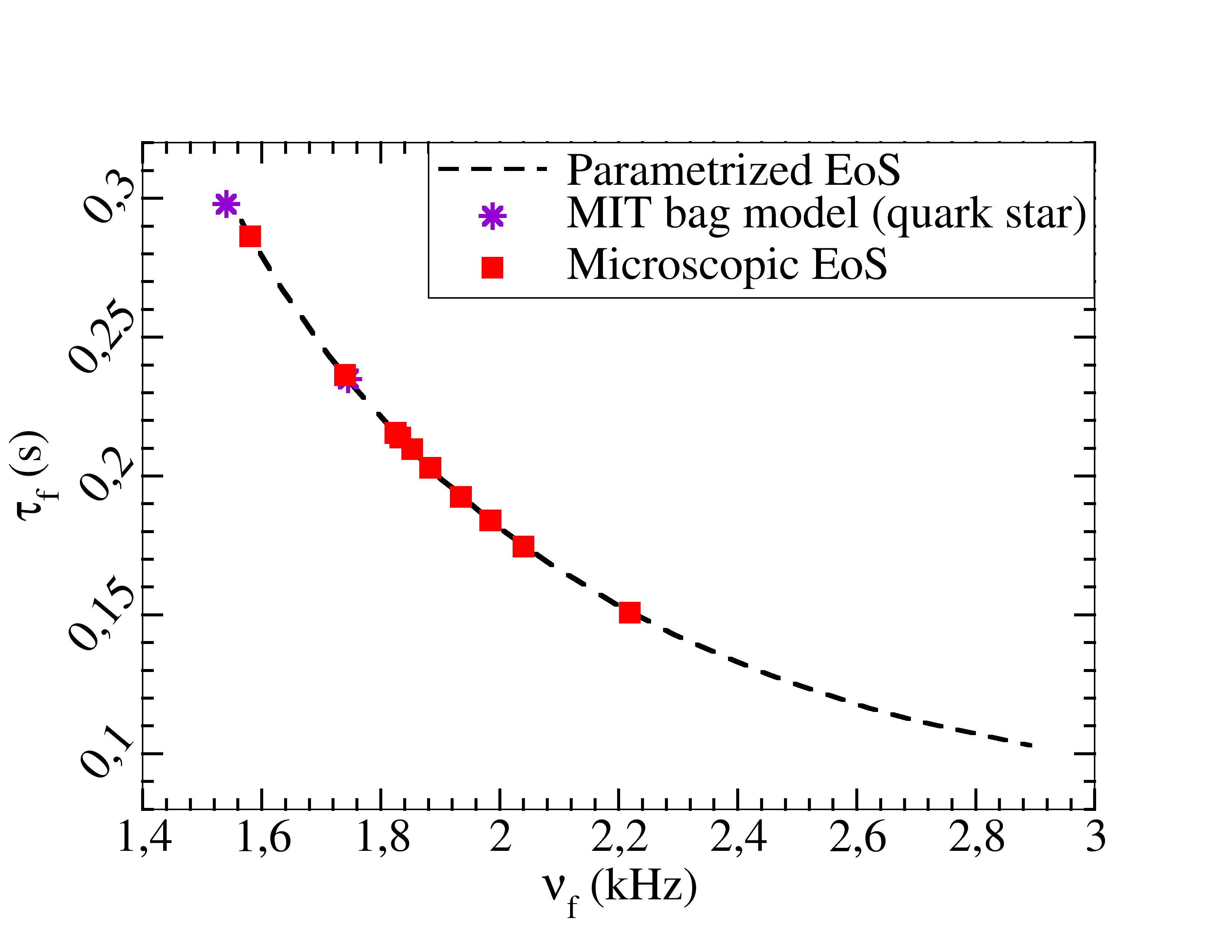}
\caption{\label{fig:corLambda-ffreq}
Left plot: The correlation of the tidal deformability with the $f$-mode frequency $\nu_f$ found by Wen {\it et al.}~\cite{Wen:2019ouw} is very clear for a large family of neutron stars, generated from 40000 parametrized nonexcluded EoS. A separation between an eventual measurement and the black line  can aid in locating very exotic stars (such as those produced with a quark EoS on the bottom left of the figure) or modifications of General Relativity.
Right plot: Correlation of the damping time with the frequency of the $f$-mode by the same authors. }
\end{center}
\end{figure}

Indeed, the tidal deformability is correlated with the $f$-mode frequency $\nu_f$ (as computed by Wen, Li, Chen and Zhang~\cite{Wen:2019ouw}) and replotted in figure~\ref{fig:corLambda-ffreq}, for the nuclear microscopic EoS (two quark stars with a bag model EoS seem to be stragglers).
The relation between the tidal deformability and $f$--mode frequency~\cite{Yang:2019kmf} reads 
$\lambda \simeq \frac{Q^2}{15\pi \nu_f^2}$ where $Q$ is a measure of the strength of the perturbation of the mass quadrupole tensor $Q_{ij}$ carefully described in~\cite{Yang:2019kmf}.

Moreover, the same authors~\cite{Wen:2019ouw}  report a strong correlation of the damping time with the frequency (right plot of figure~\ref{fig:corLambda-ffreq}) \emph{once the star mass is known} (in this example it is taken to be 1.4$M_\odot$).
so that either observable can be used in conjunction with the tidal deformability to test general relativity or the possible exoticness of the EoS. 

If General Relativity is indeed correct at the NS high densities, then the excitation modes are predictable~\cite{Blazquez-Salcedo:2013jka}, by universal relations, from the static observables of neutron stars (mass, radius, moment of inertia...) so that they become somewhat redundant. The logic of their use to impact hadron physics is then\ \
{\tt Normal modes $\xrightarrow{\text{\rm Approx.\ Universal\ Relations}}$ Global properties $\to$ EoS}.

According to Kokkotas, Apostolatos and Andersson~\cite{Kokkotas:1999mn}, the detection of the $f$-mode is more promising than either of the deeper $p$- or $w$- modes. Still, they find that with current instrumentation the necessary energy of excitation is of order $10^{-6}M_\odot$ at a distance of 10 kpc (compare with the detection of GW170817 at 40 MPc). This makes unlikely that such excitation will be visible in a supernova, so that binary NS mergers or NS-BH mergers remain the best chance. It is likely that detection needs to await the next generation of GW detectors, such as the proposed Einstein telescope. 

A clear identification of the $f$--mode in GW waves would be possible with a frequency spectral analysis of the postmerger GW signal,
if it becomes available. Because this peak is typically  in the 2-4 kHz range, current ground--based detectors are not sensitive enough (compare with the 10--100 Hz necessary to reconstruct the binary inspiral phase; or with the 200-500 Hz where the aLIGO chirps often stop, near merger). In any case, if such data became available, the postmerger signal spectrum would come to peak precisely at the $f$--mode frequency (see the review in~\cite{Bauswein:2019ybt})
\footnote{There are subdominant peaks in the frequency spectrum of the postmerger signal. A second, less prominent one but near an edge that may allow its identification corresponds to the ``inspiral'' of the tidally deformed bulges: whereas the star centres have already merged, the tidally deformed material on the outside spins as a soft bar for another couple of turns before merging too~\cite{Bauswein:2019ybt} after a few milliseconds.}.
But since $\nu_f$ strongly depends on the star's size (see figure~\ref{fig:Evolutionpostmerger} for its correlation with the tidal deformability too), one can translate the numerical computations reported there into a rough rule relating the frequency peak from the postmerger signal with a typical neutron star (of $M_{NS}=1.6M_\odot$) radius,
\begin{equation}
\nu_f({\rm kHz})\simeq 3.6-\frac{(R_{1.6}-11)({\rm km})}{2}\ .
\end{equation}
This estimate assumes that there are no phase transitions in the EoS, and then the uncertainty associated with it is typically 0.2 kHz; but if phase transitions are possible, then the correlation is significantly displaced.

\subsubsection{The inverse problem}

Assuming that General Relativity is the correct theory to address neutron star physics, the eigenvalue problem for the normal modes, Eq.~(\ref{normalmodes}) can be used in reverse to reconstruct, from experimental measurements of the eigenfrequencies $\omega_n$, the potential $V(r)$ appearing in that equation, as in the homonymous ``inverse problem'' of quantum mechanics. 

Once the potential would have been reconstructed, Eq.~(\ref{normalmodepotential}) would relate it to the EoS, as it depends on the quantity $\rho(r)-P(r)-6m(r)/r^3$.

\begin{figure}
\begin{center}
\includegraphics[width=8cm]{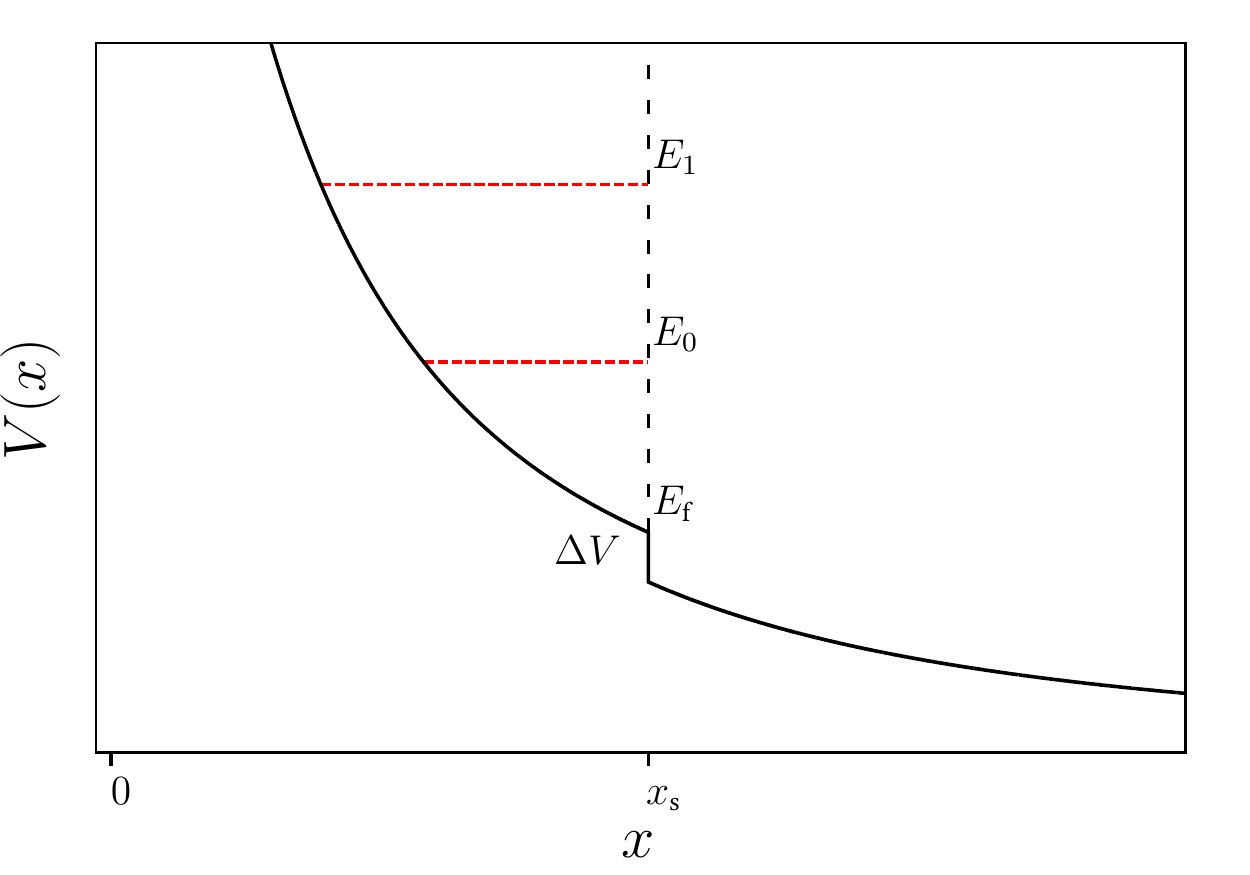}
\caption{\label{fig:inversepotential} Typical inverse potential of Eq.~(\ref{normalmodepotential}) and a few eigenvalue energies illustrating the penetration of higher modes deeper in the star, thus probing the EoS at higher densities. \emph{Adapted from~\cite{Volkel:2019gpq} with permission.}}
\end{center}
\end{figure}

Unambiguously measuring the fundamental mode would only inform us about the neutron star regions near the surface
(large $r$), but higher modes probe the potential (and thus, the EoS) deeper and deeper in the star, so that this inverse problem in terms of the normal modes might become an interesting probe of the matter at its core. This is in analogy with knowledge, in the $M(R)$-diagram inverse problem of subsection~\ref{subsec:TOV}, of the radius of  stars with very large mass.
This is illustrated in figure~\ref{fig:inversepotential}, where a typical potential of Eq.~(\ref{normalmodepotential}) is plotted together with the position of some of the normal modes.

Another possibility is to use the angular frequency $\omega$ and damping time $\tau$ to solve an inverse problem reconstructing $M$ and $I$, the star's mass and moment of inertia~\cite{Chirenti:2016xys}. This relating astrophysical data to astrophysical data without relying on nor discussing hadron physics can lead to interesting tests of General Relativity, but it is not our main focus here.

\subsubsection{w-modes}
These modes have no analogous Newtonian equivalent, and can be seen as excitations of the spacetime metric induced by its coupling to the star (recall figure~\ref{fig:modecartoon} above). 

\begin{figure}
\begin{center}
\includegraphics[width=8cm]{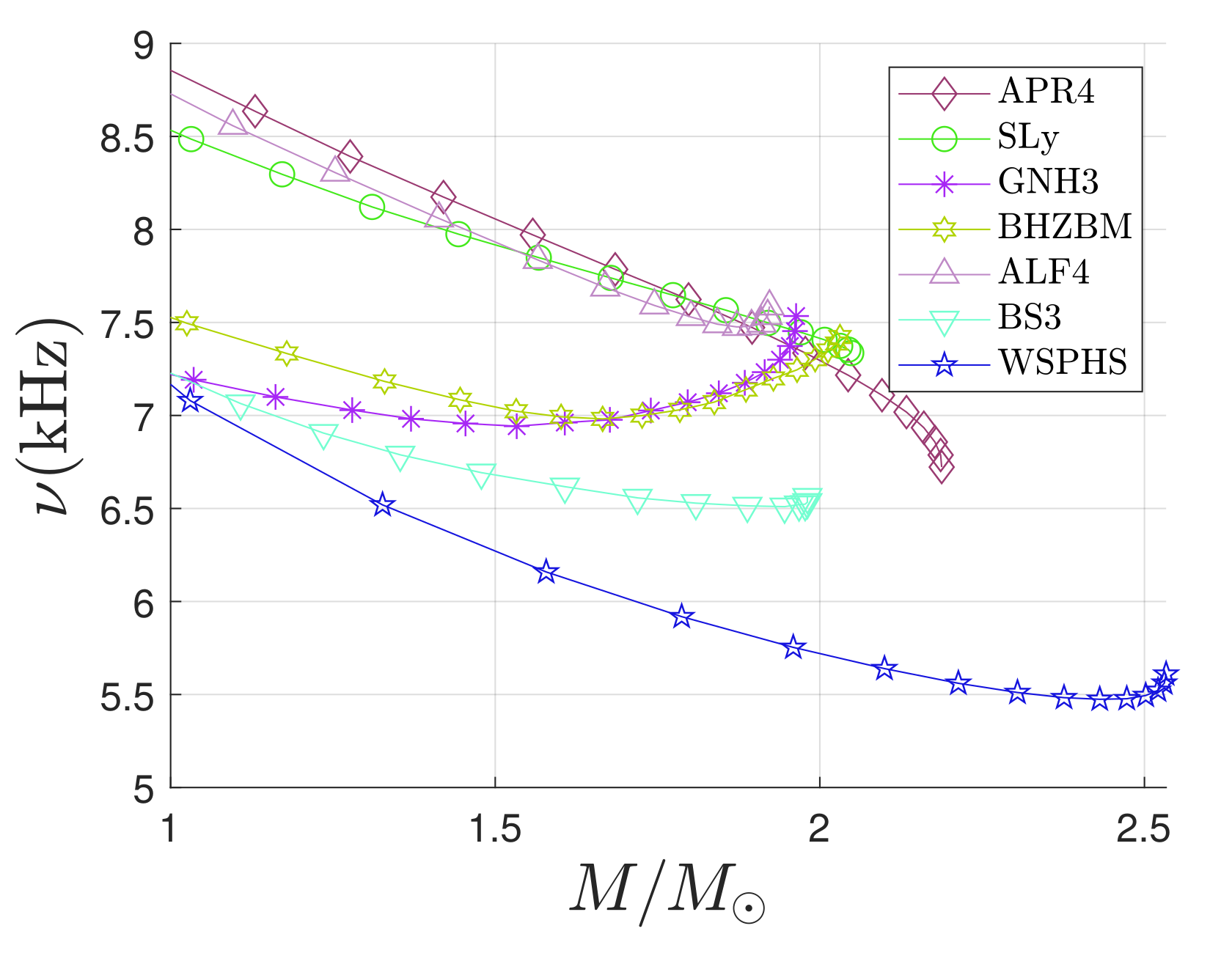}
\includegraphics[width=8cm]{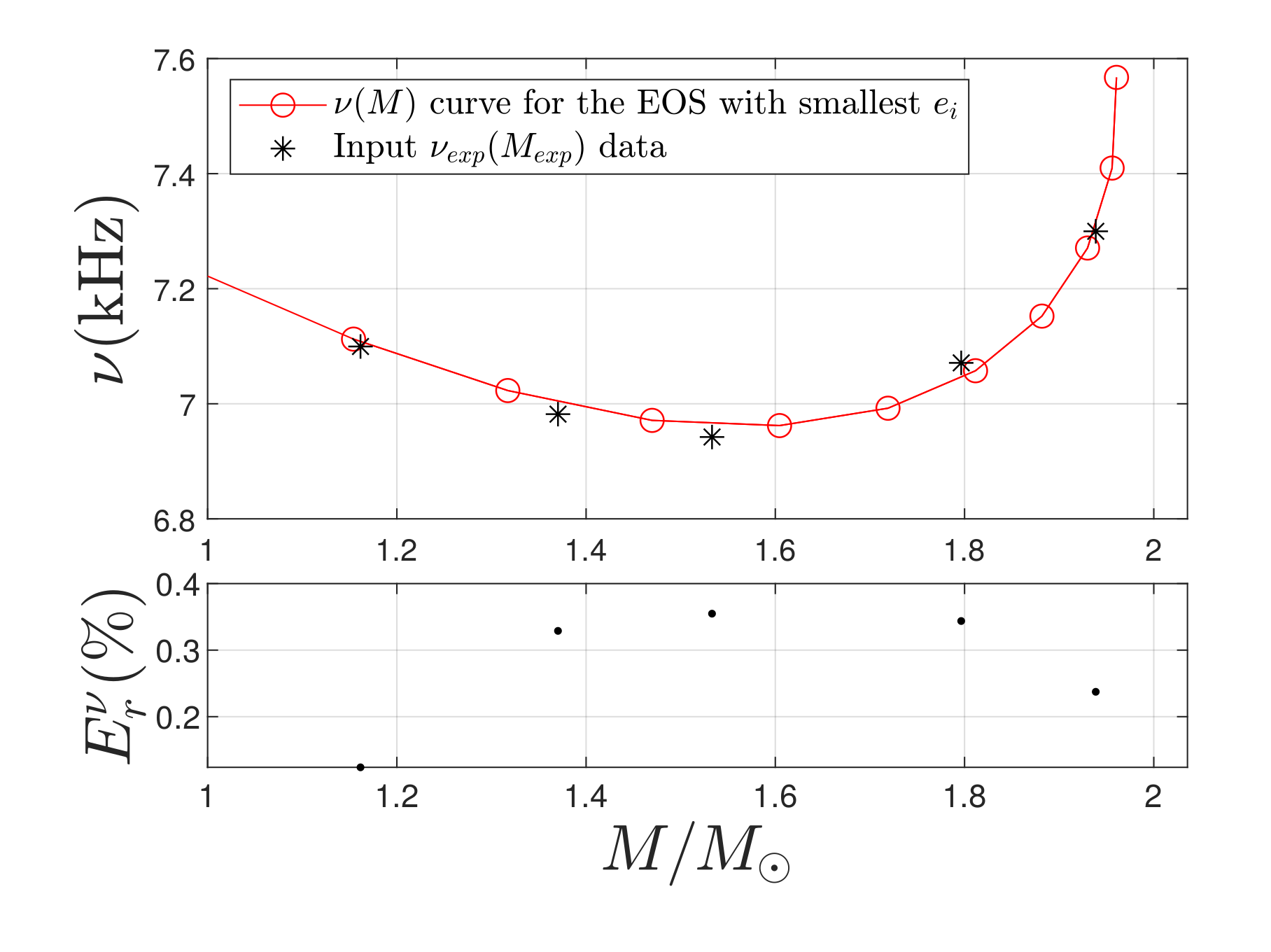}
\caption{\label{fig:wmodes} 
The quasinormal w-modes can also be used to constrain the Equation of State.
Left: frequency of the w-quasinormal modes for various equations of state. Right: reconstructed EoS (from a simulated measurement of the modes) compared to the EoS actually fed into the computation of the modes, to check the fidelity of the reconstruction. (\emph{See~\cite{Mena-Fernandez:2019irg} for details. Rerendered figure courtesy of the authors.})}
\end{center}
\end{figure}

This coupling allows, in principle, to solve the inverse problem to reconstruct the EoS. This has been demonstrated~\cite{Mena-Fernandez:2019irg} with a  piecewise-polytropic EoS that was imposed and then  also reconstructed from the resulting w-modes, with an appropriate numeric solver for the inverse problem (see figure~\ref{fig:wmodes}). 

We can see from the figure, and also for table IV in ref.~\cite{Mena-Fernandez:2019irg}, that with 5 well measured points one can reach 1\% precision in the reconstruction of the EoS in the range of interest for neutron stars.

\subsubsection{Burst oscillations of the crust (are they atmospheric oscillations?)}

The accretion of material on the neutron star projects energetic X-ray bursts (yielding super-Eddington luminosities) that pulsate. At least for some stars they correlate with the spin of the neutron star, with some frequency drift that is amenable to detailed calculation~\cite{Chambers:2018bdi} from the motion of the accreted material in the star's surface and the mode excitation thereof. 

In fact, the X-rays leaving the star in a beacon-like motion due to the rotation trace the spacetime geometry of the star's outside metric and might be usable to test General Relativity there~\cite{Silva:2019leq}.

But other authors have also related this burst oscillations to oscillations of an NS atmosphere~\cite{Bollimpalli:2018slk}. This layer, having very small mass, and pulsating radially (such as a nuclear breathing mode) around $r=r_0>R$, is not supposed to significantly feed Gravitational Waves. 
However, if their frequency can be measured, perhaps in X-ray oscillations, one can estimate the mass and radius of the neutron star. 

The oscillation modes very far from the star (large $r_0$) are undamped and have angular frequencies
\be
\omega^2_k = \left(\frac{M}{r_0^2} \right)^2 \ f_k
\ee
(with $f_k$ a $k$-dependent rational number). This is independent of the neutron star radius (as if it were a pointlike object)
As $r_0$ becomes of the order of the star radius, radiation damping activates and both the damping and the modified frequency of the mode are sensitive to $R$.

\begin{figure}
\begin{centering}
\includegraphics[width=5cm]{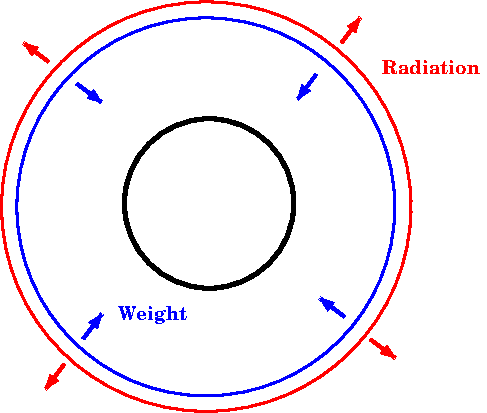}\ \ \ \ \ 
\includegraphics[width=7cm]{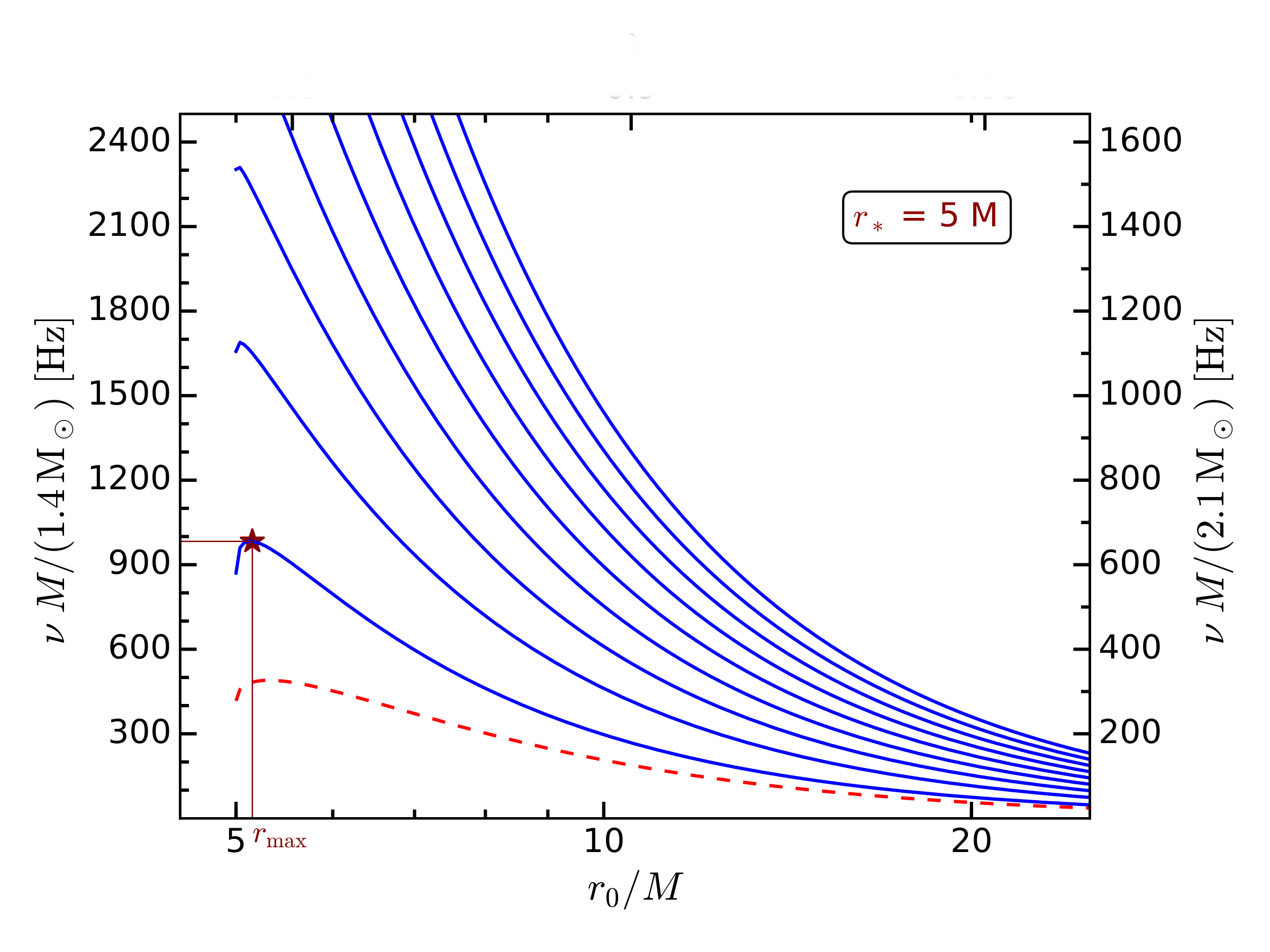}
\caption{Left plot: cartoon of the atmospheric oscillations as radial ``breathing''oscillations of a supposed atmosphere enveloping the neutron star at a (coordinate) distance $r_0$ from its center floated against its weight by the outward radiation pressure and pulsated by an accretion event. Right plot: frequency of the first ten atmospheric modes depending on the height of the atmospheric layer (the overdamped first mode is shown dashed), for a stellar radius $R=5M$. \emph{Adapted with author permission from preprint~\cite{Bollimpalli:2018slk}.}
\label{fig:atmospheric}
}
\end{centering}
\end{figure}

The lowest mode is critically damped (see figure~\ref{fig:atmospheric}), so that excited modes need to be identified. If the 300-600 Hz oscillations detected during type-I X-ray bursts have anything to do with these atmospheric oscillations, then Bollimpalli {\it et al.}~\cite{Bollimpalli:2018slk} suggest neutron star radii in excess of 13 km and extending up to 20 km, which would be a quite significant increase in the radius, and thus a quite different EoS. 
In any case, this phenomenon suggests that care needs to be exercised when trying to identify star oscillation modes purely from $X$-ray bursts.

       \subsection{Rotation phenomena} \label{subsec:rot}

Pulsars are rotating objects and their period $T$ can be inferred from the radio pulse arrival time; let us first discern when can their rotation be treated by nonrelativistic methods. 
The rotation is certainly relativistic when the velocity at the equator approaches that of light, namely $\Omega\sim \frac{c}{R_{NS}}=\frac{3\cdot 10^5\rm km/s}{12 \rm km}$, so that $T_{\rm min}=0.25$ ms.
Taking into account that we are outside the strong gravitational field, we would measure that period increased
by a redshift factor $1/\sqrt{1-R_S/R_{NS}}$ in terms of the Schwarzschild radius of about 4.2 km for a 1.4$M_\odot$ star, so that $T_{\rm min}\simeq 0.3$ ms for an external observer. (Realistic computations employing causality {\it in lieu} of an actual EoS put the limit at 0.41ms~\cite{Paschalidis:2016vmz}.)

Newly formed stars, because of angular momentum conservation for a collapsing system, have to rotate rapidly; likewise, accreting pulsars in binary systems can have their rotation reaccelerated. Thus, there is an important population of millisecond pulsars with periods $T_R\sim$ ms). The record pulsar PSR J1748-2446ad has a period~\cite{Hessels:2006ze} of 1.4 ms, almost five times larger than $T_{\rm min}$. Here, relativistic corrections are sizeable (and hence, numerical relativity is a convenient tool).

On the contrary, settled pulsars have typical periods of order seconds and their rotation can be treated considering $\Omega$ as a perturbation around a static, spherically symmetric body.

The maximum practical rotation rate is also connected to the EoS: if it is soft, the matter is compressible, which makes the star small and it can spin faster. On the other hand, stiff EoS make for larger neutron stars and thus, smaller maximum rotation rates.

\subsubsection{Perturbative analysis}
If the relativistic corrections are moderate, one can perform an expansion around $\Omega\simeq 0$.

A key angular frequency is that of Kepler, $\Omega_K$, at which a test mass at the Equator rotates with its equilibrium orbital velocity, so that any increase $\Omega>\Omega_K$ just sheds mass expelled by the centrifugal force. In a nonrelativistic analysis, for uniformly rotating polytropic stars~\cite{Paschalidis:2016vmz},
\begin{equation} \label{OmegaKepler}
\Omega_K = 
\left(\frac{2}{3}\right)^{3/2} \sqrt{\frac{GM}{R^3}}\ .
\end{equation}
With typical data from section~\ref{sec:static}, this yields a period of about 0.8 ms. Therefore, it is not clear how well the nonrelativistic approximation applies, but for the rest of this paragraph we proceed with it.

\begin{figure}
\begin{center}
\includegraphics[width=8cm]{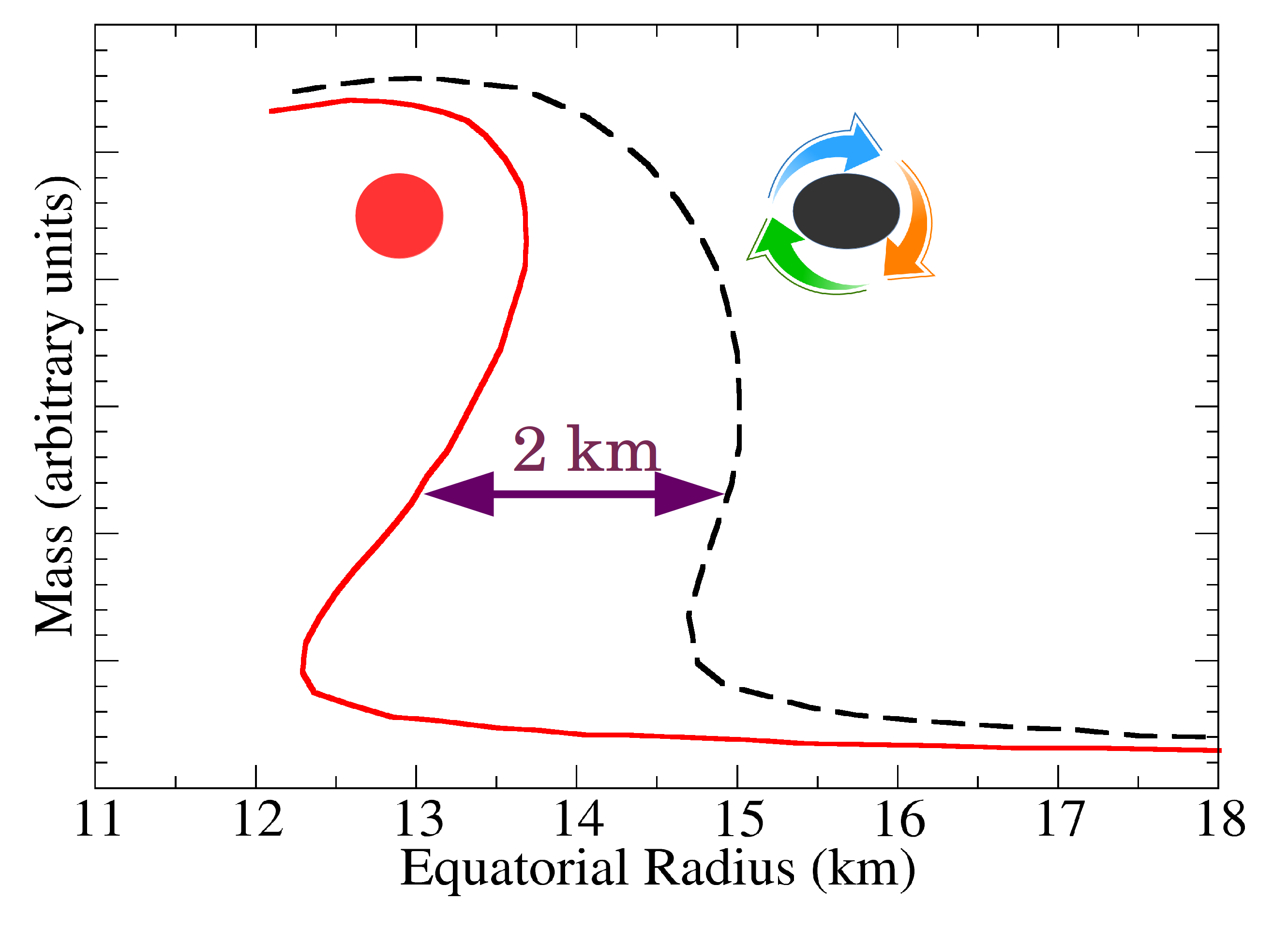}
\end{center}
\caption{\label{fig:rotation} Mass-radius diagram for static (left line) and maximally rotating stars (right line) as computed by Boshkayev~\cite{Boshkayev:2016pmk}. The difference in maximum allowed mass is small (about $0.1M_\odot$) but the difference in equatorial radius, up to 2 km, is sizeable.}
\end{figure}

Figure~\ref{fig:rotation} redraws a calculation of the mass--radius diagram for a family of rotating stars~\cite{Boshkayev:2016pmk} at the mass--shedding frequency~\footnote{We avoid specifying the EoS and discussing the largest mass, but in that work $ \Omega_{\rm max} \simeq 7800\frac{\rm rad}{\rm s} \sqrt{\frac{M}{M_\odot}\left(\frac{10\rm km}{R}\right)^3}$.}.

The difference in maximum neutron star mass is not significant, but the equatorial radius, due to the strong centrifugal force, is large. Thus, when measurements of stellar radii become available, millisecond pulsars will fall on a curve to the right of normal pulsars in this diagram.

To perform such calculations, the TOV system of Eq.~(\ref{basicTOV}) needs to be extended along the lines of~\cite{Hartle:1967he}. Starting from the static interior Schwarzschild metric with $\nu_0(r)$ and $\lambda_0(r)$ interior functions and static mass $M_0=M(\Omega=0)$, the metric for the slowly rotating star becomes~\cite{Stuchlik:2015gsa}:
\begin{equation}
ds^2 = e^{2\nu_0}(1\color{black}+2h\color{black})dt^2 -e^{2\lambda}\left(1\color{black} + \frac{e^{2\lambda_0}}{r} m(r) \color{black}\right)dr^2
+r^2( 1 \color{black} +2k\color{black}) \left( d\theta^2+ \sin^2 \theta(d\varphi
\color{black} -\omega dt\color{black})^2 \right) + O(\Omega^3).
\end{equation}

Here, $\omega(r)$ is of order $\Omega$, and the following new functions are of order $\Omega^2$ and show a quadrupolar star deformation with $\theta$ taken from the rotation axis,
\begin{equation}
h= h^{(0)}(r)+h^{(2)}(r) P_2(\cos\theta)\ ; \ \ \ m= m^{(0)}(r)+m^{(2)}(r) P_2(\cos\theta)\ ; \ \ \ 
k= k^{(0)}(r)+k^{(2)}(r) P_2(\cos\theta)\ .
\end{equation}
Likewise, the pressure also receives a perturbation $p^{(0)}$.
The system of equations~(\ref{basicTOV}) is then extended to include first order equations for these new functions and is solved to an $R$ where the pressure vanishes.
The mass of the star then receives a new contribution $\delta M= m_0(R)+J^2/R^3$. Estimating the angular momentum $J$ therein does require a couple more steps.

Very briefly, the angular velocity of the fluid relative to the local inertial frame is $\tilde\omega(r)=\Omega-\omega(r)$. The equation for $\tilde\omega(r)$ is~\cite{Boshkayev:2016pmk}
\begin{equation}
\frac{1}{r^4} \frac{d}{dr}\left( r^4 e^{-\nu_0-\lambda_0} \frac{d\tilde \omega}{dr}\right)
+\frac{4}{r}\frac{de^{-\nu_0-\lambda_0}}{dr}\tilde \omega = 0
\end{equation}
that once solved allows to evaluate 
\begin{equation}
J = \frac{R^4}{6} \left. \frac{d\tilde\omega}{r}\right|_{r=R} \ .
\end{equation}
The matching at $r=R$ with the exterior Hartle solution to the Einstein equations is too long to copy here.

\subsubsection{Numerical relativity treatment for high frequencies}

Beyond that Hartle (\& Thorne) treatment, rapidly rotating neutron stars must be 
assessed numerically~\cite{Stergioulas:1994ea}. This becomes necessary for binary neutron star mergers where mass can be shed and the angular velocity can well exceed the Keplerian limit (that also can change due to the nonspherical mass distribution, but we will ignore this point). The corresponding angular frequency~\cite{Paschalidis:2016vmz} empirically satisfies a parametric relation analogous to Eq.~(\ref{OmegaKepler}). 
\begin{equation} \label{OmegaKepler2}
\Omega_K = 0.67 \sqrt{\frac{GM_{\rm max}}{R_{\rm max}^3}}\ 
\end{equation}
in terms of the mass and radius of the heaviest neutron star on the $M(R)$ curve. This maximum velocity has a precision of order 5\%~\footnote{A more precise formula of Lasota quoted by~\cite{Paschalidis:2016vmz} in its Eq.~(79) can fit the numeric data down to 1.5\% precision by adding a term suppressed by one power of $\frac{GM_{\rm max}}{R_{\rm max}}$.}.

An interesting development brought about by GW170817 is the possibility of combining Eq.~(\ref{OmegaKepler}) with Eq.~(\ref{maxmassGW}) above; there, we saw that the ejection of mass from the merger, feeding the glow of the subsequent kilonova, provides an upper bound to the maximum mass of a neutron star. But this means that a measurement of $\Omega$ for any pulsar immediately provides an upper bound on the maximum radius of that neutron star of maximum mass, that is,
\begin{equation}
R_{\rm max}\leq 
\left( \left(\frac{\Omega}{0.67}\right)^2 \frac{1}{GM_{\rm max}}  \right)^{-1/3}\ .
\end{equation}
With the 716 Hz pulsar, $\Omega=4499$rad/s; from Eq.~(\ref{maxmassGW}) we take $M_{\rm max}\leq 2.28M_\odot$ and obtain $R_{\rm max}\leq 18.9$ km (assigning 0.2$M_\odot$ uncertainty to $M_{\rm max}$ changes the result by 3\%). 
In comparison with the bounds in table~\ref{tab:NSradii} below, this is not very stringent.
But noting that the bound on $R_{\rm max}$ depends on $\Omega_K^{-2/3}$, finding faster spinning pulsars at and above 1kHz would make the bound competitive (by pushing up $\Omega_K^{-2/3}$ and thus diminishing $R_{\rm max}$).

The numerical analysis provides a relation between the maximum mass and minimum period (again, using the stiff--most EoS allowed by causality), see Eq.~(91) of~\cite{Paschalidis:2016vmz}  
\begin{equation}
T_{\rm min}(ms) = 0.196 \left(\frac{M^{\rm max}}{M_\odot}\right)
\end{equation}

Modifications of General Relativity have also been incorporated into some numerical codes, and it appears that fast rotation enhances the difference between GR and $f(R)$ theory~\cite{Yazadjiev:2015zia}.

As for measuring the NS spin (as opposed to an upper bound) in the gravitational waves emitted by a binary neutron star system, it appears that aLIGO-Virgo will not yet achieve the needed sensitivity. Future detectors might do it, especially for large dimensionless spin $\frac{cJ}{GM^2}>0.2$ that would make precession effects appreciably large~\cite{Samajdar:2019ulq}: significant effects appear for aligned spins of merging
millisecond pulsars~\cite{Tsokaros:2019anx}. (Because mergers happen late after the formation of the Neutron Stars, such binary population has to be maintained by accreting material maintaining the angular momentum since the stars have had ample time to spin down.)
For those GW signals, the quadrupole deformation of the star ({\it e.g.} due to the centrifugal force under fast spinning) would need to be taken into account to avoid bias in the extraction of the spin.

\subsubsection{Glitches}

Pulsar periods (of order ms-s) are extremely regular in short time intervals (hours) but 
longer observations reveal that they drift down to smaller values. This is understood from the EM radiation being emitted at the expense of angular momentum.
But additionally, decade--long observations of pulsars shows ``glitches''.
These are sudden period increases, and are interpreted as a change in the moment of inertia, and indicating that neutron stars rotate differentially. 
If the cause was the infall of material with angular momentum from outside the star, the glitch would be accompanied by an X-ray burst upon burning that material. No such association is known to us in the data, so the cause should be inside the star.
Ruderman's proposal that catastrophic starquakes readjusted the star crust was disfavored by the frequent repetition of glitches in the same pulsar~\footnote{The fact that intense crustquakes are not so common makes them less attractive as sources of Gravitational Waves.}.

The commonly accepted theory~\cite{Sauls:2019ffv}
 is that inner layers of the star are disentrained from the crust (due to the low viscosity of a likely superfluid phase that has vorticity concentrated around a discrete set of pinned vortices).
Those inner layers therefore keep rotating with constant angular momentum even if the
nuclear lattice in the crust spins down electromagnetically. 

A glitch happens when the vortices of whichever superfluid phase~\footnote{There are various proposals in the literature: neutrons could condense in a $^1S_0$ BCS wavefunction at low, in a $^3P_2$ at high density, respectively, and numerous ones have been proposed for quarks.} unpin and suddenly transfer angular momentum to the crust, that speeds up.
This change in angular velocity $\Delta\Omega/\Omega$ widely varies from $10^{−11}$ to $10^{−4}$. In addition to radio waves, glitches are also observed in X-ray timing~\cite{Kaspi}.

For example, for a $^1S_0$ superfluid, the phase $\varphi$ of the order parameter $\psi({\bf r}) = \arrowvert \psi \arrowvert e^{i\varphi({\bf r})}$ entails a mass current with velocity depending on the fermion mass ${\bf v}_s = \frac{\hbar}{2m_n} \nabla \varphi$. 
Because the velocity field is derived as the gradient of a ``potential'' (the superfluid phase), it must be irrotational
$\nabla \times {\bf v}_s = {\bf 0}$ except at singular points in the fluid, which are of course the vortices~\cite{Sauls:2019ffv}.
In circulating around such a vortex, the phase must vary at most by a multiple of $2\pi$ upon returning to the same point, so that the order parameter is a well defined field. This entails that the circulation of the velocity field around the vortex is quantized,
\begin{equation} \label{vortices}
\oint_C {\bf v}_s\cdot d{\bf l} = \frac{h}{2m_n} N
\end{equation}
with $N$ an integer. Thus, the vorticity of the superfluid velocity with the origin at the nearest vortex is $\nabla\times {\bf v}_s = N \frac{h}{2m_n}\delta^{(2)}({\bf r}) \hat{\bf z}$. These are the vortices that are presumably pinned or unpinned from the crust in the process of a glitch.

It is interesting that the intensity  of the glitch does not seem to be proportional to the time since the glitch immediately before, which suggests that only a small part of the shear between the two differentially rotating layers is relaxed (it is not a build--up process from scratch between glitches).

The lesson for a nuclear/particle physicist is that there very likely exists at least one interface between two different phases of neutron star matter, with the phase on the inner side of that surface very likely a superfluid, and carrying at least 1\% of the moment of inertia which is not corotating with the pulsation~\cite{Kaspi} while the crust carries at least $\sim 10\%$ of the moment of inertia~\cite{Chamel:2016led}.

\subsubsection{Effect on normal modes; hydrodynamic instabilities}

The normal modes for rapidly rotating stars see the degeneracy in the index $m$ lifted: for each $l$--mode of the static star, there are $2l+1$ $(l,m)$ modes of the spinning one. Moreover, the direction of the mode, whether prograde ($m<0$) or retrograde ($m>0$) also affects its frequency. Additionally, the rotation couples each polar $l$--mode to an axial $l+1$--mode (and, with less intensity, to an $l-1$ one), so that energy can be transfered between them. Moreover, the rotation can cause instabilities in the flow of material.

Such hydrodynamic instabilities can lead to turbulent flow, reinforcing our thesis that turbulence in neutron stars is one of the most interesting lines of research for the future.

For a recent example, let us mention those that may arise when one considers a normal fluid and a superfluid coexisting in the rotating star~\cite{Khomenko:2019rgd}. If they are decoupled, sound propagates with dispersion relation
\begin{equation}
\omega \simeq c_i k \left( 1+ \frac{2\Omega^2}{c_i^2 k^2} \sin^2\theta \right)
\end{equation}
where the last term arises from the Coriolis force with $\theta$ the angle between the star's angular velocity $\vec{\Omega}$ and the direction of propagation $\hat{k}$.

But if the vortices are pinned to the crust, the superfluid is rotating as dictated by the discussion around Eq.~(\ref{vortices}), instead of corotating with the rest of the star. The resulting ``background flows'' can give rise to instabilities. For example, Eq.(81) of~\cite{Khomenko:2019rgd} reports a dispersion relation for sound propagating along the vortex,
(parallel to $\vec{\Omega}$ and $\hat{z}$)
\begin{equation}
\omega= \mp 2\Omega + k_zv_z + i{\mathcal B}(-2\Omega \pm k_z v_z)\ .
\end{equation}
The last term is proportional to the strength of the coupling between the two fluid components, parametrized by ${\mathcal B}$. If $k_z v_z>0$, an easily attainable condition, an instability of the flow is triggered. This supports the idea that turbulence may be playing an important role.

Next we discuss a different type of instability that leads to the emission of gravitational waves.

\subsubsection{r-mode instability} \label{subsec:r}
r-modes~\cite{Andersson:1997xt} are perturbations of the fluid of toroidal shape, analogous to the atmospheric Rossby waves on Earth, and have a velocity disturbance
\begin{equation}
\delta{\bf v} = \alpha R\Omega  \left( \frac{r}{R}\right)^l {\bf Y^B}_{lm} e^{i\omega t}
\end{equation}
where the star's angular frequency is $\Omega$ and the pulsation one $\omega$ and ${\bf Y^B}_{lm}$ is the axial vector spherical harmonic. The Coriolis force of the rotating frame acts as restoring force. Early LIGO searches bound $\alpha$ to be under $10^{-4}$.

Starting with the quadrupole, these modes can radiate gravitational waves because of the star's rotation. (For each increase in $l$, the gravitational emission decreases an order of magnitude, whereas the damping increases about 20\%. Thus, it is sufficient to consider $l=2$.) 

Now, if the disturbed fluid rotates in the direction opposite to the star in the star's rest frame, with $\omega_{\rm mode}<\Omega$, it appears to an observer at infinity to be rotating in the same sense as the star, so the angular momentum carried away by the GWs increase its retrograde velocity (in effect trying to make the quadrupole static in the frame of the observer at infinity). This means that the mode increases its retrograde angular velocity until $\omega_{\rm mode}=-\Omega$; in the reference frame of the star, the mode is accelerating backwards.

In a nonideal fluid (see subsection~\ref{subsec:transport}) however the bump in the star does not come to rest respect to infinity, but is dragged along by its coupling to the rest of the star; thus, it continues dissipating gravitational radiation extracted from the star's rotation. Beyond the $r$-mode, the generic phenomenon is called CFS-instability~\cite{Chandrasekhar:1992pr}.

\begin{figure}
\begin{center}
\includegraphics[width=8cm]{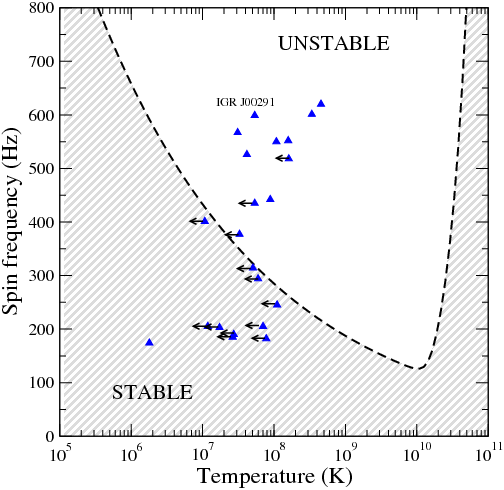}
\includegraphics[width=8cm]{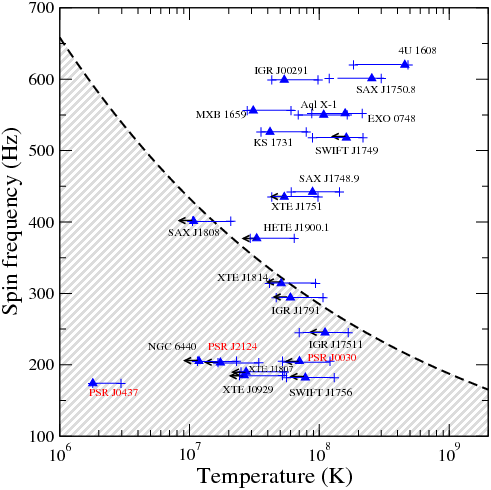}
\caption{\label{fig:rmode} Left: instability window (unshaded area above the curve). Right: detail of the left bottom area where many pulsars concentrate, with uncertainty in their temperature measurements. \emph{Reprinted from figs. 2 and 3 from ``Constraining the physics of the r-mode instability in
neutron stars with X-ray and UV observations''~\cite{Haskell:2012vg}.} }
\end{center}
\end{figure}

In an ideal fluid, the instability would be active for arbitrarily small star rotational frequency $\Omega$. But in the presence of damping, all frequencies smaller than $2\pi/\tau_{\rm damping}$ are stabilized by fluid friction.
Figure~\ref{fig:rmode}, taken from~\cite{Haskell:2012vg}, shows the window of instability (larger frequencies above the shaded area) that is limited to the right (high temperature) by the contribution of the Urca process to bulk viscosity, with characteristic time $\tau_{\zeta\nu}=2.7\times 10^{11} (M/1.4M_\odot) (10{\rm km} /R)({\rm Period}/{1\rm ms})^2 (10^9{\rm K}/T)^6 $. On the left, at low temperature, the limit is given by the shear viscosity at the boundary layer of the crust-core interface, with modeled time 
$\tau= 3\times 10^5 \sqrt{({\rm Period/ms})}(T/10^9{\rm K})$.

Figure~\ref{fig:rmode} also shows a number of millisecond pulsars and X-ray binaries whose temperature and spin rate are simultaneously known. 

Many of them lie in the instability window (unshaded area), so they should be spinning down quickly by emitting gravitational waves, which is not the case. This suggests that faster damping mechanisms are at work, and hydrodynamic turbulence might play a role. See a recent appraisal in~\cite{Roy:2019hzp}.
Alternative explanations look for phases with larger viscosity~\cite{Alford:2013pma}, suggesting that the window shifts left to lower temperatures if interacting quark matter occupies the star's core. But since significantly softening the EoS is problematic for the maximum mass discussed in section~\ref{sec:static}, exotic phases are under some tension. 

In any case, to search for the gravitational wave signal of these modes it is useful to know that the frequency of the emitted radiation is related to the rotation of the star (just like in a binary merger it is twice the orbital frequency). Indeed a reasonable approximation is $\nu_{GW\ \rm r-mode} = \frac{4}{3} (2\pi\Omega)$, though a detailed analysis~\cite{Caride:2019hcv} finds, instead of the factor of 4/3, an interval $(1.39,1.57)$ where the lower limit substantially smaller for millisecond pulsars, even down to 1.1.

\section{Outlook}

There are numerous topics that we have left outside the fence of this review, to keep it introductory, as the field is now so productive that it well deserves an encyclopedia. 

One is the gravitational wave signal of a core collapse supernova, that will certainly contain information of interest for nuclear and particle physicists. The energy emitted being of order $10^{-9}$ solar masses, it will only be detectable if the explosion sets off in our own galaxy or in its satellites~\cite{Powell:2018isq}. The $g$-modes that we have not discussed either (analogous to buoyant ocean modes on the Earth) are expected to feed them, and its characteristic frequency is 
$f_g \simeq \frac{1}{2\pi} \frac{M}{R^2} \sqrt{1.1 m_N/<E_{\bar{\nu}_e}}>(1-M/R)$, that for typical values is around 800-1000Hz and, if measured, informs us about the neutrino energy.
The shape of the GW signal is less distinct than the chirp caused by a merger, and to extract it from the data, it would be desirable to have neutrino detectors that, as for SN1987a would observe a supernova in our galaxy, advise on the time interval to extract the signal from. The amplitude peaks about 0.3s after the core collapse.

Another is the study of purely strange quark stars that are self-bound, adopting a linearized EoS  where the pressure is zero below a few times nuclear saturation density, $ \epsilon (P) = 4B^4 + \frac{1}{c_{\rm QM}^2} P$
(In bag models of very weakly interacting quarks, $c_{QM}^2\approx$ 1/3 and $B$ is the bag constant). Or, likewise, proposed boson stars or stars accreting dark matter. We think that all these proposals are for the time being removed from what can credibly be constrained with astrophysical data.

This is why we have concentrated on the equation of state at a generic level, and on signatures of phase transitions. 

When the dust settles, it appears  to us that aLIGO was lucky early on and found an NS-NS merger ahead of schedule
(subsection~\ref{subsec:howmany}) and that, while run O3 may perhaps produce some more candidates (in fact one has already been proposed, S190425z, which is under study), having good statistics on the tidal deformability of neutron stars will need to wait for a more advanced detector such as the Einstein telescope. Meanwhile, the best chance to discover or constrain a twin-star branch (an automatic telltale of a phase transition at high density, subsection~\ref{subsec:phases}) will be the simultaneous measurement of mass and radius in the traditional $M(R)$ diagram, perhaps by the X-ray studies of NICER~\cite{Ozel:2015ykl}.

The discovery of GW170817, however, brings a twist to the old question  \emph{how stiff is the Equation of State in neutron stars?}. For a few years, the standard answer has been ``very stiff'' or $2M_\odot$ stars cannot be supported. 
But the value of the tidal deformability and the observation of the following kilonova with its bound on the maximum mass (subsection~\ref{subsec:NSmasses}) have shown us that $P\sim \epsilon$ in the core of the star is in tension with observation. And at least for arbitrarily large density, we know that $c^2_s\sim 1/3$ from pQCD asymptotic freedom. 

Because the most model--independent hadronic EoS from Chiral Effective Theory at low density get increasingly stiffer with $n$, there is room for a phase transition in the core of the star that relaxes $P(\epsilon)$ and the possibility is under very active investigation.

In short, soft EoS disagree with $M_{\rm max}>2M_\odot$, and stiff EoS should be discarded by the small tidal deformation, so we now have a well determined window.

It is evident that Gravitational Waves are called to play a fundamental role in neutron matter physics since they partly emanate from the interior of the neutron star, as neutrinos do. Conversely, if solid neutron matter information is at hand, Einstein's equations can be tested in a nontrivial system with intense field an matter tensor $T$ simultaneously.

If the reader has found this short review useful to introduce her to some topics in neutron star physics, we have accomplished our goal.  It only remains to recommend, for further reading and an extension to an order magnitude more topics, a recent collection of essays~\cite{Rezzolla:2018jee} where her curiosity can be further quenched.

\appendix
\section{Units and dimensions} \label{app:units}

One textbook aspect that we have found worth spelling out, given the confusion that we perceive
when nuclear or particle physicists interact with experts in gravitation and numerical relativity
is the very different perspective that both communities have on units.

In subatomic work, the use of ``natural'' units is universal, with $\hbar=1=c$. 
The elements of the stress-energy tensor ($\epsilon$, $P$, etc.) are then commonly expressed in MeV/fm$^3$ or occasionally in an appropriate fourth power of energy such as $(100 {\rm MeV})^4$ or
$m_\pi^4$. To pass between both one uses $m_\pi\simeq 138$MeV and $1{\rm fm} \simeq \frac{1}{197\rm MeV}$ ($\hbar c=1=197$ MeV fm). In this system, length and energy have opposite dimension, $[L] = [E^{-1}]= [M^{-1}]$.

At the scale of neutron stars, Planck's constant vanishes for all practical purposes, $\hbar=0$. 
The natural choice of units is then that of geometrodynamics, $G=1=c$, where Cavendish's constant is set to unity. Thus, in this system mass and length have equal dimension, $[L]=[E]=[M]$: the mass of a star can be measured in kilometers, which is practical for computer codes addressing, for example,
the $M(R)$ mass/radius diagram. Then it can be converted to solar masses ($M_\odot \simeq 1.99\times 10^{30}$kg $\simeq 1.477$ km). That is, a solar mass corresponds to about one and a half kilometers, so that the Schwarzschild radius of the sun is about $R_S= 2M_\odot\simeq 3$ km.

To convert between the two systems we can construct, for example, the quantity proportional to the energy density given by $\epsilon \frac{G}{c^4}$: in the geometrodynamic system, this directly gives $\epsilon$ expressed in $km^{-2}$ since it is an inverse length squared. Expressing $\epsilon$ in MeV/fm$^3$ and multiplying by the Cavendish constant and $c^{-4}$ as indicated, one finds the conversion factor $1$MeV/fm$^3\to 1.32\times 10^{-6}$ km$^{-2}$. 

Finally, we should comment on the not uncommon usage in astrophysics where the cgs system (!) is adopted. Thus, for example, 1$M_\odot = 2\times 10^{33}$ g. 
To convert densities to it, just use 1g/cm$^3=  5.61\times$ 10$^{-13}$ MeV/fm$^3$.

\section{Compilation of recent constraints on the neutron star radius and tidal deformability}
\label{app:tablas}

Under this header we collect, in tables~\ref{tab:NSradii} and~\ref{tab:tidaldefs}, several recent constraints on neutron star radii and tidal deformabilities, for ease of reference.
From several works \cite{Piekarewicz:2018sgy,Kalaitzis:2019dqc} we know that the effect of the crust on the second Love number is small, of order 1\%.

\begin{table}
\caption{\label{tab:NSradii} We collect various recent bounds on and estimates of the neutron star radius (in those where some sort of statistical analysis has been carried out, the interval $(r_-,r_+)$ is understood to be the 1$\sigma$ uncertainty band 
).}
\begin{center}
\begin{tabular}{|c|cccc|}\hline
NS mass ($M_\odot$)     & $r_-$(km) & $r_+$(km) & Method                   & Reference \\ \hline
1.4                     & 9.0       &  13.6     &  EFT-based nuclear EoS+CMM ($n_{tr}$=$n_0$) & \cite{Tews:2019cap}
\\
1.4                     & 11.3      &  13.6     & EFT-based nuclear EoS+MM ($n_{tr}$=$n_0$) &
\cite{Tews:2019cap}
\\
1.4                     & 9.2       &  12.5     &  EFT-based nuclear EoS+CMM ($n_{tr}$=2$n_0$) & \cite{Tews:2019cap}
\\
1.4                     & 11.3      &  12.1     & EFT-based nuclear EoS+MM ($n_{tr}$=2$n_0$) &\cite{Tews:2019cap} \\
1.4                     & 9.7       &  13.9     & EFT-based nuclear EoS & \cite{Kruger:2013kua}\\
1.4                     & 10.8      &  12.8     & EFT-based nuclear EoS+polytropes & \cite{Sammarruca:2018whh} \\
1.4                     & 11.0      &  13.2     & Multimessenger EM+GW     &\cite{Radice:2018ozg} \\
1.3                     & 11.8      &  14       & Multimessenger EM+GW     & \cite{Margalit:2017dij}\\
1.4                     & >10.5     &  >12.8    & Cooling tail X-ray burst & \cite{Nattila:2015jra} \\
1.6                     & 10.6      &  ----     & Multimessenger EM+GW     & \cite{Bauswein:2017vtn} \\
2.1?                    & 9.6       &  ----     & Multimessenger EM+GW     & \cite{Bauswein:2017vtn} \\
1.17-1.60               & 9.8       &  13.2     & Binary $\tilde{\Lambda}$+fixed chirp mass ${\mathcal M}$+no phase transitions & \cite{Raithel:2019uzi} \\
1.3-1.4                 & 10.7      & ----      & Grav. waves & \cite{Bauswein:2019skm}    \\
1.4                     & 13        &  19       & Atmospheric oscillations & \cite{Bollimpalli:2018slk} \\
2.1                     & 16        &  20       & Atmospheric oscillations & \cite{Bollimpalli:2018slk} \\
1.4                     & ----      & +1.7      & Due to star rotation     & \cite{Boshkayev:2016pmk} \\
2.0                     & ----      & +1.1      & Due to star rotation     & \cite{Boshkayev:2016pmk} \\
\hline
\end{tabular}
\end{center}
\end{table}

\begin{table}
\begin{center}
\caption{We collect several recent determinations of the tidal deformability $\Lambda$ of an individual neutron star of standard mass 1.3-1.4$M_\odot$  or of a binary system composed of two standard neutron stars, $\widetilde{\Lambda}$, see Eq.~(\ref{joint:tidal}). Generically, the possibility of a phase transition has not been incorporated into the Equation of State entering some of these calculations. \label{tab:tidaldefs}}
\begin{tabular}{|cccc|} \hline
Reference             & $\Lambda$ & $\tilde{\Lambda}$ & Comment \\ \hline
\cite{Abbott:2018wiz}  & & $70-720$            & Gravitational waves \\
\cite{Abbott:2018exr}  & & $70-580$            & Grav. waves, two NS assuming same EoS \\
\cite{Abbott:2018exr} & $190^{+390}_{-120}$   & & \\
\cite{Radice:2018ozg} & & $490^{+290}_{-160}$------ & Multimessenger: combines GW and light curve \\
\cite{Coughlin:2018fis}& & $280-820$          & Bayesian multimessenger \\
\cite{Han:2018mtj} & & $350-370$              & Sharp phase transition ($M=1.3M_\odot$) \\
\cite{Han:2018mtj} & & $160-185$              & Sharp phase transition ($M=1.0M_\odot$) \\
\hline
\end{tabular}
\end{center}
\end{table}

\subsection*{Acknowledgements}

Work supported by grants MINECO:FPA2016-75654-C2-1-P (Spain); Universidad Complutense de Madrid
under research group 910309 and the IPARCOS institute;
and the EU COST Action ``PHAROS: The multi-messenger physics and astrophysics of neutron stars'' (CA16214).

\newpage


\end{document}